\documentclass[aps,reprint,superscriptaddress,amsmath,amssymb,longbibliography]{revtex4-2}
\usepackage{graphicx}
\usepackage{dcolumn}
\usepackage{bm}
\usepackage[dvipsnames]{xcolor}
\begin{document}


\title{Prediction of chaotic dynamics and extreme events:\\ A recurrence-free quantum reservoir computing approach}

\author{Osama Ahmed}
\email{o.ahmed22@imperial.ac.uk}
\affiliation{Imperial College London, Department of Aeronautics, Exhibition Road, London SW7 2BX, United Kingdom} 

\author{Felix Tennie}
\affiliation{Imperial College London, Department of Aeronautics, Exhibition Road, London SW7 2BX, United Kingdom} 

\author{Luca Magri}%
\email{l.magri@imperial.ac.uk}

\affiliation{Imperial College London, Department of Aeronautics, Exhibition Road, London SW7 2BX, United Kingdom} 
\affiliation{The Alan Turing Institute, London NW1 2DB, UK}
\affiliation{Politecnico di Torino, DIMEAS, Corso Duca degli Abruzzi, 24 10129 Torino, Italy}

\date{\today}

\begin{abstract}
In chaotic dynamical systems, extreme events manifest in time series as unpredictable large-amplitude peaks. Although deterministic, extreme events appear seemingly randomly, which makes their forecasting difficult. By learning the dynamics from observables (data), reservoir computers can time-accurately predict extreme events and chaotic dynamics, but they may require many degrees of freedom (large reservoirs). In this paper, by exploiting quantum-computer ansätze and entanglement, we design reservoir computers with compact reservoirs and accurate prediction capabilities.  
First, we propose the recurrence-free quantum reservoir computer (RF-QRC) architecture. By developing ad hoc quantum feature maps and removing recurrent connections, the RF-QRC has quantum circuits with smaller depths. This allows the RF-QRC to scale well with higher-dimensional chaotic systems, which makes it suitable for hardware implementation. Second, we forecast the temporal chaotic dynamics and their long-term statistics of low- and higher-dimensional dynamical systems. We find that RF-QRC requires smaller reservoirs than classical reservoir computers for higher-dimensional systems and the same predictability. Third, we apply the RF-QRC to the time prediction of extreme events in a model of a turbulent shear flow with turbulent bursts. We find that the RF-QRC has longer predictability than the classical reservoir computer for extreme events forecasting. The results and analyses indicate that quantum-computer ansätze offers nonlinear expressivity and computational scalability, which are useful for forecasting chaotic dynamics and extreme events.  This work opens new opportunities for using quantum machine learning on near-term quantum computers.
\end{abstract}

\maketitle

\section{\label{sec:introduction}Introduction:\protect }
Data-driven prediction of chaotic dynamical systems has gained significant interest in the last decade \cite{cheng2015time,ghadami2022data}. A data-driven model learns the solution of a dynamical system from data with the goal of predicting its temporal evolution. Weather and climate predictions \cite{holmstrom2016machine,rolnick2022tackling}, financial time-series forecasting \cite{SEZER2020106181}, thermoacoustics \cite{huhn_stability_2020}, turbulence \cite{srinivasan_predictions_2019,doan2019physicsaware}, among many others, are examples of chaotic dynamical systems. The use of data-driven methods for predicting nonlinear systems has been motivated mainly by two reasons: (a) the availability of large amounts of data for these systems, and (b) the difficulty of predicting temporal dynamics due to chaos. In chaotic systems, a small change in the initial conditions can drastically change the dynamical system's solution.

Because chaotic dynamics is a sequential-data problem,  Recurrent Neural Networks (RNNs) are a natural choice for chaotic forecasting. RNNs are machine learning models that introduce recurrence in conventional neural network architectures. RNNs are employed in time-series forecasting, modeling nonlinear dynamical systems and chaos, and extreme event predictions \cite{vlachas_backpropagation_2020,srinivasan_predictions_2019,racca_statistical_2022,10.1007/978-3-030-22747-0_15,margazoglou_stability_2023, RACCA2021252}. On the one hand, the computational capabilities of RNNs are excellent. On the other hand, RNNs are difficult to train as they require backpropagation through time at each time step \cite{58337}. Because RNNs are designed to learn correlations using internal hidden states and long-lasting time dependencies, the training through backpropagation can be difficult for long time-series forecasting \cite{vlachas_backpropagation_2020}.

Reservoir Computing (RC) bypasses the problem of backpropagation by introducing a static recurrence---the reservoir---and training with ridge regression. This avoids backpropagation at each time step. Reservoir computing is a unified computing framework that was introduced as 'Echo State Networks' \cite{jaeger__2001} and 'Liquid State Machines' \cite{10.1162/089976602760407955}. In this study, we work with Echo State Networks, which are a type of reservoir computing approach, because of their potential in forecasting chaotic dynamics \cite{10.1007/978-3-030-22747-0_15,vlachas_backpropagation_2020,boedecker_information_2012}, extreme events predictions \cite{doan2019physicsaware,racca_data-driven_2022,doan2021short} and the stability properties \cite{margazoglou_stability_2023,huhn_stability_2020}. (For brevity, we will refer to echo state networks as reservoir computing in this work). Despite their excellent forecasting abilities, the prediction capabilities of reservoir computers are sensitive to the choice of hyperparameters and the reservoir size, which is constrained by the computational resources and memory of current classical computers.

Improving the performance of reservoir computing typically requires higher computational resources. Quantum computers hold the potential to provide exponential speed-up over classical computers for certain computational tasks \cite{feynman2018simulating}. This speed-up is also known as 'quantum advantage'. There are already algorithms that have theoretically proven this concept. For example, the prime factorization problem, which is the basis of most modern encryption and is NP-hard classically, can be solved in polynomial time using quantum computing principles \cite{365700}. The second example is the matrix inversion problem for solving linear systems of equations, which is a key step in many numerical solvers. Proposed quantum algorithms for matrix inversion have the potential to provide quadratic speed-up over conventional classical methods \cite{Harrow_2009,bravoprieto2020variational}. The underlying phenomena that enable the quantum advantage are entanglement, superposition, and interference effects in quantum computers. 
Recently, quantum computing has been applied in the area of machine learning. The current quantum hardware, however, is limited by the maximum number of qubits in the order of $\sim$ $10^2-10^3$ with environmental noise and decoherence. In the current Noisy Intermediate Scale Quantum (NISQ)-era \cite{bharti2022noisy},  hybrid quantum-classical methods are one of the leading candidates with prospects of achieving a quantum advantage by employing methods such as Variational Quantum Algorithms and Quantum Circuit Learning \cite{cerezo_variational_2021, Mitarai_2018}. The hybrid quantum-classical methods that can be used for the prediction of chaotic dynamics and extreme events are inspired by classical machine learning architectures such as RNNs. 

Quantum Reservoir Computing (QRC) is a type of Quantum Machine Learning model that combines both frameworks, including Quantum Computing and Reservoir Computing. The idea of QRC is to use quantum dynamics to enhance the reservoir implementation \cite{fujii_harnessing_2017,Fujii2021,chen2022reservoir}. Some numerical implementations of QRC have shown that the quantum systems of 5-7 qubits have comparable computational capabilities to the classical reservoir sizes of 100-500 \cite{fujii_harnessing_2017}. Previous works have proposed hybrid QRC architectures for time series predictions and chaos modeling \cite{pfeffer_hybrid_2022,suzuki_natural_2022,fry2023optimizing,wudarski2023hybrid}. In hybrid QRC, the input data is encoded in the form of Bloch-sphere rotation angles in individual qubits along with reservoir states, which are the probability amplitudes. The resulting quantum states undergo a unitary operation at each time step \cite{pfeffer_hybrid_2022}. In another approach, a reservoir state is associated with the density operator of the encoded quantum state, and reservoir states correspond to measured expectation values for each qubit \cite{fry2023optimizing, dudas_quantum_2023}. Although it has been conjectured that quantum reservoir computers are potential candidates for providing a quantum advantage in the near-term NISQ era, only a few studies have been conducted on analyzing higher-dimensional chaotic systems \cite{tran_higher-order_2020}, and long-term statistical predictions using QRC. The {performance of classical reservoir computers is often limited by classical computers' memory \cite{vlachas_backpropagation_2020}, which constrains the predictability of these models. Reservoir computers require a high dimensional feature space, which increases with the nonlinearity and the dimensions of  the physical system.} Consequently, using QRC to predict high dimensional nonlinear dynamical systems that require a large reservoir size is key to realizing a quantum advantage of QRC in predicting chaos.

In this work, we build up on the concept of hybrid QRC that encodes reservoir states as probability amplitudes of the quantum system \cite{pfeffer_hybrid_2022}. We use gate-based quantum computing to implement this architecture and use it for high-dimensional chaotic time-series forecasting and extreme-event predictions. More specifically, we investigate the three-dimensional Lorenz-63, ten-dimensional Lorenz-96 systems \cite{75462} and the MFE (Moehlis, Faisst, and Eckhardt) \cite{moehlis_low-dimensional_2004} model for time-accurate, statistical predictions, and extreme events forecasting. Extreme events are sudden and unmitigated changes in the observables of chaotic flows, while long-term statistical prediction is also an important metric in forecasting chaos. 

We benchmark our classical and quantum reservoir networks by studying them for low-order models and then extend our analysis to incorporate higher-dimensional dynamics. Both classical and quantum networks are used for extreme event predictions in a reduced order model of a shear flow between infinite plates subjected to a sinusoidal body forcing, also referred to as 'Moehlis, Faisst, and Eckhardt (MFE)' \cite{moehlis_low-dimensional_2004}. We assess the performance using multiple performance metrics: first, by considering short-term time series prediction capabilities that are quantified using the Predictability Horizon (PH) and Valid Prediction Time (VPT) \cite{vlachas_backpropagation_2020, BOFFETTA2002367}, and, second, by comparison of long-term statistics that are quantified using statistical measures such as Probability Density Function (PDF) and F-Score \cite{margazoglou_stability_2023, RACCA2021252}. 

This paper is structured as follows. In Sec.~\ref{sec:methods}, we present the classical and proposed quantum reservoir architectures and their comparison. Different quantum reservoir architectures refer to the different ansätze chosen for forming a reservoir. Second, we apply both classical and quantum reservoir architectures on three-dimensional Lorenz-63 and ten-dimensional Lorenz-96 systems. We increase the dimensionality and complexity of the dynamical systems progressively for the analysis. Third, we compare the results of the best quantum feature map with the best classical reservoir applied to the MFE model to analyze the extreme-event prediction capabilities of both architectures. The numerical models of different dynamical systems and the results for the reservoir predictions are presented in Sec.~\ref{sec:results}. Finally, in Sec.~\ref{sec:conclusion}, we conclude our findings by highlighting the potential of QRC in chaotic time series forecasting along with its associated challenges and the direction of future works.

\section{\label{sec:methods}Background:\protect }

In this section, we explain and compare classical and quantum reservoir computing algorithms and their schematic representations. To be self-contained, we also outline the basic mathematical framework and building blocks of quantum computers.

\subsection{\label{sec:CRCM}Classical Reservoir Computing (CRC)}

Unlike other RNNs, reservoir computing is a type of RNN framework that does not require backpropagation through time, which represents a major computational advantage \cite{jaeger__2001, RACCA2021252}. As a result, the training cost is substantially reduced and the objective function is minimized through a simple linear ridge regression. 
\begin{figure}[!ht]
\includegraphics[width=1.0\linewidth]{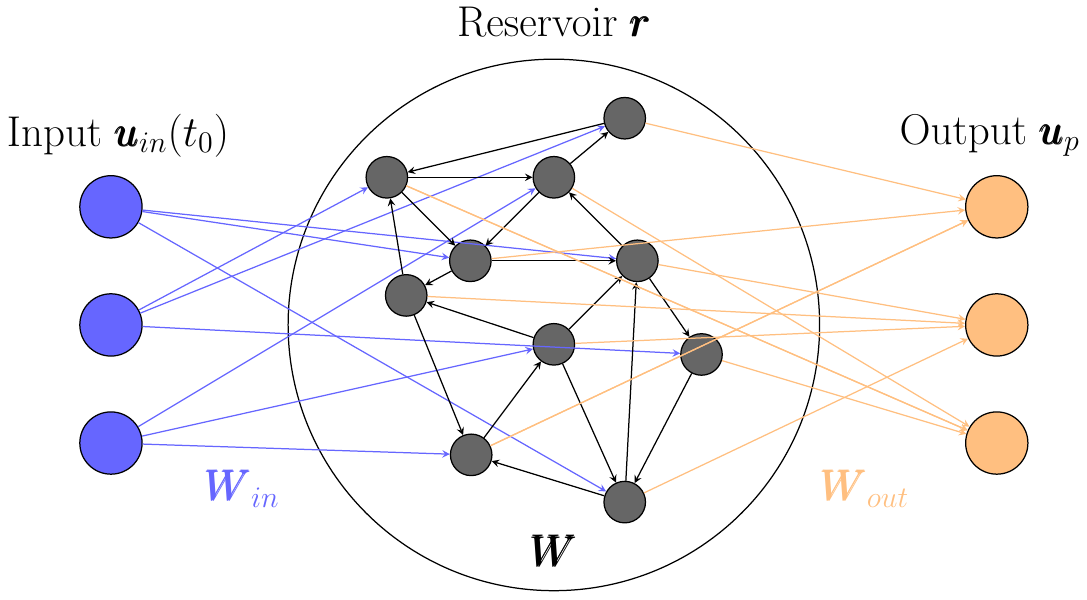}%
\caption{\label{fig:C_Reservoir}Schematic representation of a classical reservoir computer \cite{jaeger__2001}. The input data $\pmb{u}_{in}$ is mapped to the reservoir matrix via $\pmb{W}_{in}$. The reservoir neuron connections governed by $\pmb{W}$ matrix allow the flow of information between neurons. The linear readout layer using the trained $\pmb{W}_{out}$ matrix is used to make output predictions $\pmb{u}_{p}$.}
\end{figure}
\begin{figure}[!t]
\includegraphics[width=0.7\linewidth]{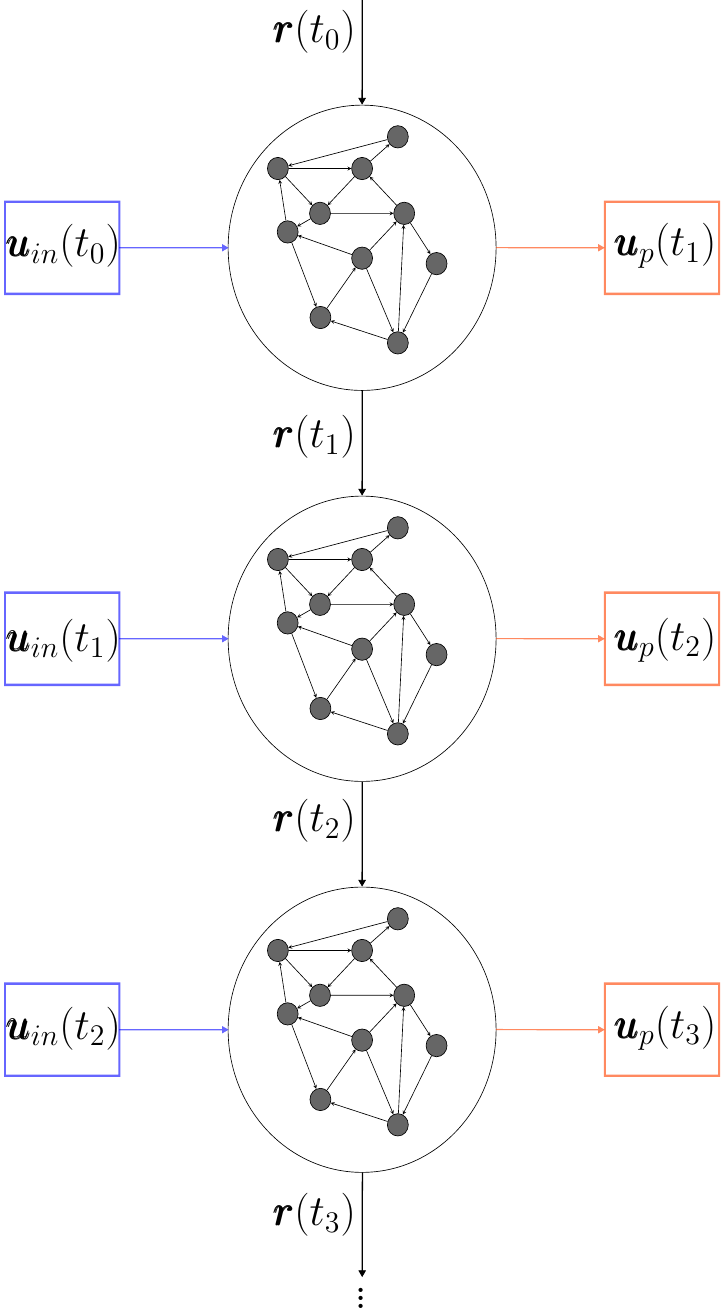}%
\caption{\label{fig:OpenClosed_1}Schematic representation of reservoir computers. Open-loop training phase.}
\end{figure}
\begin{figure}[!t]
\includegraphics[width=0.7\linewidth]{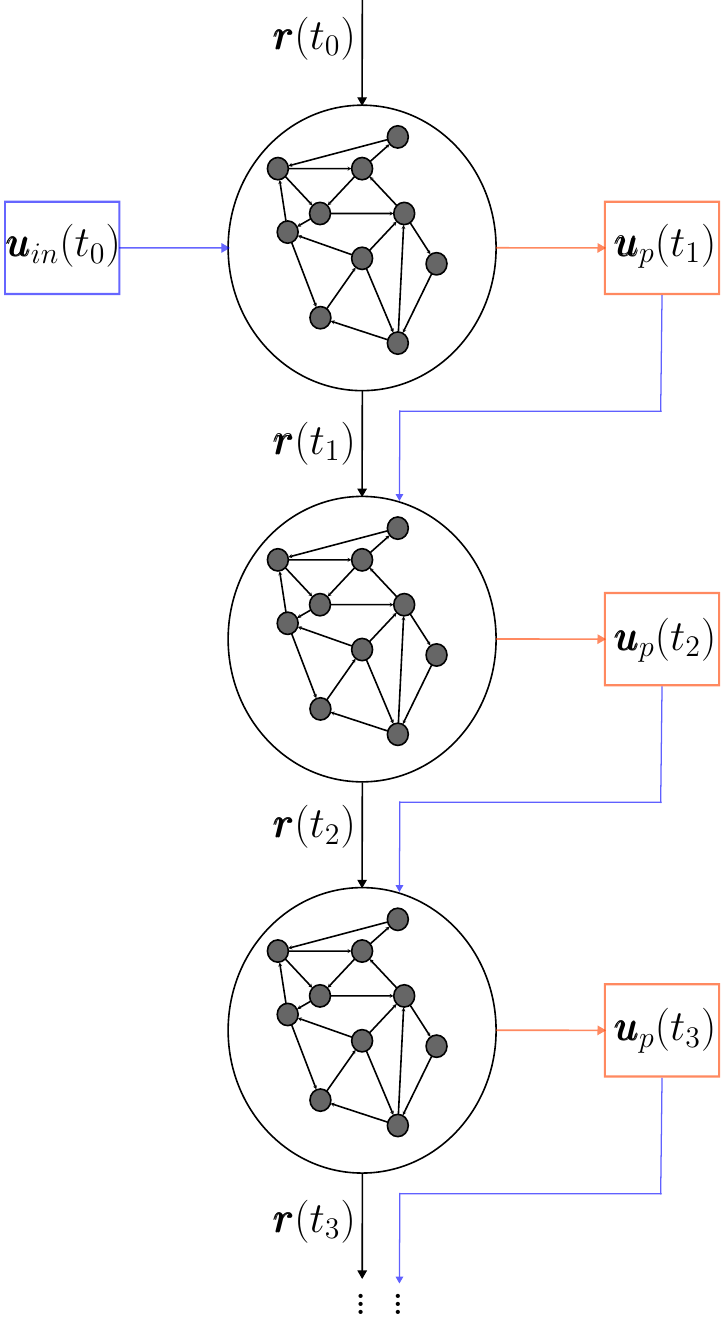}%
\caption{\label{fig:OpenClosed}Schematic representation of reservoir computers. Closed-loop prediction phase.}
\end{figure}
\begin{figure*}
\includegraphics[width=1.0\linewidth]{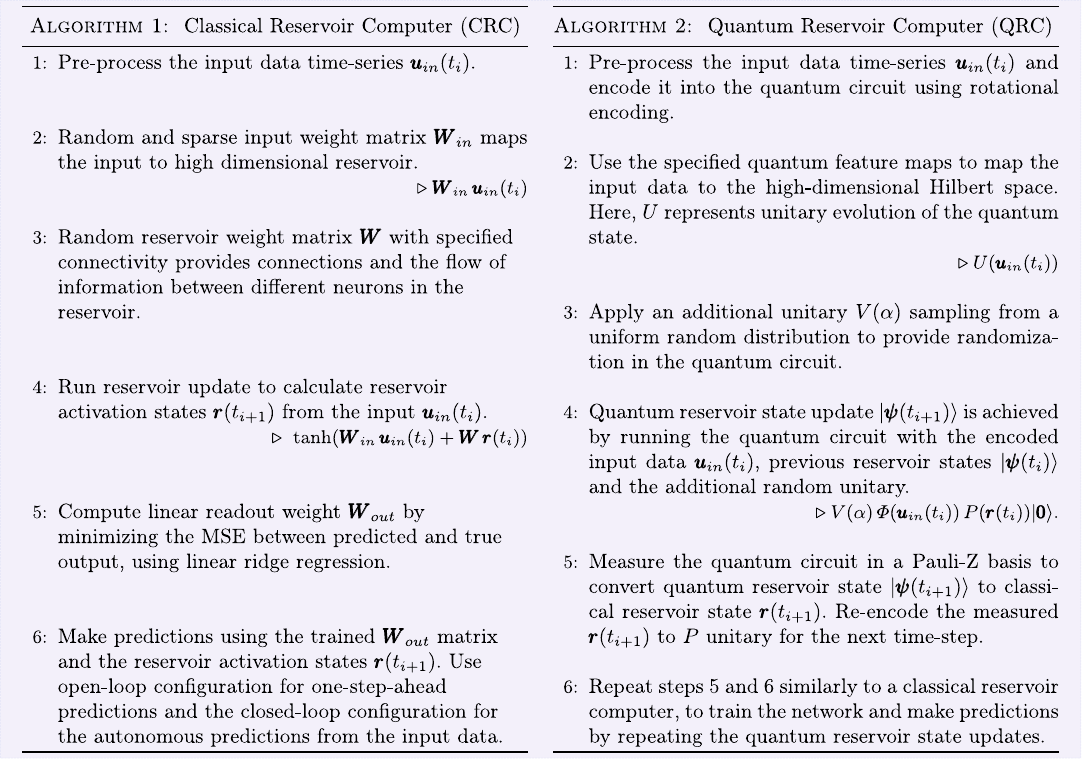}%
\caption{\label{fig:algorithm}Classical and Quantum Reservoir Computing algorithms.}
\end{figure*}

Let $\{\boldsymbol{\hat{u}}(t_0),\boldsymbol{\hat{u}}(t_1),\ldots, \boldsymbol{\hat{u}}(t_{N_{tr}})\}$ be a training data set of a dynamical state $\hat{\pmb{u}} \, \epsilon \,\, \mathbb{R}^{{N}_{u}} $ , which is known at $N_{tr}+1$ steps in time. Generally, it is recommended to choose a set of variables that is (re-)scaled by the range component-wise, which is indicated by ( $\hat{}$ ) on the training data set \cite{lukosevicius_practical_2012}. In reservoir computing, these states are mapped onto higher-dimensional reservoir states of an $N_r$-dimensional vector space. For a randomly chosen initial reservoir state $\boldsymbol{r}(t_0)$, we recursively compute a set of $N_{tr}$ reservoir vectors $\pmb{r}(t_1),\ldots,\pmb{r}(t_{N_{tr}})$ as
\begin{align}
\pmb{\hat{r}}(t_{i+1}) &= \tanh \,(\pmb{W}_{in}  \pmb{\hat{u}}_{in}(t_{i})+\pmb{W} \pmb{r}(t_{i})), \label{eq:1} \\
\pmb{r}(t_{i+1}) &= (1-\varepsilon) \, \pmb{r}(t_{i}) + \varepsilon \, \pmb{\hat{r}}(t_{i+1}).\label{eq:2}
\end{align}
Here, the state $\hat{\pmb{r}} \, \epsilon \,\, \mathbb{R}^{{N}_{r}} $  represent a state-vector containing reservoir activation states, $\pmb{W}_{in} \, \epsilon \,\, \mathbb{R}^{{N}_{r} \, \times {N}_{u}} $ is the input matrix and $\pmb{W} \, \epsilon \,\, \mathbb{R}^{{N}_{r} \, \times {N}_{r}}$ is the reservoir weight matrix \cite{jaeger__2001}. The $\tanh$ function is applied component-wise to provide a nonlinear activation. The matrices $\pmb{W}_{in}$ and $\pmb{W}$ are pseudo-randomly generated and constant throughout the training and prediction phase. The elements of $\pmb{W}_{in}$ are sampled from a uniform distribution in $[-{\sigma_{in}},{\sigma_{in}}]$, where ${\sigma_{in}}$ is the input-scaling. The reservoir weight matrix $\pmb{W}$ is an Erdös-Rényi matrix (Erdös-Rényi is a model for generating random graphs or the evolution of a random network in the field of graph theory), with average connectivity between each neuron varied by input reservoir density $D$. The spectral radius of the reservoir state matrix $\rho$ is given by its maximal eigenvalue. The matrix $\boldsymbol{W}$ is rescaled such that $\rho\leq 1$ \cite{lukosevicius_practical_2012,jaeger__2001}. Effectively, this guarantees a rapid decay of temporal correlations between successive reservoir states and helps to avoid overfitting. Whereas Eq.~\eqref{eq:1} is the nonlinear activation step, Eq.~\eqref{eq:2} combines the linear memory of the reservoir with the nonlinear activation, as parametrized by the leaking rate $\varepsilon$. This additional post-processing step is known as memory non-linearity tradeoff \cite{lukosevicius_practical_2012}. 

Reservoir computing networks can be run in either open-loop or closed-loop configurations (Figs.~\ref{fig:OpenClosed_1} and \ref{fig:OpenClosed}). In the open-loop, we use the input data at each time step and compute the corresponding reservoir dynamics $\pmb{r}(t_{i})$ according to Eqs.~\eqref{eq:1} and \eqref{eq:2}. Because of the initialization process with a randomly chosen reservoir state, it is necessary to discard a small number $N_w$ of the initial transients to satisfy the Echo State Property (ESP) \cite{jaeger__2001}, which leads to dynamics that are less sensitive to the initial random choice. The process of discarding the initial transients is also called the washout phase. After the washout interval, reservoir dynamics at each time step $\pmb{r}(t_{i})$ are collected to form a reservoir state matrix $\pmb{R} \,\, \epsilon \,\, \mathbb{R}^{{N}_{r} \, \times (N_{tr} -N_W)}$. 
Reservoir computing in the closed-loop prediction mode requires an output matrix $\pmb{{W}}_{out}$ for forecasting the dynamics of the learned chaotic system.
The training of the output matrix $\pmb{{W}}_{out}$ involves minimizing the mean square error between input and output data over the training data set, consisting of $N_{tr}$ number of training steps. The simplicity of the reservoir network allows us to achieve this by solving a linear ridge regression problem
\begin{equation}\label{eq:3}
(\pmb{R}\pmb{R}^{T}+\beta\pmb{I}) \, \pmb{W}_{out} = \pmb{R} \, \pmb{U}^{{\textnormal T} }_{d},
\end{equation}
where $\beta$ is a user-defined Tikhonov regularization parameter to prevent overfitting, $\pmb{I}$ is the identity matrix, and $\pmb{{U}}^{{\textnormal T}}_{d} \,\, \epsilon \,\, \mathbb{R}^{{N}_{u} \, \times {N}_{tr}} $ is the horizontal concatenation of the output data. Once the output matrix has been computed from Eq.~\eqref{eq:3}, it can be used to predict the evolution of dynamical variables by 
\begin{equation}\label{eq:4}
\pmb{u}_{p}(t_{i+1}) = [\pmb{r}(t_{i+1})]^{{\textnormal T} } \, \pmb{W}_{out}.
\end{equation}
In the closed-loop configuration, we recursively predict the system's dynamics from Eq.~\eqref{eq:4} without additional samples from a training data set. Given an initial condition, this allows for an autonomous prediction on the unseen data set. 
Because reservoir computers have a symmetry (Eq.~\eqref{eq:1}), we add a constant output bias of 1 to the reservoir activation states to break the inherent symmetry of the reservoir architecture \cite{10.1007/978-3-030-50433-5_10,herteux2020breaking}, effectively replacing $\boldsymbol{r} \rightarrow [\boldsymbol{r},1]$.
A summary of the reservoir computing procedure is shown in  Fig.~\ref{fig:algorithm}, and compared with its quantum counterpart, which will be introduced in Sec.~\ref{sec:QRCM}. The performance of reservoir computers critically depends on the set of chosen hyper-parameters. We, therefore, use grid search and Bayesian optimization to tune the hyperparameters \cite{snoek2012practical}. We have used the $scikit-learn$ library \cite{scikit-learn} and recycle validation techniques \cite{RACCA2021252} for hyperparameter tuning. The hyperparameters and performance metrics for different chaotic systems are reported in Sec.~\ref{sec:results}.

\subsection{\label{sec:qubit}Qubits and quantum states}

To be self-contained, we provide a brief outline of the key elements of quantum computers. A detailed introduction can be found in Ref.~\cite{9781107002173}.
In quantum computers, the internal state of the machine can be represented by a complex vector in a Hilbert space. The time evolution during computation is described by a matrix multiplication of this vector with a unitary operator. 

In most quantum hardware architectures, the elementary building blocks, known as qubits, are microscopic systems that can be described by a complex-valued, normalized, two-dimensional {vector in a Hilbert space,} in which the computational basis states $|0\rangle$ and $|1\rangle$ (kets) can be associated with the Cartesian basis vectors, and the complex-valued amplitudes $\alpha$ and $\beta$ are normalized as $|\alpha|^2 + |\beta|^2 = 1$
\begin{align}\label{eq:5a}
   |  \psi  \rangle \ = \alpha |0\rangle + \beta |1\rangle \hat{=} \begin{pmatrix}
           \alpha \\
           \beta \\
         \end{pmatrix} \in \mathbb{C}_2.
\end{align}
States of $n$ qubits occupy the tensor product Hilbert space of the individual two-dimensional Hilbert spaces
\begin{align}\label{eq:5}
   |  \psi_n \rangle \ &=  \begin{pmatrix}
           a_{1} \\
           a_{2} \\
           \vdots \\
           a_{2^n}
         \end{pmatrix} \in \, \mathbb{C}_2^{\otimes n},
\end{align}
where the computational basis states are tensor products of the single qubit basis states. The state-vector dimension scales exponentially with the number of qubits $n$. Unentangled pure states can be written as a single product of individual qubit states
\begin{widetext}
\begin{equation}\label{eq:6}
   | \phi_{n} \rangle = \, (\alpha_0 |0_0\rangle + \beta_0 |1_0\rangle) \otimes (\alpha_1 |0_1\rangle + \beta_1 |1_1\rangle) \otimes\ldots \otimes (\alpha_n |0_n\rangle + \beta_n |1_n\rangle).
\end{equation}
\end{widetext}
Most quantum states [Eq.~\eqref{eq:5}] are entangled and therefore cannot be decomposed into such a product. 
Unitary evolution amounts to a matrix multiplication, i.e.~the final quantum state after a computation is
\begin{equation}
    |\psi_f\rangle = U |\psi_i\rangle .
\end{equation}
Quantum hardware architectures provide elementary sets of unitary operators, known as gates that can be used to assemble any arbitrary desired unitary operation. 
The final output of the computation is revealed by measurements of the quantum state, which provide estimates of the modular squared amplitudes $(|a_1|^2, |a_2|^2, \ldots)$.

\subsection{Hybrid quantum-classical reservoir computing}\label{sec:QRCM}

Hybrid quantum-classical reservoir computing \cite{pfeffer_hybrid_2022} is based on principles that are similar to classical reservoir computing introduced in Sec.~\ref{sec:CRCM}. In QRC, the exponential size of $n$-qubit Hilbert spaces is utilized to encode and process classical reservoir state vectors, $\boldsymbol{r}(t_i)$. Thus, the classical reservoir is replaced by a quantum counterpart.
More specifically, within a single loop of the training phase, the reservoir state $\boldsymbol{r}(t_i)$ is mapped onto an $n$-qubit unitary operator $P(\boldsymbol{r}(t_i))$, that takes $|0\rangle^{\otimes n}$ to the state $|\psi(t_i) \rangle = P(\boldsymbol{r}(t_i))|0\rangle^{\otimes n}$. The explicit form of $P$ is given by the quantum circuit architecture that characterizes the specific QRC framework. Circuits are generally parametrized by single-qubit rotation angles and a number of entangling CNOT gates (Appendix \ref{sec:ansatz}).

\begin{figure*}
    \centering
    \includegraphics[width=0.8\linewidth]{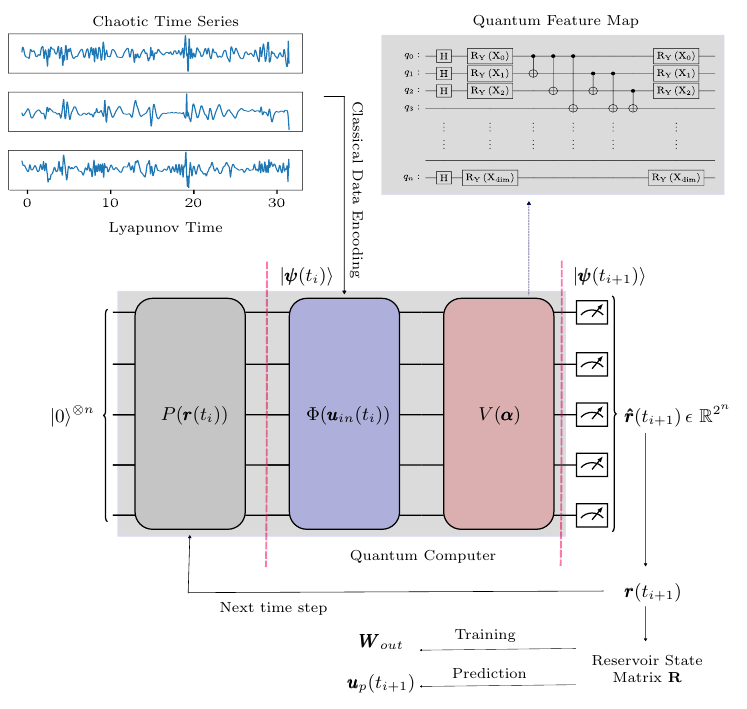}
    \caption{Schematic representation of gate-based quantum reservoir computer. The first two unitaries are used to encode previous reservoir states $\pmb{r}(t_{i})$ and input time series $\pmb{u}_{in}(t_{i})$, the third additional unitary provide randomization. Each unitary consists of a specific feature map. The quantum circuit execution starts from the ground state $ | \pmb{0} \rangle $ and measures the reservoir state activation for the next time step $\hat{\pmb{r}}(t_{i+1})$, which is post-processed classically before feeding the next time-step. Each reservoir activation state $\pmb{r}(t_{i+1})$ is saved classically to use for training and predictions. In RF-QRC, we remove the feedback loop and the first unitary ${\pmb{P}}$. }
    \label{fig:QESN_prob}
\end{figure*}

For a training data set $\{\boldsymbol{\hat{u}}(t_0),\boldsymbol{\hat{u}}(t_1),\ldots, \boldsymbol{\hat{u}}(t_{N_{tr}})\}$, the input time-series is encoded in the quantum circuit using single-qubit rotation angles. With this encoding, a second unitary operator $\Phi(\boldsymbol{u_{in}}(t_i))$ is applied  to the quantum state $|\psi(t_i) \rangle$. Thereafter, a third unitary operator $V(\alpha)$ given by a random parameterized circuit with $n$ parameters $\alpha$ is applied. The combined action for each time step is in Fig.~\ref{fig:QESN_prob} 
\begin{equation}\label{eq:11}
   {|\psi(t_{i+1}) \rangle } =  V(\alpha) \, \Phi(\pmb{u}_{in}(t_{i})) \, P(\pmb{r}(t_{i})) | 0 \rangle^{\otimes n}.
\end{equation}
Similarly to classical reservoir computing, we derive a pre-processed reservoir state vector $\hat{\boldsymbol{r}}(t_{i+1})$ from measured probabilities of $|\psi(t_{i+1})\rangle$ in the computational basis. This requires sampling and measuring the state $|\psi(t_{i+1})\rangle$ multiple times for each time stepping. 

Following the post-processing of $\hat{\boldsymbol{r}}(t_{i+1})$ to $\boldsymbol{r}(t_{i+1})$ [Eq.~\eqref{eq:2}], the reservoir state matrix $\pmb{R}$ is formed by concatenating the reservoir state vectors. As in classical reservoir computing, the initial reservoir state is chosen randomly, which in turn, necessitates the washout of a few initial time-stepped reservoir vectors. 
\begin{figure*}
    \centering
    \includegraphics[width=0.8\linewidth]{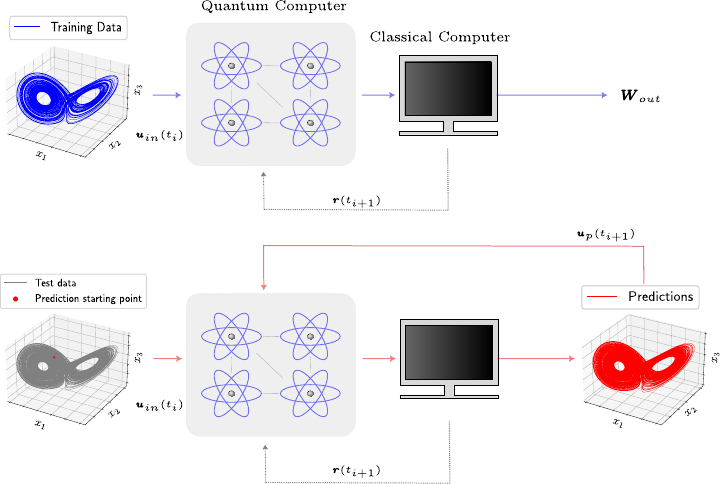}
    \caption{Quantum-classical reservoir architecture schematic representation (a) open-loop training to produce $\pmb{W}_{out}$ matrix (b) closed-loop autonomous predictions starting from an arbitrary point in the unseen data set. The dotted line in both figures indicates the recurrence involved in conventional reservoir architectures. In RF-QRC architecture, we remove this recurrent feedback layer of reservoir states.}
    \label{fig:QESN_schemm}
\end{figure*}
In the training phase, QRC is run in an open-loop and the output matrix $\boldsymbol{W}_{out}$ is computed by a linear ridge regression [Eq.~\eqref{eq:3}]. When operated in a closed-loop, QRC is used to predict the system dynamics [Eq.~\eqref{eq:4}].
We compare the elementary steps of classical and quantum reservoir computing algorithms in Fig.~\ref{fig:algorithm}. The open-loop and closed-loop configurations of the hybrid quantum-classical architecture are presented in Fig.~\ref{fig:QESN_schemm}. Each reservoir update $\pmb{r}({t}_{i})$ in both the open-loop and closed-loop requires the input of the previous reservoir state $\pmb{r}({t}_{i-1})$, similarly to the classical reservoir update (dotted feedback loop in Fig.~\ref{fig:QESN_schemm}). In Sec.~\ref{sec:proposed}, we propose a quantum-classical architecture, which is independent of this feedback loop, thereby allowing for parallelization of the training phase of QRC. 

Although QRC exploits the exponential size of the state space of the quantum register, there is a practical limitation due to the limited connectivity of the employed quantum processors (e.g.~neighbour-neighbour interactions only on 2D superconducting qubit processors). The performance of QRC generally depends on the degree of the entanglement present in the circuit \cite{PhysRevA.108.052427}. Therefore, we investigate the performance of structurally different architectures. 

\section{\label{sec:proposed}Recurrence-Free Quantum Reservoir Computing (RF-QRC):\protect }

\begin{table*}
\caption{Quantum reservoir computer - Different ans\"{a}tze designs. }
\begin{ruledtabular}
\begin{tabular}{llll}
&Reservoir states ($P$)&\mbox{Input ($\Phi$)}&\mbox{Variation ($V$)}\\
\hline
QRC-C1 & Linearly entangled & \mbox{Fully entangled} & \mbox{Fully entangled symmetric} \\
QRC-C2 & Linearly entangled & \mbox{Linearly entangled} & \mbox{Linearly entangled} \\
QRC-C3 & - &\mbox{Linearly entangled (x2)} & \mbox{Linearly entangled} \\
QRC-C4 (RF-QRC) & - & \mbox{Fully entangled (x2)} & \mbox{Fully entangled symmetric} \\
QRC-C5 & - & \mbox{Product states (x2)} & \mbox{Linearly entangled} \\
\end{tabular}
\label{tab:table_ansatz}
\end{ruledtabular}
\end{table*}

The gate-based quantum reservoir computing framework of \cite{pfeffer_hybrid_2022} (Sec.~\ref{sec:QRCM}) is in principle equivalent to its classical counterpart. It has been used to study thermal convection flows and the turbulent Rayleigh-Bernard flow \cite{pfeffer_hybrid_2022,pfeffer_reduced-order_2023}. In this section, we propose an ans\"{a}tze for the circuit architecture in QRC. To do so, a specific quantum feature map must be chosen and we therefore start by briefly reviewing a number of common choices. Following that, we present the RF-QRC architecture, which will be employed to study low and high-dimensional chaotic flows in Sec.~\ref{sec:results}. 

We consider four different feature maps for QRC. These are the (a) Linearly {entangling} feature map, used in Ref.~\cite{pfeffer_hybrid_2022} for QRC applications, (b) Product {state} feature map, which was proposed in Ref.~\cite{havlicek_supervised_2019}, and has been shown to offer a high expressivity \cite{abbas_power_2021}, (c) Fully {entangling} feature map, in which CNOT gates entangle pairs of qubits, (d) Symmetric {fully entangling} feature map, which only differs from (c) by an additional data encoding layer following the CNOT gates. These feature maps do not require anywhere-to-anywhere connectivity (which would increase the computational overhead on current superconducting quantum processors). Schematic circuit diagrams of all four feature maps are shown in Appendix \ref{sec:ansatz}. When the number $\Theta$ of encoded parameters of a string of data (e.g.~for $\boldsymbol{r}$, $\Theta = 2^n$), exceeds the number of qubits, the feature map encoding is applied multiple times. In particular, this leads to exponentially growing circuit depths of the unitary $P$ and motivates the exploration of new efficient QRC architectures, as displayed in Tab.~\ref{tab:table_ansatz}.

\begin{figure}
    \centering
    \includegraphics[width=1.0\linewidth]{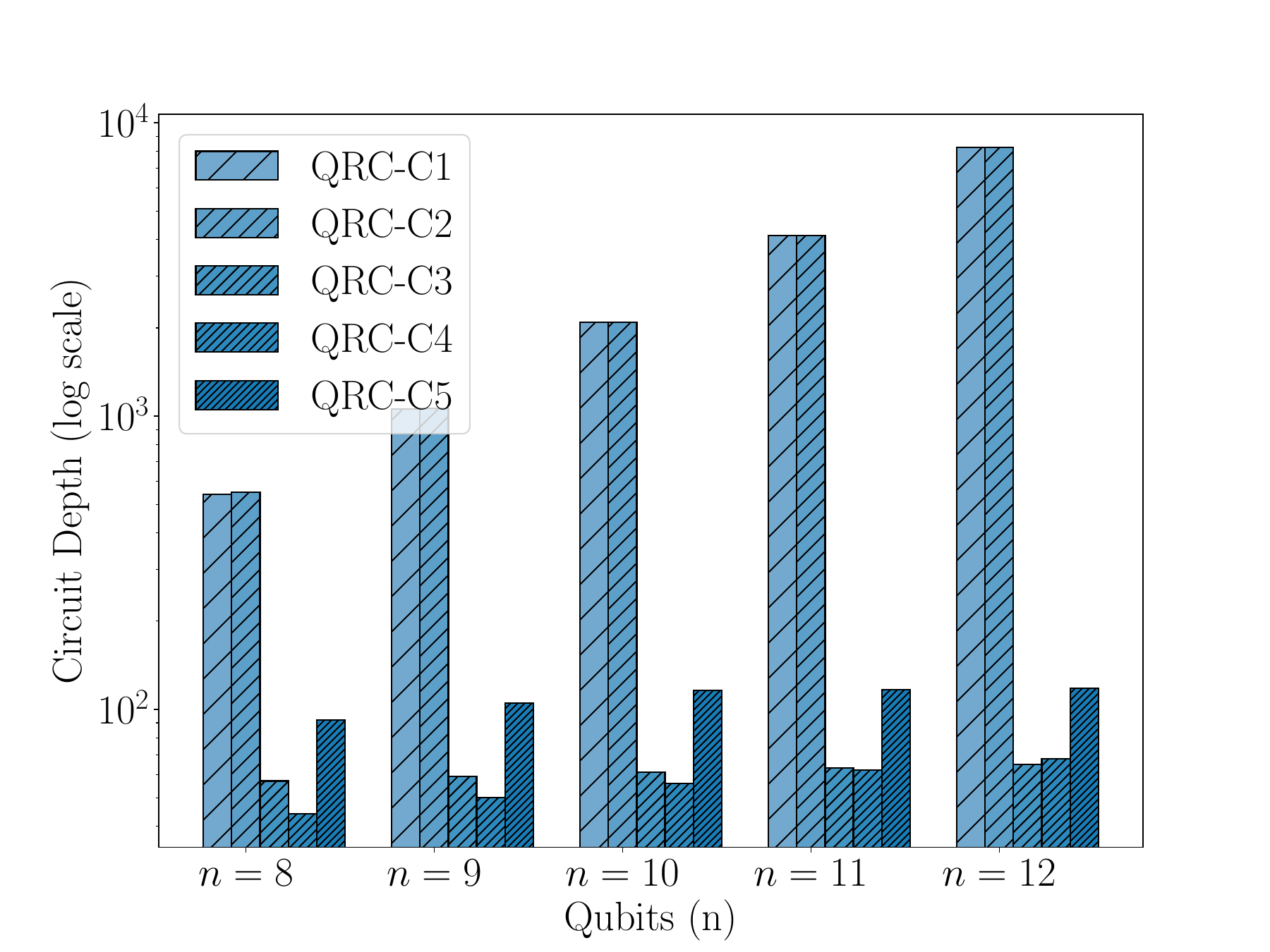}
    \caption{Circuit depth of different ans\"{a}tze for the ten-dimensional Lorenz-96 input.}
    \label{fig:L96_CircuitDepth}
\end{figure}

An example comparison of circuit depths corresponding to the five architectures is shown in Fig.~\ref{fig:L96_CircuitDepth}. Removing the recurrence, i.e. by setting $P$ to the identity, and by applying $\Phi$ twice for stimulating the reservoir dynamics, yield circuits depths that are independent of the reservoir size. The suitability of this choice is justified \textit{a posteriori} by an improved performance of QRC in a range of prototypical model systems. Recurrence-free QRC is different than Quantum Extreme Learning Machines (QELMs) \cite{mujal_opportunities_2021}, because of the use of a linear combination of the memory and nonlinearity of previous reservoir activation states, as expressed by Eq.~\eqref{eq:2} (also known as the leaky-integral reservoir computing approach \cite{JAEGER2007335}).

\section{Numerical Results:}\label{sec:results}

In this section, we analyze the dynamics of a number of chaotic systems, such as Lorenz-63, Lorenz-96, and MFE using CRC and QRC. Thereby, we can assess the prediction capabilities of various architectures. More specifically, we evaluate the short-term time-accurate and long-term statistical predictions with performance measures. The sets of training, validation, and test data are generated by a $4^{th}$-order Runge-Kutta numerical scheme. To emulate the quantum circuits, we have used the \textit{Qiskit} \cite{Qiskit} software package to compute the noise-free evolution of the quantum register.

 {In a physical implementation, at each time step between $10^{4}$ to $10^{5}$ shots are required to reconstruct the reservoir activation signal. A number of shots between $10^{4}$ to $10^{6}$ is common in variational algorithms and other QML applications \cite{mujal_time-series_2023}. Mitigating the requirement of a large number of shots in quantum computing is an area of active research, which is beyond the scope of this paper. (QRC and RF-QRC could be combined with the concept of weak and projective measurements \cite{mujal_time-series_2023}, classical shadows \cite{huang2020predicting}, Resolvable Expressive Capacity (REC) \cite{hu2023tackling}, or artificial memory restriction \cite{vcindrak2024enhancing} to potentially reduce the number of required shots.)}

\subsection{Performance measures}\label{sec:metrics}

The dynamics of chaotic systems can be characterized by the leading Lyapunov {exponent} $\Lambda_{1}$ \cite{BOFFETTA2002367}. This exponent corresponds to the average exponential rate of divergence for initially nearby trajectories. The Lyapunov {exponent} also provides a time scale to assess the time-accurate prediction of the chaotic systems. We have re-scaled our time units to the inverse of the Lyapunov {exponent} ($\Lambda_{1}$), which is called the Lyapunov Time ($1 LT = \Lambda_{1}^{-1}$). 

In this work, we choose the Valid Prediction Time (VPT) \cite{vlachas_backpropagation_2020} as a performance measure of short-term time-accurate predictions. For a given threshold value $\varepsilon$, VPT is defined as the time for which the Normalized Root Mean Square Error (NRMSE) between the predicted and true values is less than $\varepsilon$. In this work we take $\varepsilon$ = 0.5, as in Ref.~\cite{vlachas_backpropagation_2020}
\begin{equation}\label{eq:16}
    NRMSE = \sqrt{||{\frac{\pmb{y}_{t}-\pmb{\hat{y}}_{t}}{\sigma^{2}}}||},
\end{equation}
\begin{equation}\label{eq:17}
    VPT = \frac{1}{\Lambda_{1}} \, \text{argmax} \, [\,{t_{f} |\,\, NRMSE < \, \varepsilon=0.5 , \,\, \forall t \leq t_{f}}\,],
\end{equation}
where $\pmb{y}_{t}$ is the true value and $\pmb{\hat{y}}_{t}$ is the corresponding predicted value, and $\sigma$ is the standard deviation of the time series, $t_{f}$ is the largest value of the time step at which NRMSE is smaller than the threshold $\varepsilon$. High VPT values indicate a high predictability. To assess the long-term statistical prediction capabilities of our reservoir networks, we evolve the trained networks autonomously in a closed-loop configuration from an arbitrary point on the attractor and compare the state variable distributions with the actual distributions. For each configuration, we run simulations with five different sets of optimal hyperparameters, which corresponds to different random seeds. 

\subsection{Three-dimensional Lorenz-63 model}\label{sec:LOR63}
As a first test case, we start with the analysis of the three-dimensional Lorenz-63 system \cite{75462}. The Lorenz-63 system is a reduced-order model of a thermal convection flow, in which the fluid is heated uniformly from below and cooled from the top. This model is defined mathematically by
\begin{equation}\label{eq:3A}
    \frac{dx_{1}}{dt} = \sigma \, (x_{2}-x_{1}), \, \frac{dx_{2}}{dt} = x_{1} \, (\rho-x_{3}) - x_{2}, \, \frac{dx_{3}}{dt} =  x_{1}x_{2}-\beta{x_{3}},
\end{equation}
where [$\sigma$ , $\rho$ , $\beta$] = [10, 28, 8/3] results in a chaotic behavior of the system. For the Lorenz-63 system, $\Lambda=0.9$ and $1 LT = 1/0.9$, \cite{75462}. The time series data set is derived by a Runge-Kutta method for $dt=0.01s$. The training time series comprises data points over a total time of 20 LT for both the classical and quantum reservoir networks. For the training, both networks are executed in an open loop to evolve the reservoir states and calculate the $\pmb{W}_{out}$ matrix. For predictions, both classical and quantum networks evolve dynamically in the closed-loop configuration from ensembles of points in phase space, sampled on the attractor randomly, to quantify the time accuracy. These points correspond to the points in the unseen test data set. 

In Tab.~\ref{tab: param_LOR63}, we compare the hyperparameters of both quantum and classical reservoirs. Values inside square brackets indicate that the parameters are optimized within this range, multiple values indicate that we repeated the hyperparameter search for each of these values and then selected the best VPT out of them. The random rotation angles ($\alpha$) in the third unitary block can also be treated as a hyperparameter. Alternatively, one could derive them from a uniform random distribution, which we shall do in this paper. {For each ensemble, we sample $V(\alpha) \, \, \epsilon \,\, \mathbb{R}^{n}$ from a uniform distribution interval $[0,4\pi]$ with a predefined seed, which we keep it fixed throughout the training and prediction of a particular realization. This is similar to the classical input $\pmb{W}_{in}$ and reservoir weight matrices $\pmb{W}$, which are also pseudo-randomly generated and fixed for any specific realization, as explained in Sec.~\ref{sec:CRCM}.}

\begin{table*}
\caption{Parameters for three-dimensional Lorenz-63 system}
\begin{ruledtabular}
\begin{tabular}{llll}
Parameters & Symbol & Classical RC & Quantum RC \\ \hline
Time step & $dt$ & 0.01s & 0.01s \\
Input scaling & $\sigma_{in}$ & [0 , 1] & - \\
Spectral radius & $\rho$ & [0.1 , 1] & 1 \\
Tikhonov regularization & $\beta$ & 1$\times10^{-6}$,1$\times10^{-9}$,1$\times10^{-12}$ & 1$\times10^{-6}$,1$\times10^{-9}$,1$\times10^{-12}$ \\
Leak rate & $\epsilon$ & [0.05, 1] & [0.05, 0.3] \\ 
Resevoir density & $D$ & 0.1,0.6,0.9 & Configurations C1-C5 \\ 
\end{tabular}
\label{tab: param_LOR63}
\end{ruledtabular}
\end{table*}

\subsubsection{Time-accurate predictions}

\begin{figure}
    \includegraphics[width=0.98\linewidth]{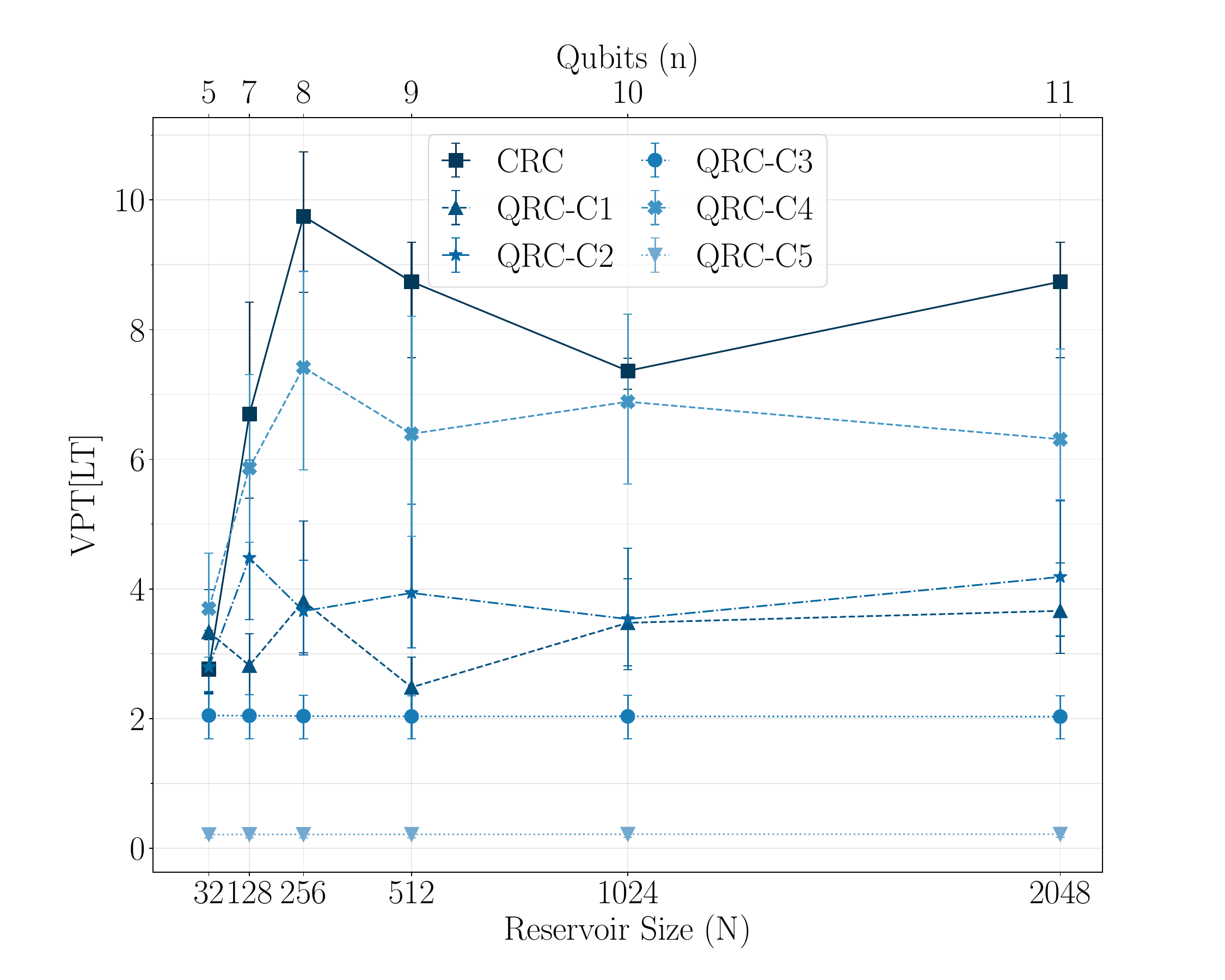}
    \caption{Lorenz-63 system. VPT vs reservoir sizes for both classical and quantum reservoir configurations.}
    \label{fig:VPT_L63a}
\end{figure}

In Fig.~\ref{fig:VPT_L63a}, we present the VPT values derived with classical reservoir computing and compare them with different configurations of quantum reservoir computers from Tab.~\ref{tab:table_ansatz}. The error bars at each value indicate the variation arising from different random seeds. We have tested five different seeds for each value. These seeds correspond to randomly generated weight matrices $\pmb{W}$ and $\pmb{W}_{in}$ for the classical reservoir and variation of ${V(\alpha)}$ values for quantum reservoir architectures. For each of these choices, after training, we initialize the closed-loop predictions from 20 different initial points for quantifying VPT and its mean value.

\begin{figure}
    \includegraphics[width=0.98\linewidth]{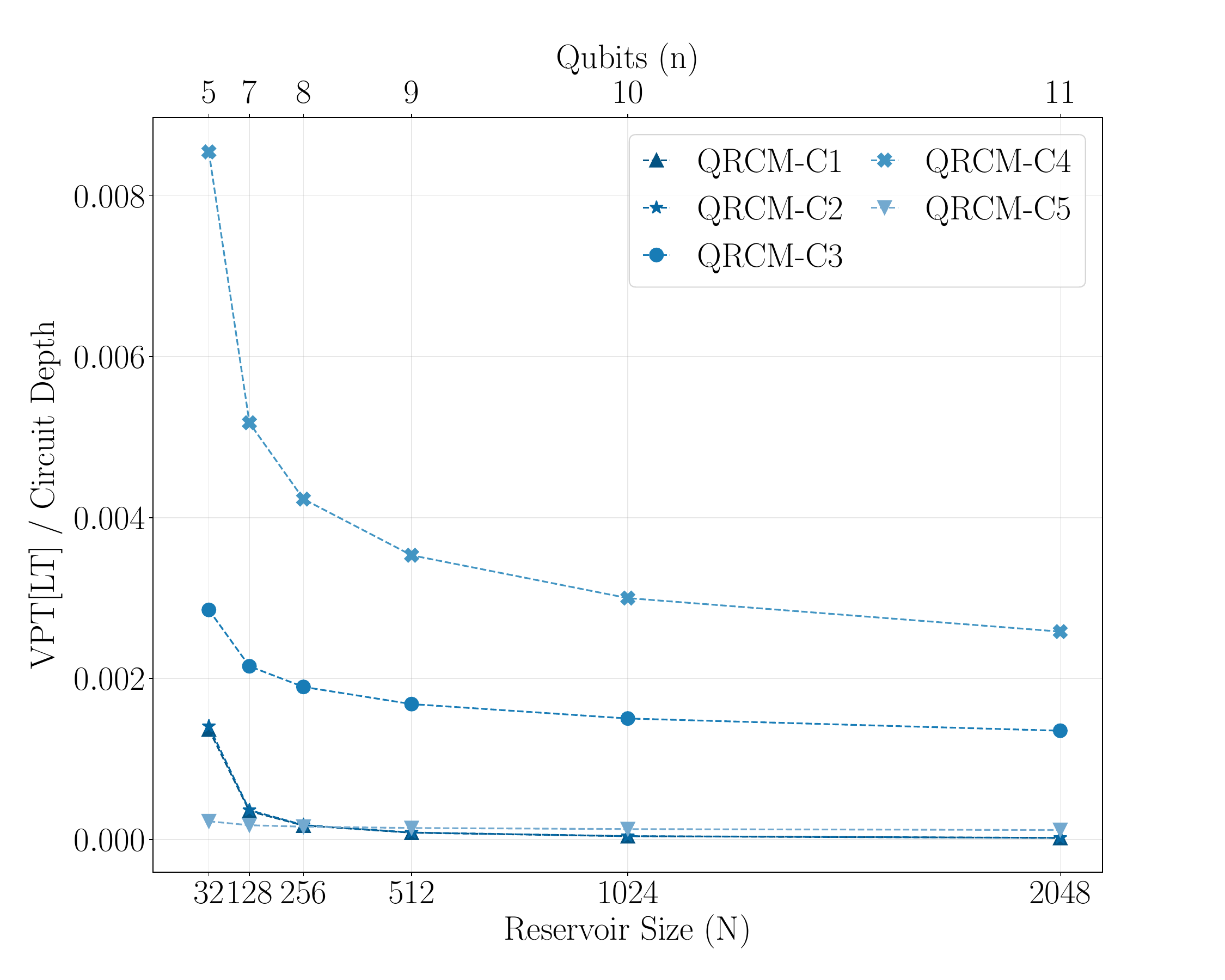}
    \caption{Lorenz-63 system. VPT/circuit depth vs. reservoir sizes for different quantum reservoir configurations.}
    \label{fig:VPT_L63b}
\end{figure}

The performance of different reservoir networks varies with the reservoir sizes. We find that CRC outperforms all the QRC architectures, for the same reservoir sizes. The reservoir size corresponds to the state-vector dimension, which scales exponentially with the number of qubits ($n$). The best-performing quantum reservoir architecture, QRC-C4 with 9-11 qubits ($n$), can predict VPT values similar to the classical reservoir sizes of 500-2000. 

We emphasize that the three QRC configurations C3-C5 do not have a feedback loop. These configurations evolve only from the input time series and the information on reservoir activation states from previous time steps is provided via the classical update Eq.~\ref{eq:2}. In contrast, architectures QRC-C1 and QRC-C2 involve an additional active feedback loop that encodes reservoir states to the quantum circuit at each time step, which leads to an additional overhead in the quantum-classical layer. 

The prediction performance of the two RF-QRC configurations, QRC-C3 and QRC-C5, do not improve with the reservoir sizes. The QRC-C4 architecture, however, shows comparable performance to the reservoirs with feedback loops (QRC-C1 and QRC-C2). These results highlight the importance of carefully chosen feature maps for QRC. 

\begin{figure}
    \centering
    \includegraphics[width=1\linewidth]{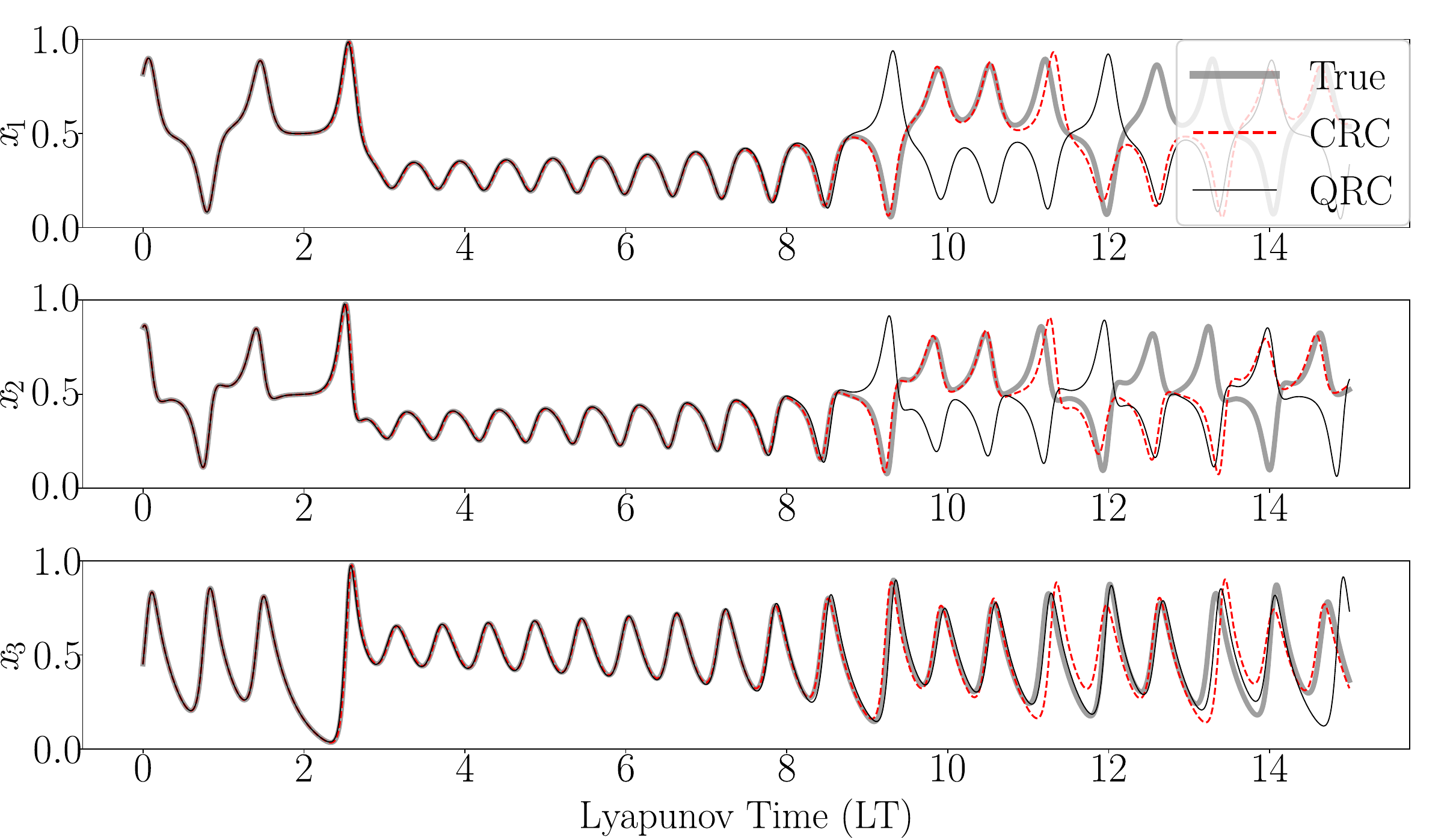}
    \caption{Lorenz-63 system. Closed-loop time-series predictions. Comparison of true predictions with classical reservoir computing (CRC) and emulated quantum reservoir computing (QRC).}
    \label{fig:time_L63}
\end{figure}

In Fig.~\ref{fig:VPT_L63b}, the VPT values normalized by the circuit depth are compared between different QRC architectures. We find that the proposed QRC-C4 architecture outperforms all other architectures by demonstrating the best prediction capabilities while requiring a smaller circuit depth. 

Time-accurate predictions for the Lorenz-63 system are shown in Fig.~\ref{fig:time_L63}. We display the best-performing CRC outcomes (reservoir size of 512), as well as predictions made by the QRC-C4 architecture using 9 qubits. Notice that both the quantum and classical reservoirs can accurately predict up to $\sim$ 8.5 LTs. Beyond that, the prediction diverges from the true solution due to the inherent chaotic nature of the system. 

\subsubsection{Long-term statistical predictions}

\begin{figure}[h]
    \centering
    \includegraphics[width=0.98\linewidth]{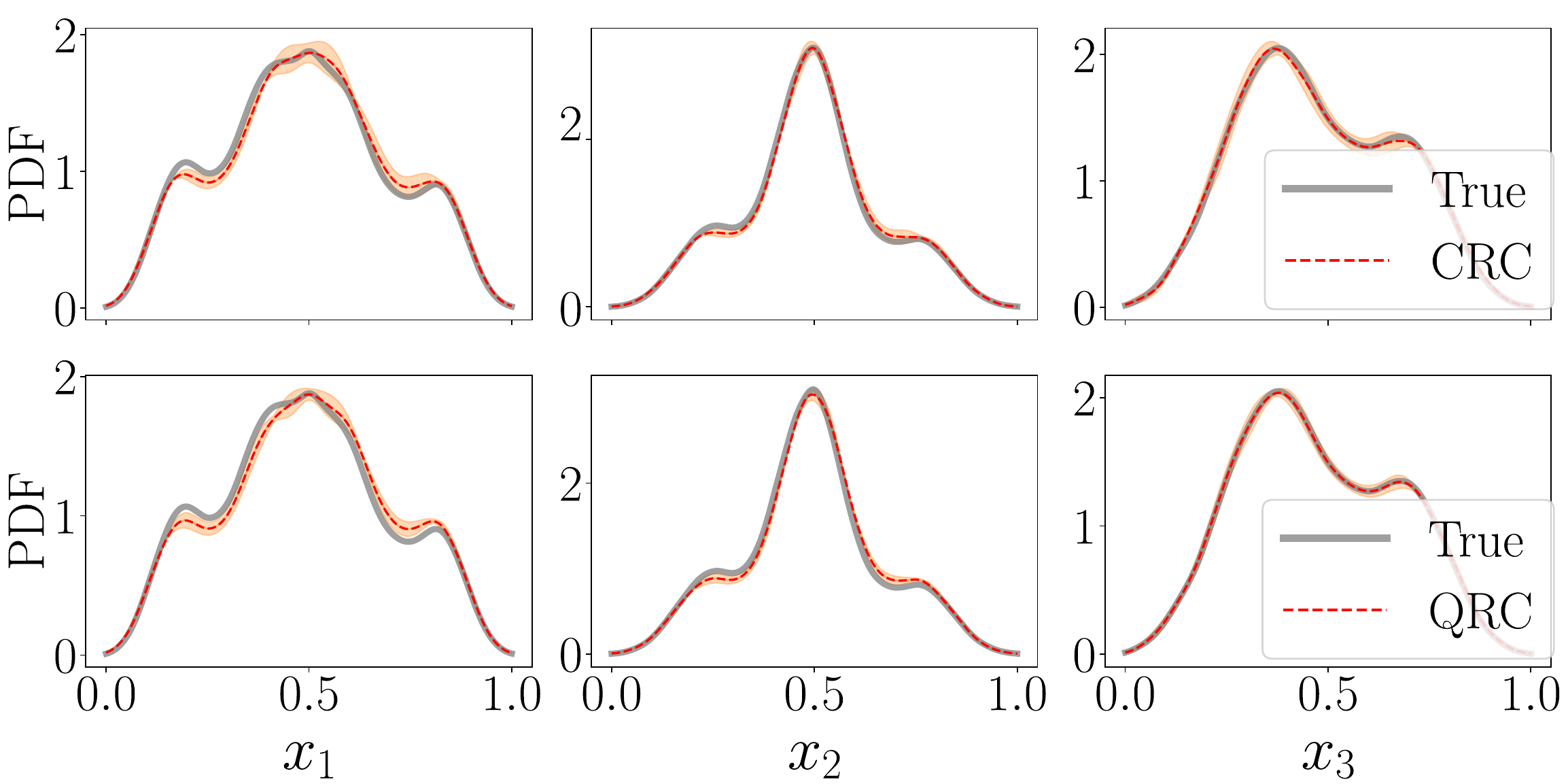}
    \caption{Lorenz-63 system. Closed-loop statistical predictions. The top panels present the result of the true and classical (CRC) network. The bottom panels compare the true with the QRC predictions. The shaded region in both figures indicates the uncertainty associated with different random seeds in both CRC and QRC.}
    \label{fig:stat_L63}
\end{figure}

We select the best-performing networks for which the short-term predictions are shown in Fig.~\ref{fig:time_L63}. We evolve the networks autonomously in a closed-loop for 250 LT, starting from an arbitrary point. The resulting statistics are shown in Fig.~\ref{fig:stat_L63} for each state variable. Both quantum and classical reservoir architectures are able to predict long-term statistics accurately. 

\subsection{Higher-dimensional Lorenz-96 model}\label{sec:LOR96}

In this section, we extend our comparison of classical and quantum reservoirs to higher-dimensional chaotic systems. We follow the same procedure defined in Sec.~\ref{sec:LOR63} and analyze the Lorenz-96 model \cite{75462}.

\begin{equation}\label{eq:18}
    \frac{dx_{i}}{dt} =  (x_{i+1}-x_{i-2}) \, x_{i-1}-{x_{i}} + F  , \quad \quad \quad i = 1, \dotsc ,m
\end{equation}
where $F$ is the external body forcing term which we set to $F$ = 8, to ensure chaotic behavior \cite{vlachas_backpropagation_2020}. We apply periodic boundary conditions, i.e. $x_{1} = x_{m+1}$, and study the reduced order model of Lorenz-96 with ten dimensions ($m=10$). The training set covers an evolution time of $200$ LT for the Lorenz-96 system with a leading Lyapunov {exponent} value of $\Lambda_{1}=1.2$. In Tab.~\ref{tab: param_LOR96}, we present the hyperparameters for the ten-dimensional Lorenz-96 system. The set of hyperparameters is derived by using the same procedure as discussed in Sec.~\ref{sec:methods}.

\begin{table*}
\centering
\caption{Parameters for ten-dimensional Lorenz-96 system.}
\begin{ruledtabular}
\begin{tabular}{llll}
Parameters & Symbol & Classical RC & Quantum RC \\ 
\hline
Time step & $dt$ & 0.01s & 0.01s \\
Input scaling & $\sigma_{in}$ & [0 , 1] & - \\
Spectral radius & $\rho$ & [0.1 , 1] & 1 \\
Tikhonov regularization & $\beta$ & 1$\times10^{-6}$,1$\times10^{-9}$,1$\times10^{-12}$ & 1$\times10^{-6}$,1$\times10^{-9}$,1$\times10^{-12}$ \\
Leak rate & $\epsilon$ & [0.05, 1] & [0.05, 0.3] \\ 
Resevoir density & $D$ & 0.1,0.6,0.9 & Configurations C1-C5 \\ 
\end{tabular}
\label{tab: param_LOR96}
\end{ruledtabular}
\end{table*}

\subsubsection{Time-accurate predictions}

\begin{figure}[h]
\centering
    \includegraphics[width=\linewidth]{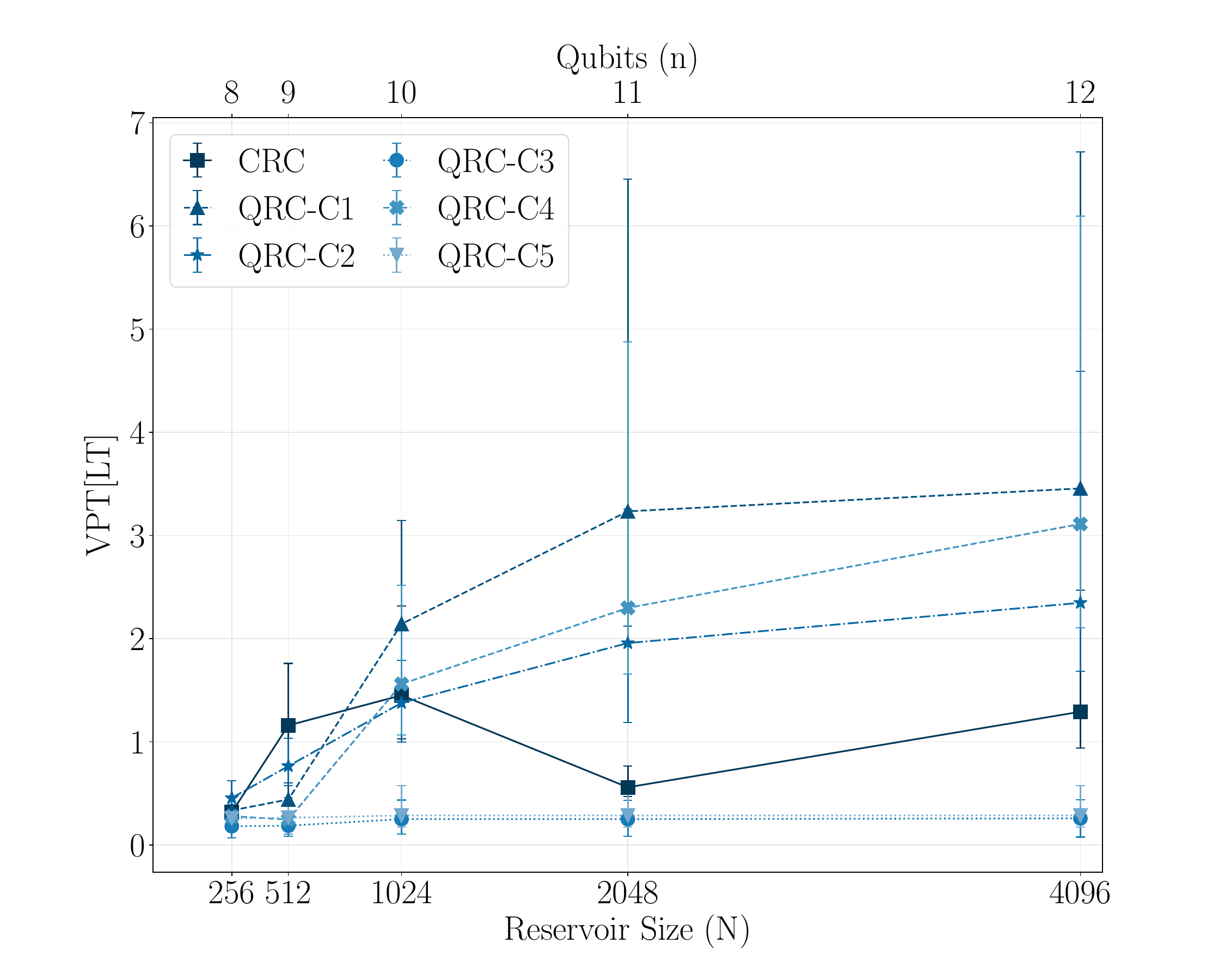}
    \caption{Lorenz-96 system with ten dimensions. VPT {vs} reservoir sizes for both classical and quantum reservoir configurations.}
    \label{fig:VPT_L96a}
\end{figure}

We analyze the time-accurate prediction capabilities of CRC and different QRC architectures for a ten-dimensional reduced-order model of Lorenz-96. Fig.~\ref{fig:VPT_L96a} shows the VPT values of both networks. The QRC architectures C1, C2, and C4 outperform the classical reservoir for increasing reservoir sizes. To study the robustness of the model, the results are averaged over five different sets of optimal hyperparameters and 20 different starting points. For a particular realization of hyperparameters and a starting point, at reservoir size 4096 the VPT value for CRC can be as large as $\sim$ 10 LT. However, the performance is not robust and degrades quickly for other hyperparameter choices and different initial points. Improved performance scaling and robustness indicate a potential quantum advantage in quantum reservoir computers. 

Similar to the results in the Lorenz-63 system, the choice of a feature map and quantum reservoir architectures is critical for the performance. In Fig.~\ref{fig:VPT_L96a}, QRC-C4 and C1 have comparable VPT values at different reservoir sizes {while QRC-C2 underperforms slightly. The performance of  QRC-C3 and C5 do not improve with the reservoir size because there is no feedback loop, and because the linear input feature map only utilises a fraction of state space. By contrast, QRC-C4 (RF-QRC), which also does not have a recurrent feedback loop, has a fully connected input feature map that enriches the reservoir dynamics and  has a comparable VPT to QRC-C1 and C2.} 

\begin{figure}[h]
    \includegraphics[width=\linewidth]{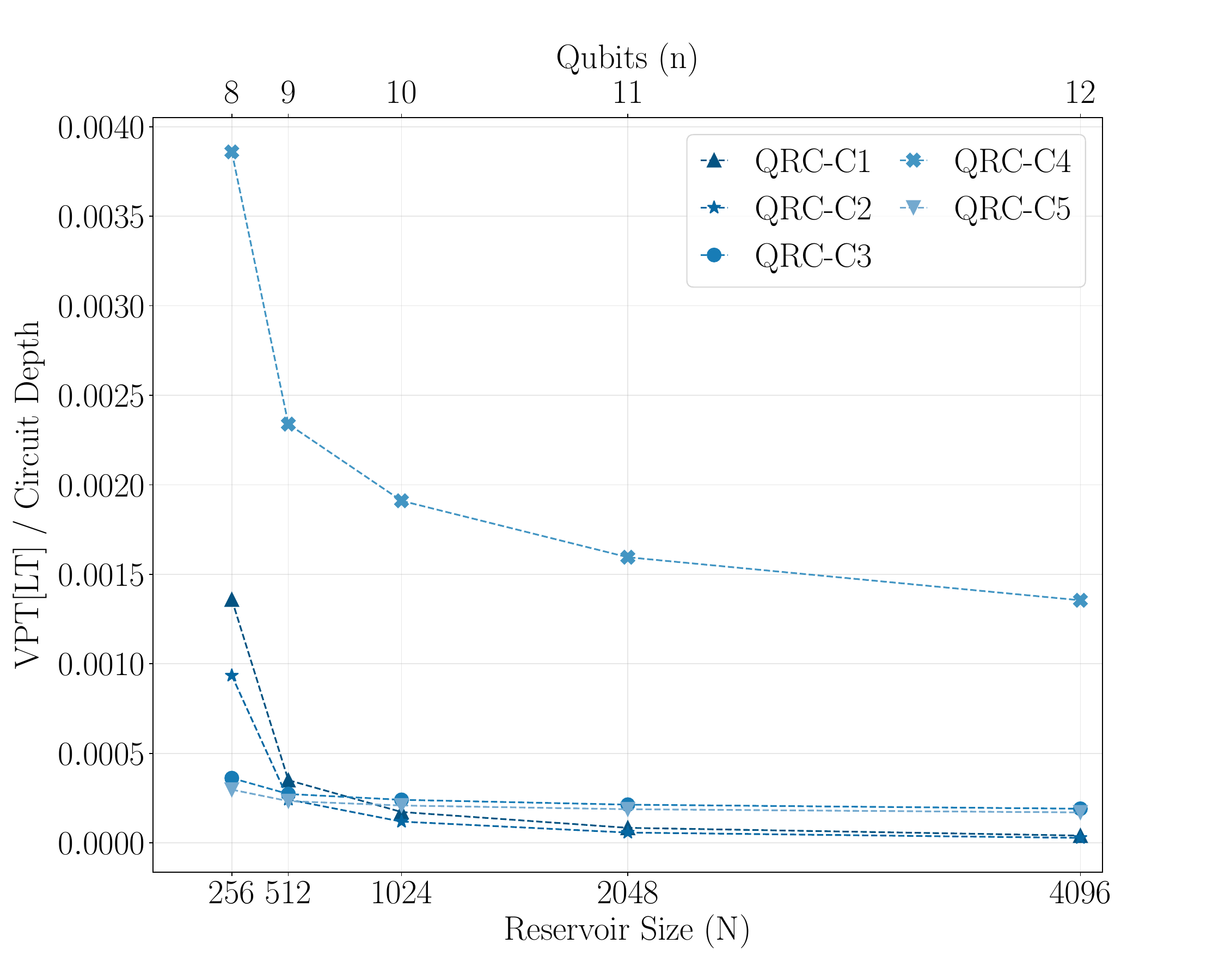}
    \caption{Lorenz-96 system with ten dimensions. VPT/circuit depth vs reservoir sizes for different quantum reservoir configurations.}
    \label{fig:VPT_L96b}
\end{figure}

As compared to QRC with recurrent connections, QRC-C4 achieves the highest VPT to circuit depth ratio for all of the reservoir sizes as shown in Fig.~\ref{fig:VPT_L96b}. This is due to a lack of recurrence. In comparison, QRC-C1 and C2 require exponentially higher circuit depths at increasing reservoir sizes to achieve similar performance. 

\subsubsection{Long-term statisitical predictions}

\begin{figure}
    \centering
    \includegraphics[width=\linewidth]{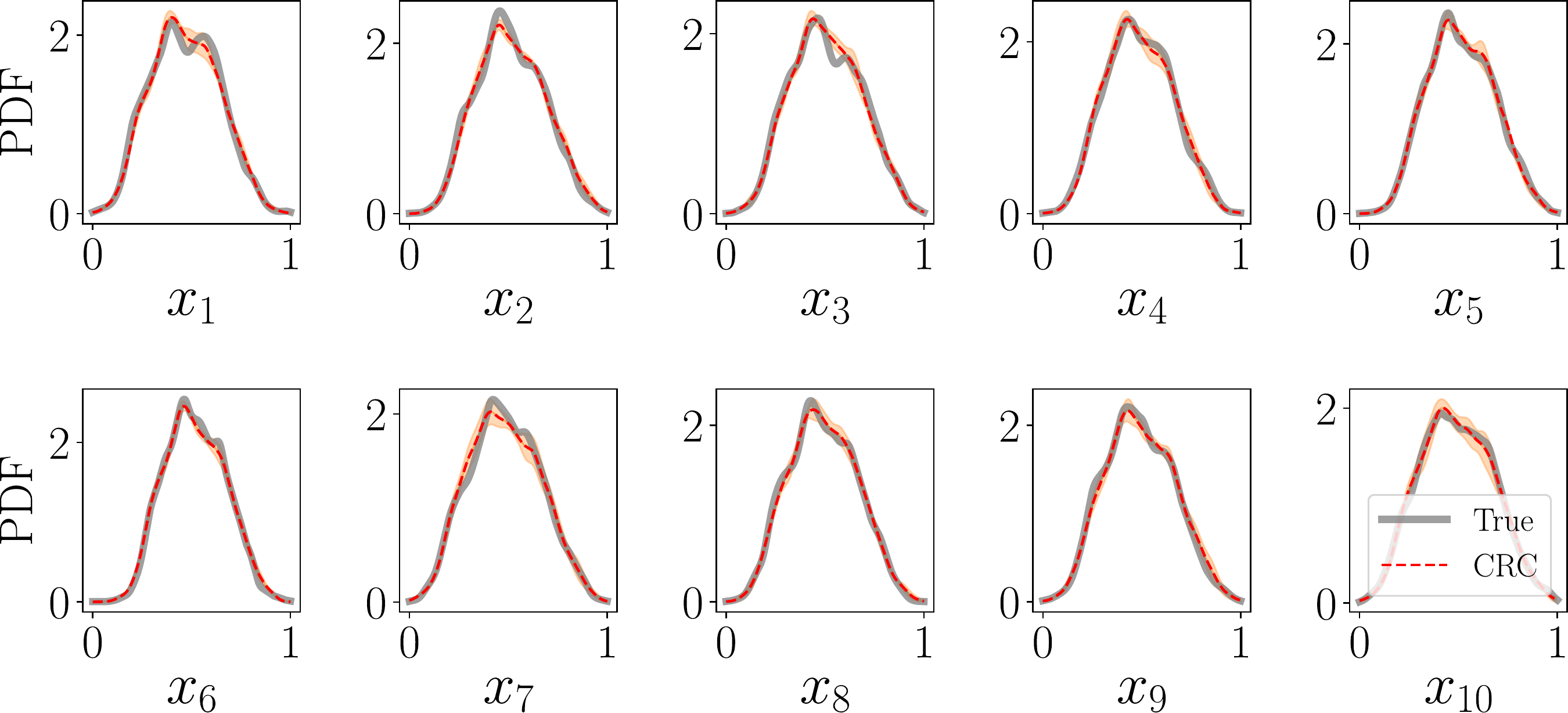}
    \caption{Lorenz-96 system with ten dimensions. Closed-loop statistical predictions using CRC. The shaded region indicates the uncertainty associated with different random seeds in CRC.}
    \label{fig:stat_L96C}
\end{figure}

\begin{figure}
    \centering
    \includegraphics[width=\linewidth]{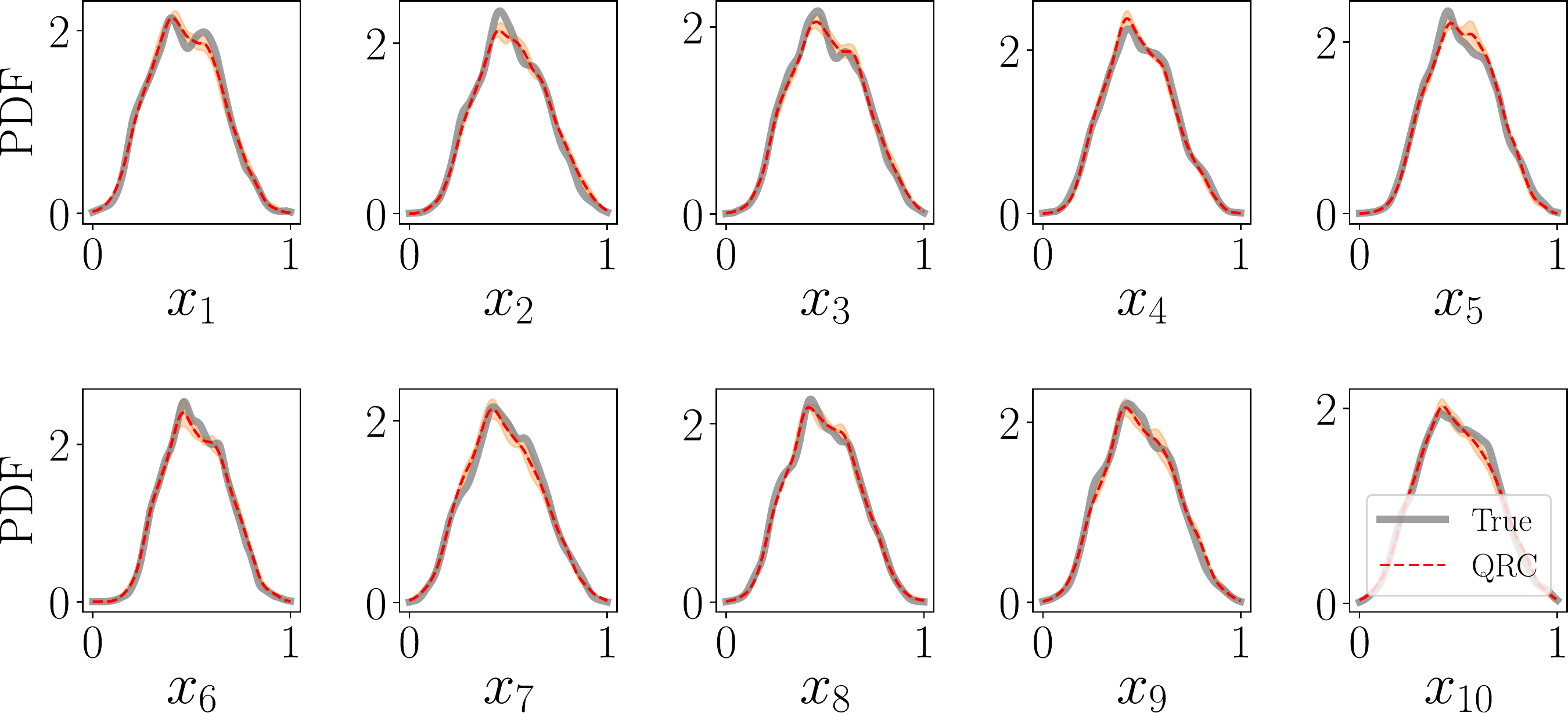}
    \caption{Lorenz-96 system with ten dimensions. Closed-loop statistical predictions using QRC. The shaded region indicates the uncertainty associated with different random seeds in QRC.}
    \label{fig:stat_L96Q}
\end{figure}

In Figs.~\ref{fig:stat_L96C} and \ref{fig:stat_L96Q}, we compare the long-term statistical predictions. The displayed results show the best-performing classical reservoir and quantum reservoir network (QRC-C4) outputs. We evolve these networks for 400 LT in a closed-loop, for the reservoir size of 1024. Both quantum and classical reservoir networks can recover long-term statistics of the chaotic model. The shaded region highlights the uncertainty of different realizations. We find that QRC-C4 predicts the statistical variables with smaller uncertainties when compared to the classical reservoir with equal reservoir sizes. 


\subsection{Chaotic shear flow model for extreme events forecasting}\label{MFE}

\begin{figure}
    \centering
    \includegraphics[width=0.8\linewidth]{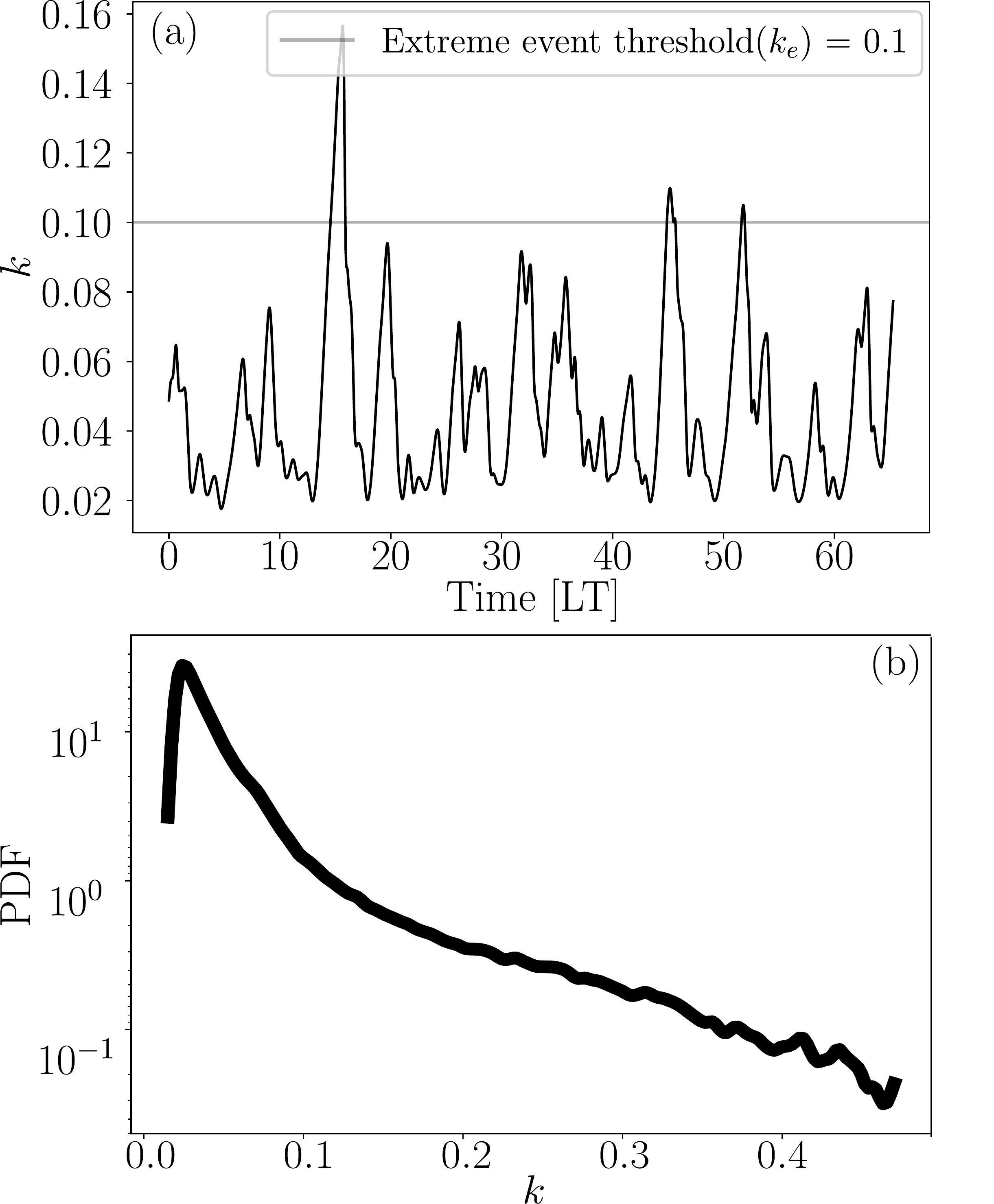}
    \caption{MFE time series kinetic energy and Probability Density Function (PDF) (a) single time series in which the Kinetic Energy exceeds the extreme event threshold $k_{e}$, highlighting the presence of an extreme event (b) probability distribution of kinetic energy calculated over an ensemble of time series  }
    \label{fig:MFE_True}
\end{figure}

In fluid mechanics, extreme events are sudden and unmitigated changes of observables. The forecasting of extreme events is the first step towards the control and suppression of these violent bursts. Recurrent Neural Networks and reservoir computers are used to study these events qualitatively and quantitatively \cite{racca_data-driven_2022,srinivasan_predictions_2019,doan2019physicsaware}. We consider a qualitative low-order model of turbulent shear flows, which is based on Fourier modes and describes a self-sustained turbulent process. This model is also known as the MFE model.
The MFE model is nonlinear and it captures the relaminarization and turbulent bursts \cite{moehlis_low-dimensional_2004}. Owing to the nonlinear nature of this model, the MFE model has been employed to study turbulence transitions and chaos predictability \cite{doan2021short,srinivasan_predictions_2019}. Mathematically, the MFE system can be described by the non-dimensional Navier-Stokes equations for forced incompressible flow
\begin{equation}\label{eq:19}
    \frac{{\partial}\pmb{v}}{{\partial} t} =  - (\pmb{v} \, . \, \pmb{\nabla} )\,\pmb{v} \,-\pmb{\nabla} \, {p}+ \frac{1}{Re} \Delta \pmb{v}  + \pmb{F}(y) , \quad \quad \pmb{\nabla}.\pmb{v} = 0
\end{equation}
where $\pmb{v} = (u, v, w)$ is the three-dimensional velocity vector, $p$ is the pressure, $Re$ is the Reynolds number, $\nabla$ is the gradient, and $\Delta$ is the Laplacian operator. $\pmb{F}(y)$ on the right-hand side is the sinusoidal body forcing term and equals $\pmb{F}(y) = \sqrt{2} \pi^{2}/(4 Re) \, \text{sin}(\pi y/2) \, \pmb{e_{x}} $. The body forcing term is applied along the $x,y$ direction of the shear between the plates. We consider a three-dimensional domain of length $L_{x}, \, L_{y},\, L_{z}$ = $[4\pi,2,2\pi]$ and apply free slip boundary conditions at $y = L_{y}/2$, periodic boundary conditions at $x = [0 ; L_{x}]$ and $z = [0; L_{z}]$. The set of PDEs can be converted into ODEs by projecting the velocities onto Fourier modes as given by Eq.~(\ref{eq:20})
\begin{equation}\label{eq:20}
    \pmb{v}(\pmb{x} , t) = \sum_{i=1}^{9} \, a_{i}(t) \, \hat{\pmb{v}_{i}}\,(\pmb{x}).
\end{equation}
These nine decompositions for the amplitudes ${a_{i} (t)}$ are substituted into Eq.~(\ref{eq:19}) to yield a set of nine ordinary differential equations as in Ref.~\cite{moehlis_low-dimensional_2004}. The MFE system displays a chaotic transient, which in the long term converges to a stable laminar solution. We want to predict the turbulent burst of kinetic energy and chaotic transients, which are extreme events. Fig.~\ref{fig:MFE_True} shows the evolution of kinetic energy [$ k(t) = \frac{1}{2} \sum_{i=1}^{9} \, a_{i}^{2}(t) \,$] and the associated extreme event, and the long-term statistical distribution of the kinetic energy. An extreme event occurs when the kinetic energy $k(t)$ value exceeds the threshold ($k(t) \geq k_{e})$. In this work, we take $k_{e} = 0.1$, as in Ref.~\cite{racca_data-driven_2022}.

We solve the MFE system ODEs using an RK4 solver with $dt = 0.25 s$. The leading Lyapunov exponent is $\Lambda = 0.0163$ for MFE model \cite{racca_data-driven_2022}. Unlike the Lorenz-63 and Lorenz-96 models, generating a single long-time series for washout, training, and test sets is not feasible for MFE because a single long-time series eventually laminarizes and gives limited information about the chaotic transients. To address this issue, first, we generate an ensemble of 2000 time series from different random initial points. We take the length of a single time series equal to 65 LT or 16000 time steps. Second, we discard the time series whose maximum kinetic energy is larger than the laminarization threshold $k_{l} = 0.48$. This threshold is selected to be close to the (asymptotic) laminarization value of $k = 0.5$. Out of the 2000 time series, $27\%$ of the series laminarizes. We divide the remaining 1441 time series into washout, training, and test sets. We train our network on a 25-time series, each of length $20 LT$. The test set consists of 500 time series. In Secs.~ \ref{sec:MFE_time} and \ref{sec:MFE_stats}, we compare the time-accurate and statistical predictions for the classical reservoir method against the best-performing quantum reservoir QRC-C4. 

\begin{table*}
\caption{{Parameters for the MFE system.}}
\begin{ruledtabular}
\begin{tabular}{llll}
Parameters & Symbol & Classical RC & Quantum RC \\ \hline
Time step & $dt$ & 0.25s & 0.25s \\
Input scaling & $\sigma_{in}$ & [0 , 1] & - \\
Spectral radius & $\rho$ & [0.1 , 1] & 1 \\
Tikhonov regularization & $\beta$ & 1$\times10^{-6}$,1$\times10^{-9}$,1$\times10^{-12}$ & 1$\times10^{-6}$,1$\times10^{-9}$,1$\times10^{-12}$ \\
Leak rate & $\epsilon$ & [0.05, 1] & [0.05, 0.3] \\ 
Resevoir density & $D$ & 0.1,0.6,0.9 & Configuration C4 (RF-QRC)\\ 
\end{tabular}
\label{tab: param_MFE}
\end{ruledtabular}
\end{table*}

\subsubsection{Time-accurate predictions}\label{sec:MFE_time}

In the MFE model, we are concerned about quantifying the extreme event prediction capabilities of our reservoir networks. These capabilities are not accurately quantified by VPT, because extreme events are difficult to predict due to the chaotic nature of the attractor. This means that the network can show good predictability in low kinetic energy regions while showing smaller VPT values near extreme events. This may produce an inaccurate representation of the predictability of our reservoir networks. 

\begin{figure}[h]
    \centering
    \includegraphics[width=\linewidth]{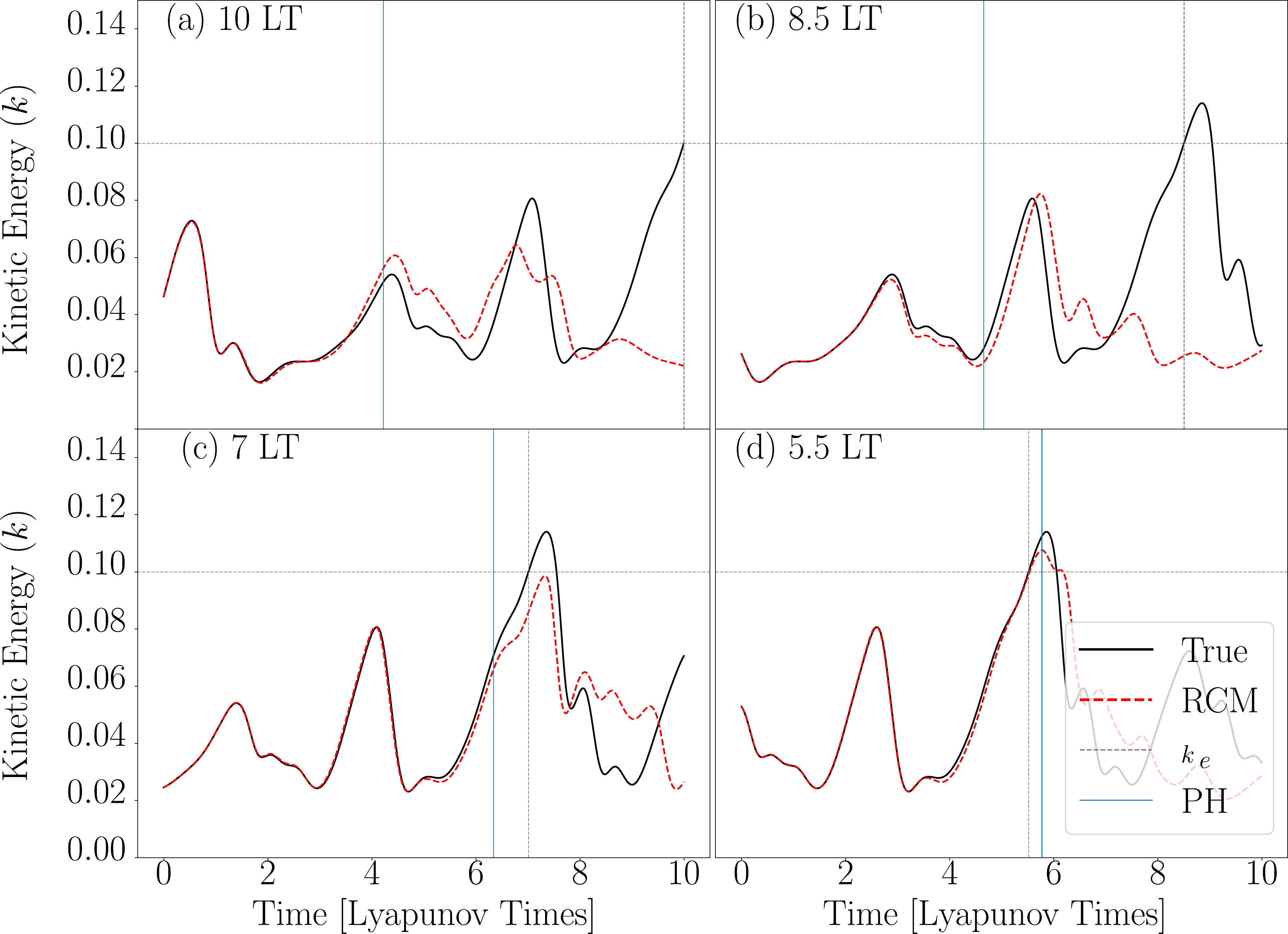}
    \caption{Visualization of the prediction of extreme events for both Reservoir Computing Methods (RCMs) i.e. both CRC and QRC (PH= 5.89LT). }
    \label{fig:PH_visual}
\end{figure}

To quantify extreme event prediction capabilities, we use the Predictability Horizon (PH) \cite{racca_data-driven_2022,doan2021short,10.1063/1.5028373}, which is defined as the time interval during which the predicted kinetic energy $k_{pred}$ and true kinetic energy $k_{true}$ are bounded by

\begin{equation}\label{eq:22}
    \left| \frac{k_{pred}(t)-k_{true}(t)}{k_{e}-\Bar{k}} \right|  < 0.2, 
\end{equation}
where $k_{e}$ is the extreme event threshold, which is equal to 0.1 here as in Ref.~\cite{racca_data-driven_2022}. $\Bar{k}$ is the time average of the kinetic energy, and 0.2 is the user-defined error threshold \cite{racca_data-driven_2022}. To quantify the PH, we sample 100 different extreme events from the unseen test set data. For each extreme event, we start the {predictions 12 LT before the extreme event and discard the transients in the washout phase for 2 LT by running reservoir networks in an open loop. This gives $\delta_{k_{e}}$ of 10 LT. After this, we start autonomous closed-loop predictions with $\delta_{k_{e}} = 10$}. The value of $\delta_{k_{e}}$ represents the time difference from the start of closed-loop predictions to the extreme event. If the PH value is greater than $\delta_{k_{e}}$ we take PH equal to $\delta_{k_{e}}$. If the prediction diverges before the extreme event occurs, or the PH value is less than $\delta_{k_{e}}$, we decrease the $\delta_{k_{e}}$ value by a factor of $\tau_{e}$. We take $\tau_{e}$ = 0.5 LT and the new $\delta_{k_{e}}$ value equals to 9.5 LT. For the next step, we repeat the closed-loop predictions for the extreme event occurring at 9.5 LT, i.e., $\delta_{k_{e}}$ = 9.5 LT. We repeat the same method decreasing $\delta_{k_{e}}$ by $\tau_{e}$ until the PH is greater than $\delta_{k_{e}}$ (Fig.~\ref{fig:PH_visual}). That is, the extreme event is predicted correctly by the reservoir computers by the value of PH in the future. 

\begin{figure}[h]
    \centering
    \includegraphics[width=0.8\linewidth]{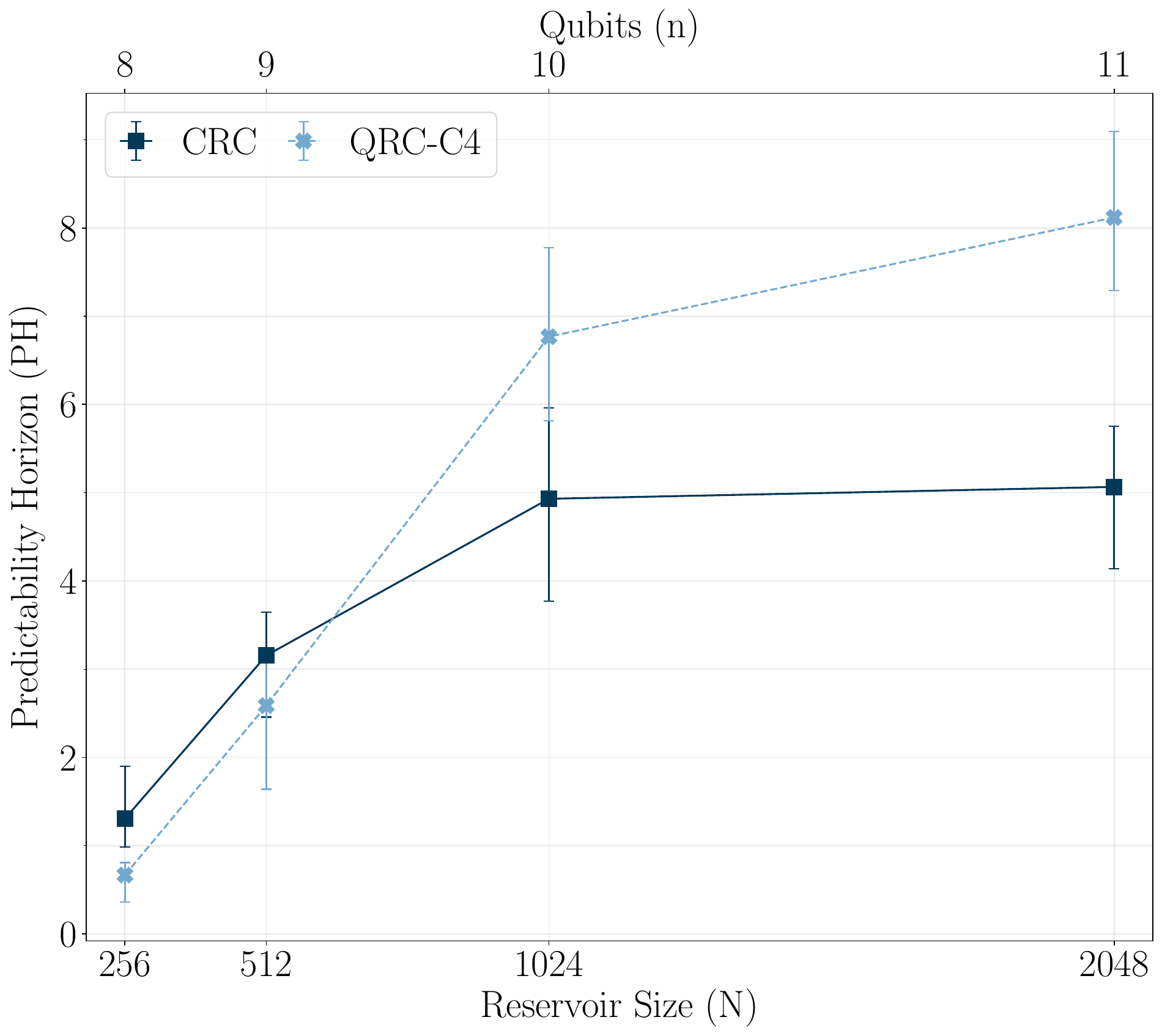}
    \caption{MFE system. Predictability Horizon (PH) for predictions with different reservoir sizes {for the best performing CRC and QRC-C4 (RF-QRC) networks}. }
    \label{fig:PH_MFE}
\end{figure}

We calculate the PH for a set of reservoir sizes for both quantum and classical reservoir networks.  For a classical reservoir, the number of neurons is varied from $N = 256$ to $N = 2048$. We tune the hyperparameter for each reservoir and calculate the corresponding PH value. We also perform the same analysis for the quantum reservoir with qubits $n=8$ to $n=11$, which, in principle, corresponds to the classical reservoirs of $N=256$ to $N=2048$. The performance of the quantum reservoir also improves with the reservoir size. In Fig.~\ref{fig:PH_MFE}, the comparison of classical and quantum reservoirs for various reservoir sizes is shown. Both reservoir networks have comparable performances for reservoir sizes $N = \{256, 512\}$, with the classical reservoir out-performing quantum reservoir by a small margin. However, the PH value of the classical reservoir converges with a median value of around 5.5 LT. This convergence of PH value with increasing reservoir sizes is due to performance saturation in CRC and it is consistent with the results in Ref.~\cite{racca_data-driven_2022}. In contrast, the quantum reservoir outperforms the classical reservoir for increasing reservoir sizes and has a median PH value of 8.09LT with 11 qubits. 

A second statistical measure to assess the extreme event prediction capabilities is the $F$-Score \cite{10.1007/978-3-540-31865-1_25}, which measures the model's accuracy by combining metrics such as precision and recall. To calculate the $F$-score, we measure three different occurrences: (a) an extreme event occurs in the true data set and the model also predicts the extreme event (True Positive, TP) (b) there is no extreme event in the true data but the model predicts an extreme event (False Positive, FP) (c) an extreme event occurs in the true data set but the model does not predict an extreme event, (False Negative, FN) \cite{racca_data-driven_2022}. These three metrics are combined to calculate Precision ($p$), Recall ($r$), and $F$-score ($F$) from 1000 different starting points, which span 20 different time series

\begin{equation}\label{eq:23}
    p = \frac{TP}{TP+FP} , \qquad  r = \frac{TP}{TP+FN} , \qquad  F =  \frac{2}{p^{-1}+r^{-1}} , 
\end{equation}

\begin{figure}
    \includegraphics[width=\linewidth]{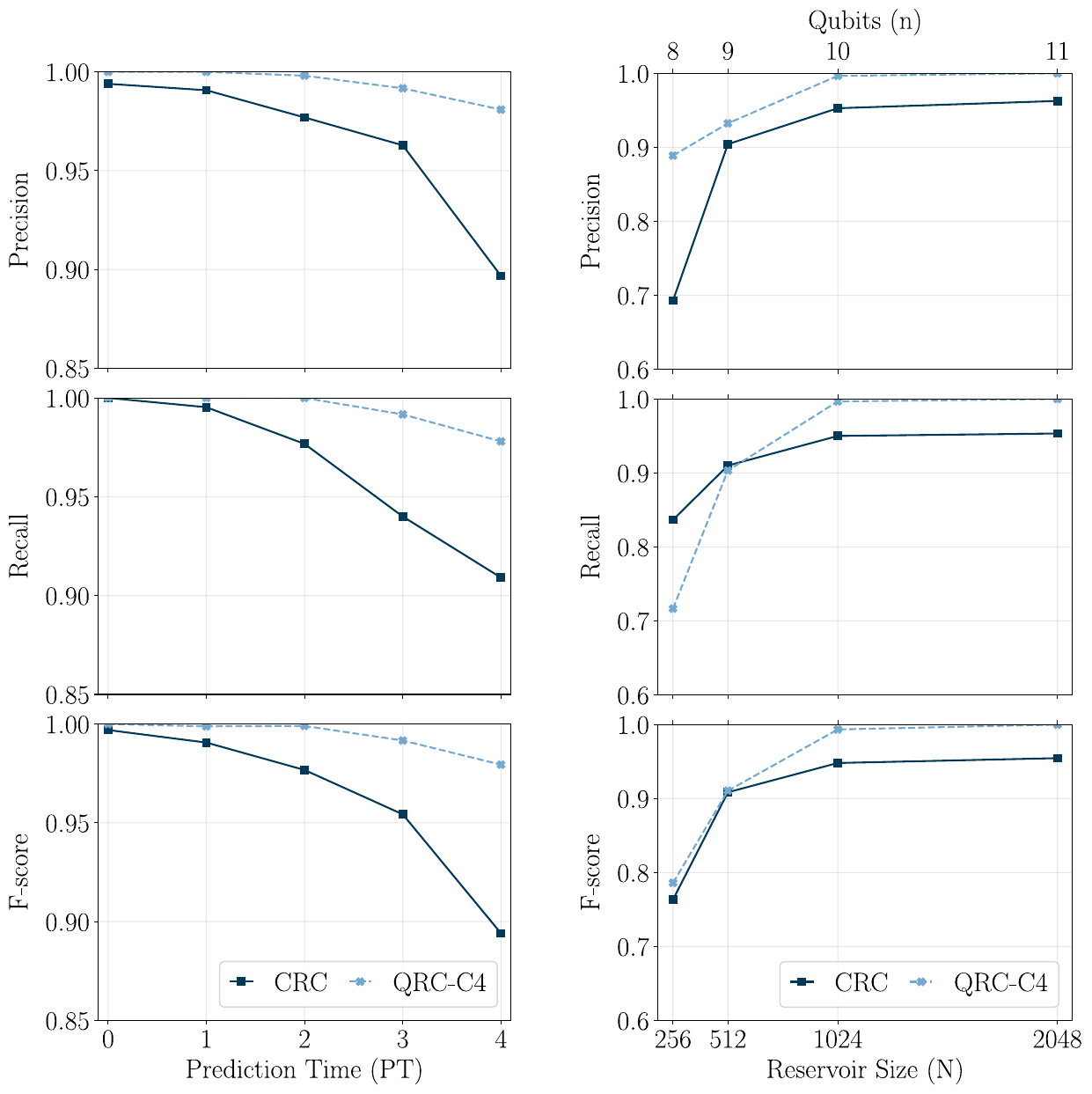}
    \caption{MFE system. $75^{th}$ percentile of Precision, Recall, F-score vs On the left side, Prediction Time (PT) for a constant reservoir size $(N=1024)$. On the right side, Reservoir Size for a constant Prediction Time $(PT=3LT)$.}
    \label{fig:fscore_MFE}
\end{figure}

We compare quantum and classical reservoirs for two different cases. First, we fix the reservoir size for $N = 1024$, which is equals to 10 qubits for the quantum reservoir and varies Prediction Time (PT). The Prediction Time (PT) is defined as the time interval between the extreme event and the start of the prediction. The PT value of 0 LT indicates that the prediction interval is between 0 and 1 LT, and PT = 3 indicates the prediction interval of 3 - 4 LT. The lower PT values have an $F$-score value closer to 1 because the network predicts nearly all the extreme events accurately, which are located within 1 LT of the start of predictions. As expected, the performance deteriorates with increasing the PT value due to the inherent chaotic nature of the model. The proposed quantum reservoir architecture (QRC-C4) achieves a higher F-score as compared to the CRC for different PT intervals as shown in Fig.~\ref{fig:fscore_MFE}

Second, we fix the prediction time to 3 LT (PT = 3LT), which indicates the model is predicting extreme events between 3 - 4 LT. After fixing the PT value, we vary the reservoir sizes from $N = 256$ to $N = 2048$ and compare the performance metrics. The performances are similar for both quantum and classical reservoirs with reservoir sizes of $\{256,512\}$. Similar to the results shown for the predictability horizon previously, the quantum architecture has a higher $F$-score value for 10,11 qubits as compared to the classical reservoirs of the same sizes (Fig.~\ref{fig:fscore_MFE}). All the results are computed for 5 different ensembles of networks and for 1000 different starting points. 

\subsubsection{Long-term statistical prediction}\label{sec:MFE_stats}

For the classical reservoir, we chose the reservoir size of $N = 512$. For the quantum reservoir, we chose a reservoir that corresponds to $n=9$ qubits. We evolve the networks from different starting points in the test set of 500 time series, in which each starting point corresponds to a different time series. As previously mentioned, we perform washout for each time series and discard the time series whose kinetic energy exceeds the laminarization threshold $k_{l} > 0.48$. Finally, we calculate the Probability Density Function (PDF)

\begin{figure}[h]
    \centering
    \includegraphics[width=\linewidth]{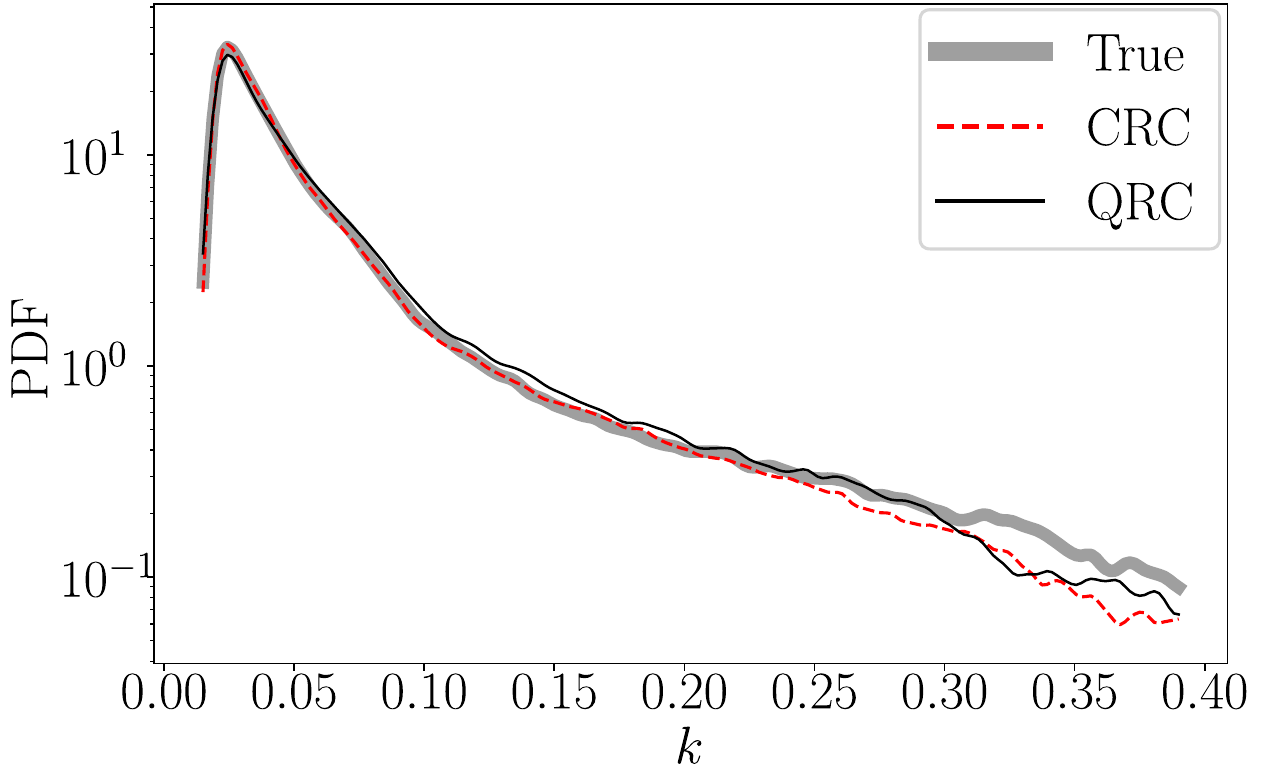}
    \caption{MFE system. Long-term statistical predictions with classical and quantum reservoir computing.}
    \label{fig:stats_MFE}
\end{figure}

Figure \ref{fig:stats_MFE} presents the results for the statistical predictions against the true value. Both reservoirs can predict long-term statistics with high accuracy for the smaller kinetic energy values. Particularly, for $k < 0.15$ the prediction agrees with the true value of the statistics. This analysis provides a statistical measure of the prediction of extreme events. Both quantum and classical reservoirs can predict most of the extreme events ($k < 0.10$) in the test set. For the higher kinetic energy values, the log scale amplifies the difference between true value and predictions. The accuracy in the prediction of extreme events could be improved by considering a larger reservoir size, which scales better for QRC as compared to CRC, as shown previously in Figs.~\ref{fig:PH_MFE} and \ref{fig:fscore_MFE}. 

\begin{figure}[h]
    \centering
    \includegraphics[width=\linewidth]{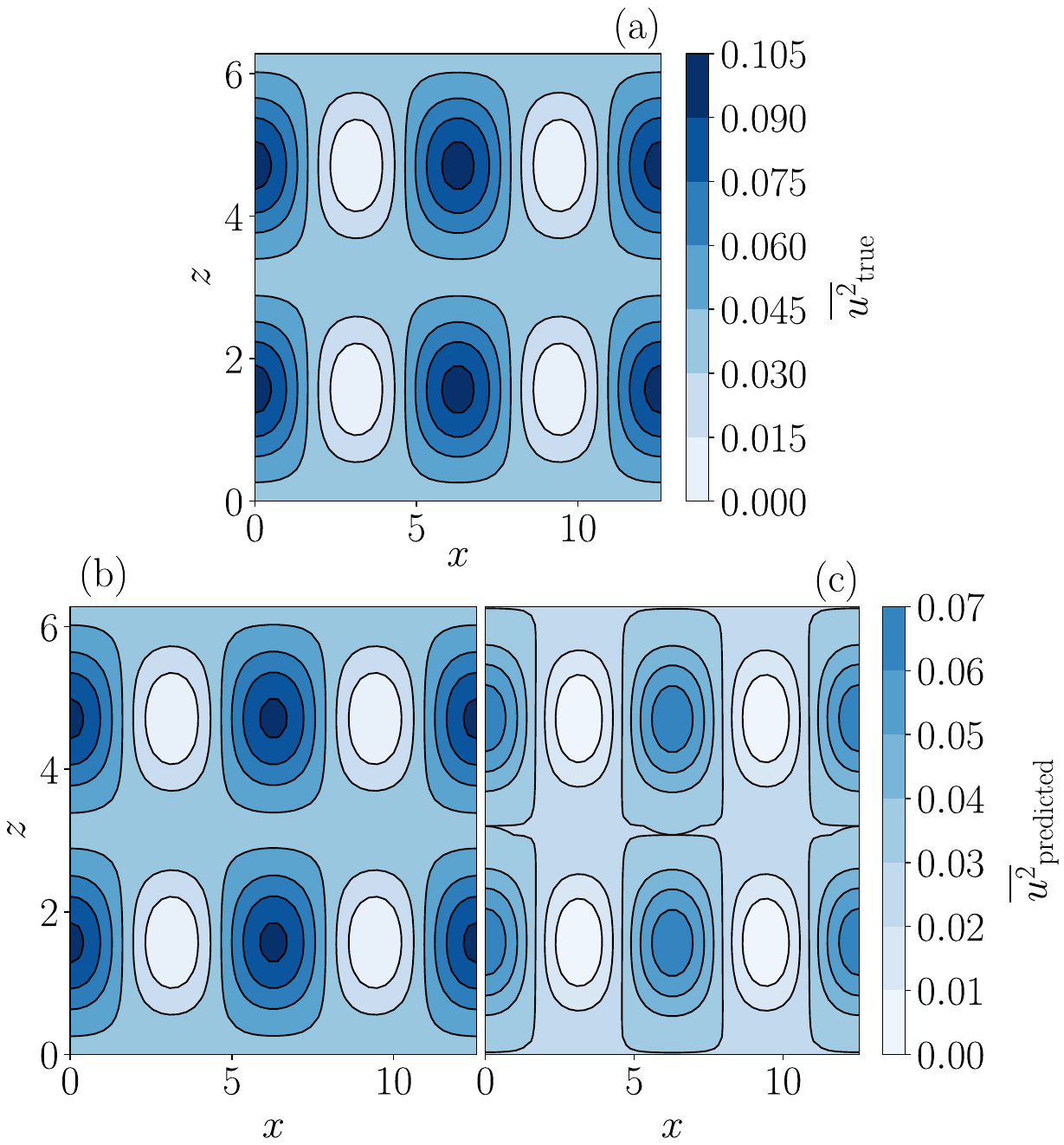}
    \caption{{MFE system. The time average of the square of the midplane velocity along x. (a) True data (b) QRC-C4 predictions (c) CRC predictions for the best-performing networks with a reservoir size of 512 (9 qubits)}}
    \label{fig:statsflow_MFE}
\end{figure}

{To further visualize the statistical predictions and compare CRC and RF-QRC, we predicted the velocity components of the flow fields from the predicted Fourier modes. In Fig.~\ref{fig:statsflow_MFE} the results of the average streamwise velocities are shown for the true data, and compared against classical and quantum reservoir predictions on the test data set. The predictions of RF-QRC approximate the true flow field better than the predictions from CRC, which have larger deviations. We conjecture that RF-QRC can be used to reconstruct flowfields with better accuracy than CRC with similar reservoir sizes. The prediction of flowfields with a better accuracy also allows the RF-QRC model to predict turbulent bursts of relaminarization and thus kinetic energies (extreme events) with better prediction horizon as previously shown in Figs.~\ref{fig:PH_MFE} and \ref{fig:fscore_MFE}}.

\section{Conclusion}\label{sec:conclusion}

By exploiting quantum-computer {ansätze} and entanglement, we analyze and design reservoir computers to predict chaotic dynamics and extreme events from data. 
First, we show that the design of a feature map, which encapsulates entanglement and data-encoding layers, is key to the reservoir's performance. 
Second, we design the Recurrence-Free Quantum Reservoir Computer (RF-QRC), which has a small circuit depth that does not scale with the number of qubits (reservoir size), unlike previous proposals of QRC. RF-QRC also does not have recurrent connections at each time step. This enables the training of the network without an additional classical feedback loop, which makes the implementation suitable to current noisy intermediate-scale quantum devices. 
Third, we numerically analyze quantum reservoir computers and the proposed RF-QRC on prototypical chaotic dynamical systems, from low- to higher-dimensional. We show that 
(a) the forecasting capability of state-of-the-art classical and quantum reservoir computers increases with the reservoir size until saturation; 
(b) for low-dimensional chaotic systems, classical reservoir computers have larger predictability than quantum reservoir computers for the same degrees of freedom; (c) for higher-dimensional chaotic systems, RF-QRC has larger predictability than classical reservoir computers for the same degrees of freedom; and (d) RF-QRC can predict extreme events while scaling better than state-of-the-art reservoir computers. The RF-QRC requires smaller reservoir sizes as compared to classical reservoir computers for the same performance. The RF-QRC can be used to encode classical and quantum data on a quantum computer to make time-series predictions. This work opens new opportunities for the prediction of chaotic dynamics and extreme events with quantum reservoir computing. Current work is focused on the analysis of hardware and environmental noise, and finite-sampling errors.   

\begin{acknowledgments}
The authors acknowledge financial support from the UKRI New Horizon grant EP/X017249/1.
L.M. is grateful for the support from the ERC Starting
Grant PhyCo 949388, and the grant EU-PNRR YoungResearcher TWIN ERC-PI 0000005. F.T. acknowledges support from the UKRI AI for Net Zero grant EP/Y005619/1.
\end{acknowledgments}

\appendix

\section{Quantum Circuits for different ans\"{a}tze}\label{sec:ansatz}

\begin{figure}[h]
\centering
    \includegraphics[width=1\linewidth]{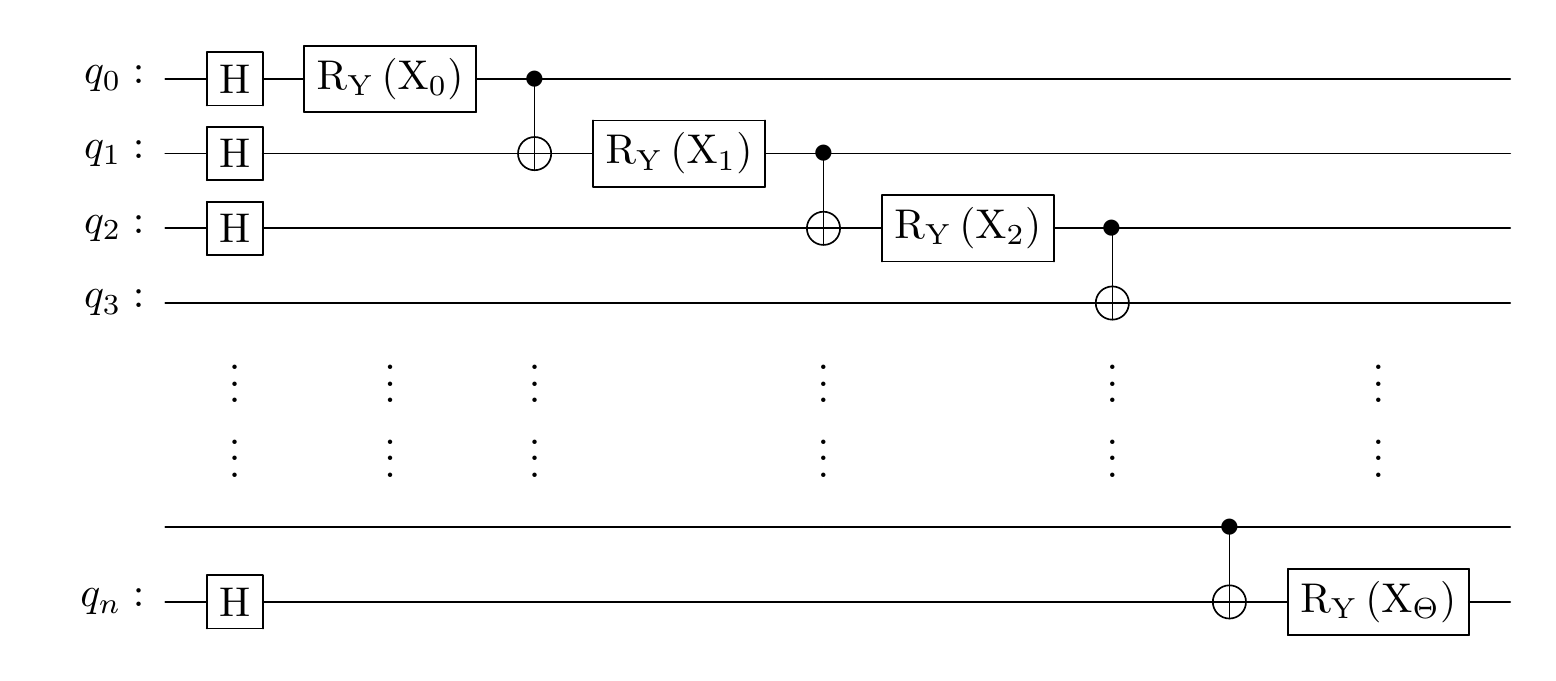}
    \caption{Quantum Circuit for linearly entangled qubits}
    \label{fig:LE}
\end{figure}

\begin{figure}[h]
\centering
    \includegraphics[width=1\linewidth]{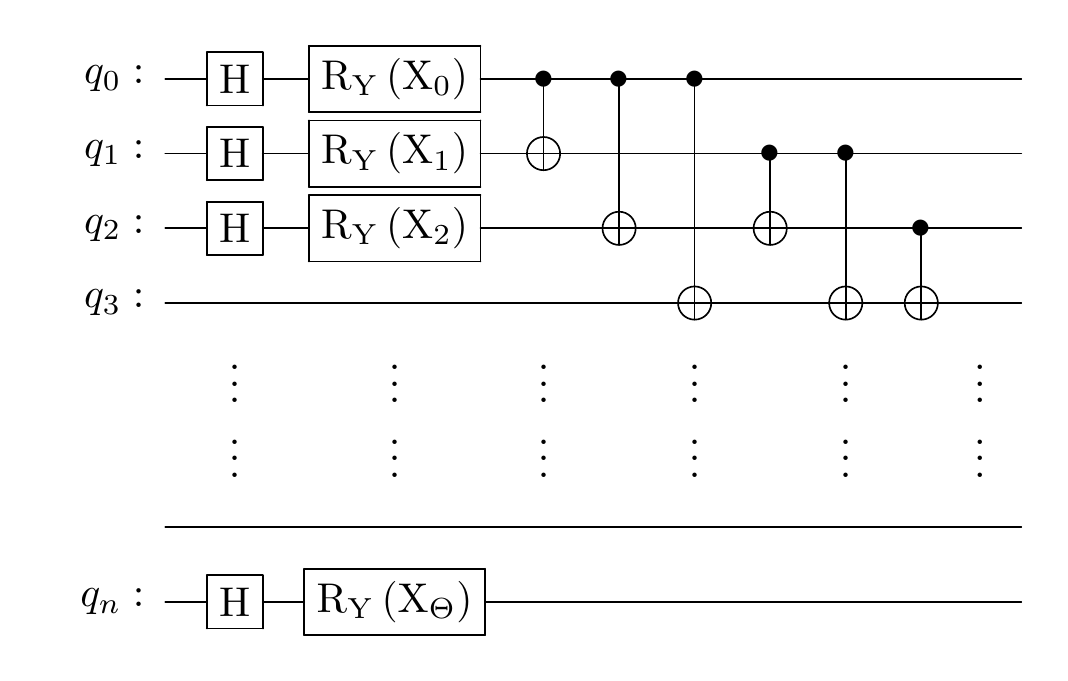}
    \caption{Quantum Circuit for fully entangled qubits}
    \label{fig:FE}
\end{figure}

\begin{figure}[h]
\centering
    \includegraphics[width=1\linewidth]{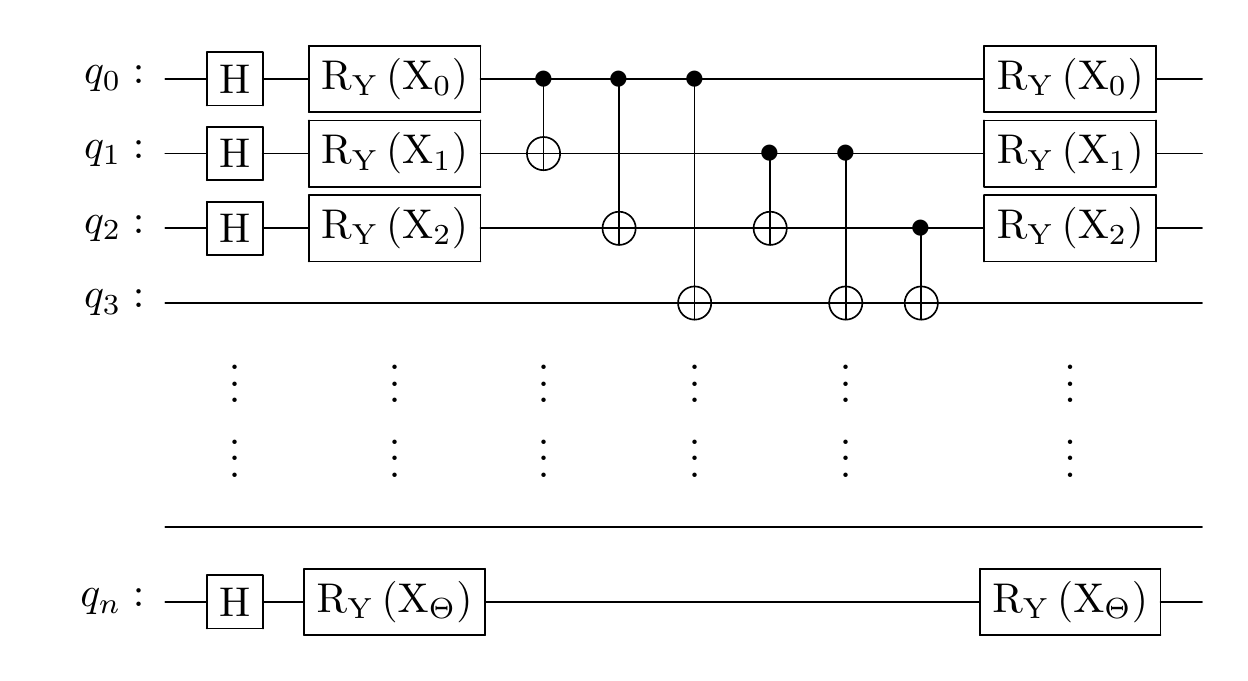}
    \caption{Quantum Circuit for fully entangled symmetric qubits}
    \label{fig:FES}
\end{figure}

\begin{figure*}
\centering
    \includegraphics[width=0.9\linewidth]{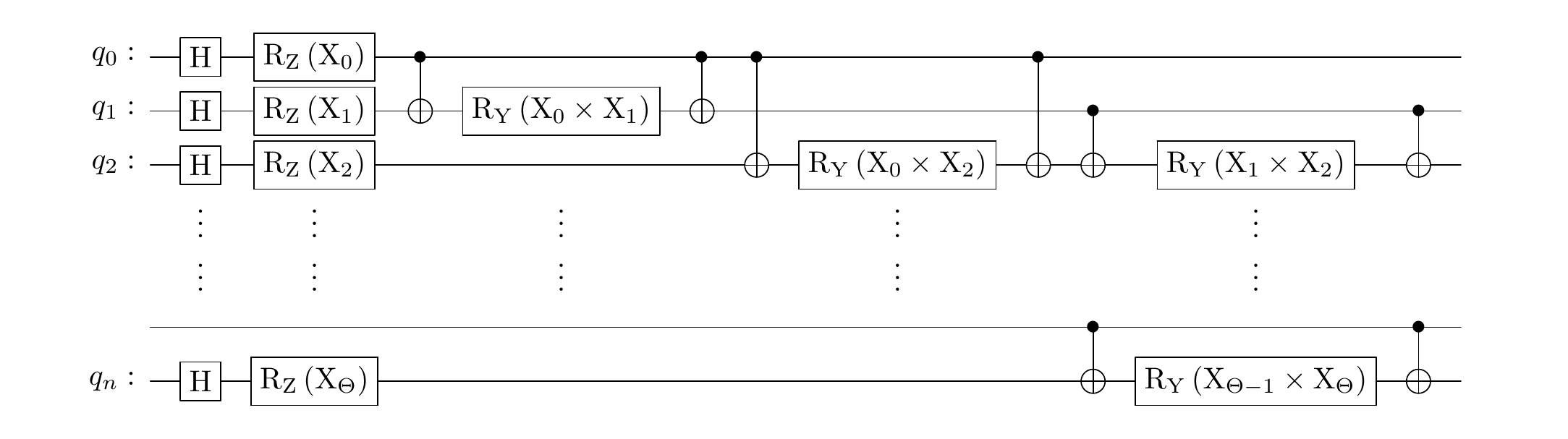}
    \caption{Quantum Circuit for product feature map qubits}
    \label{fig:FM}
\end{figure*}

The Figs.~\ref{fig:LE}-\ref{fig:FM} represent the quantum circuits for different ans\"{a}tze presented in Tab.~\ref{tab:table_ansatz}. The rotation angles $X_\Theta$ represent the mapped classical data rescaled to the interval $[0,2\pi]$.

\section{Memory of the Reservoir Networks}\label{sec:STM}

In this section, we discuss the empirical metrics including fading memory and linear Memory Capacity (MC) \cite{boyd_fading_1985,carroll_optimizing_2022,farkas_computational_2016,han2021revisiting,Jaeger_2001} for different reservoir architectures. In order to efficiently reconstruct the reservoir dynamics, reservoir computers must satisfy the Echo State Property (ESP) \cite{jaeger__2001,lukosevicius_practical_2012}.  In classical reservoir networks, ESP is enforced by scaling the hyperparameter $\rho$ (Sec.~\ref{sec:CRCM}). In quantum reservoir networks, unitary evolution is norm-preserving, i.e. $(U^{\dag}U=I)$, and therefore no tuneable hyperparameter $\rho$ exists. In both, quantum and classical reservoir computing, we perform a washout interval by discarding a limited number of initial reservoir states.

The Short-term Memory Capacity (MC) is a performance measure that is used to characterize the linear memory capacity of reservoir computers \cite{Jaeger_2001}. MC quantifies the maximally possible linear correlations between the current reservoir states and previous input data, $\textbf{u}(t-d)$, and can be computed as 
\begin{equation}\label{eq:24}
       MC = \sum_{d=1}^{d_{max}} MF_{d} , \quad \quad {MF}_{d} = \frac{cov^{2}(\pmb{u}_{d}(t),{\pmb{u}(t-d)})}{\sigma^{2}(\pmb{u}_{d}(t)) \sigma^{2}({\pmb{u}(t)})} ,
\end{equation}
\begin{figure}[h]
\centering
    \includegraphics[width=1\linewidth]{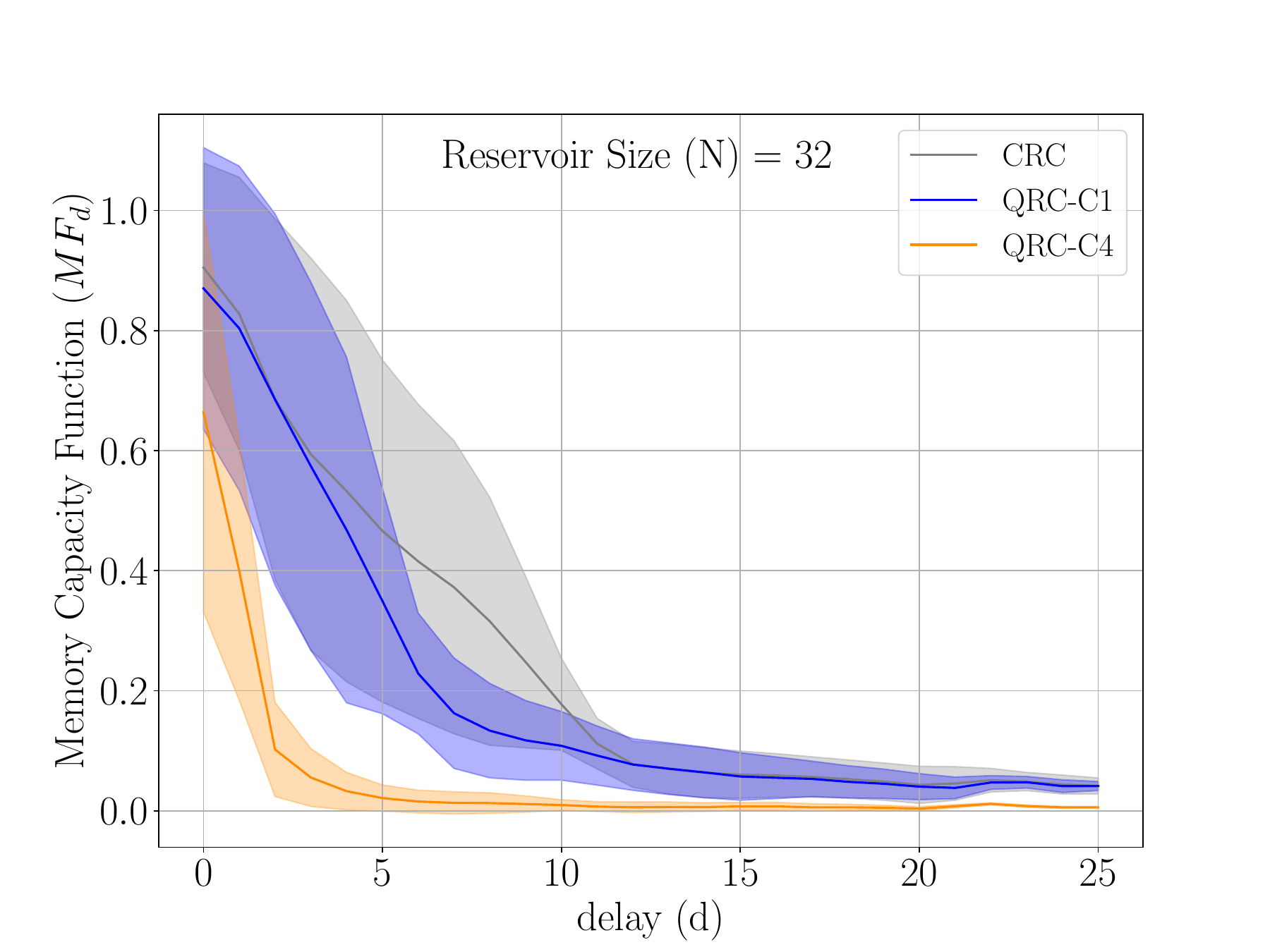}
    \caption{Linear Memory Capacity (MC) of reservoir networks, mean and standard deviation with hyperparameters. For a summary of hyperparameters cf.~Tab.~\ref{tab: param_LOR63}. For a reservoir size $(N) = 32$.}
    \label{fig:QRC_MC32}
\end{figure}
\begin{figure}[b]
    \includegraphics[width=1\linewidth]{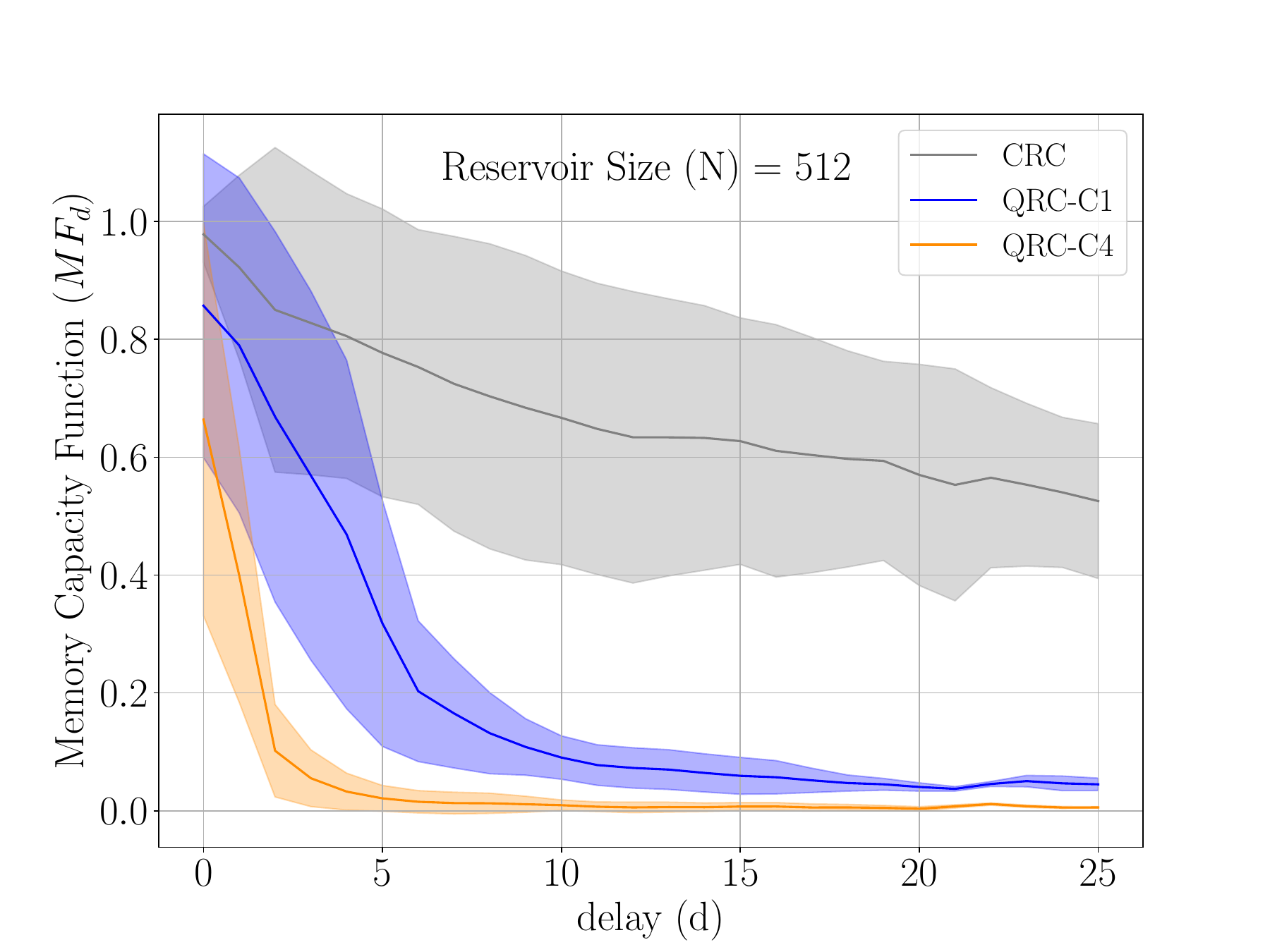}
    \caption{Linear Memory Capacity (MC) of reservoir networks, mean and standard deviation with hyperparameters. For a summary of hyperparameters cf.~Tab.~\ref{tab: param_LOR63}. For a reservoir size $(N) = 512$.}
\label{fig:QRC_MC512}
\end{figure}

where $cov(\cdot,\cdot)$ is the covariance, $\sigma^2(\cdot)$ is the variance, $d$ is the delay, $\pmb{u}(t-d)$ is the delayed prediction for a particular $d$ value and $\pmb{u}_{d}(t)$ is the current true input value. Although the total memory capacity (MC) comprises infinitely many terms ($d_{max}= \infty$), for practical reasons it is sufficient to truncate the sum at a finite $d_{max}$. For a comparison of the memory capacities of CRC and QRC we choose $d_{max}=25$. While the MC quantifies the linear memory of the reservoir, the dynamical system itself has some correlations that can increase the MC. To quantify the memory of the reservoir itself, we train our quantum and classical reservoir networks on statistically independent (i.i.d) uniform distributions. In Figs.~\ref{fig:QRC_MC32} and \ref{fig:QRC_MC512}, we compare the linear memory capacity of the classical reservoir with QRC-C1 (architecture with recurrence) and QRC-C4  (recurrence-free architecture). The global MC is bounded by the size of the reservoir and is sensitive to the choice of hyperparameters \cite{carroll_optimizing_2022}. We compute our MC for various hyperparameter values and reservoir sizes ($N=32,512$).  

The results in Fig.~\ref{fig:QRC_MC32} and \ref{fig:QRC_MC512} demonstrate that CRC exhibits the highest MC value followed by QRC-C1 and then QRC-C4. A high memory capacity in CRC highlights that the reservoir contains linear information about the input time series for a larger delay value $d$. In contrast, QRC-C4 does not have any recurrent or feedback loop. Due to the removal of the recurrence reservoir memory is lost. We emphasize that the higher MC value is not indicative of better predictability. This is because of two reasons: (a) The short-term memory capacity quantifies only the linear memory of the reservoir. (b) Maximizing the linear memory and nonlinear functional approximation are mutually exclusive and are referred to as the memory non-linearity trade-off \cite{storm_constraints_2022}. We find comparable performances in short and long-term prediction tasks, irrespective of the MC value, for both classical and quantum reservoirs.

\bibliography{References}

\begin{thebibliography}{63}%
\makeatletter
\providecommand \@ifxundefined [1]{%
 \@ifx{#1\undefined}
}%
\providecommand \@ifnum [1]{%
 \ifnum #1\expandafter \@firstoftwo
 \else \expandafter \@secondoftwo
 \fi
}%
\providecommand \@ifx [1]{%
 \ifx #1\expandafter \@firstoftwo
 \else \expandafter \@secondoftwo
 \fi
}%
\providecommand \natexlab [1]{#1}%
\providecommand \enquote  [1]{``#1''}%
\providecommand \bibnamefont  [1]{#1}%
\providecommand \bibfnamefont [1]{#1}%
\providecommand \citenamefont [1]{#1}%
\providecommand \href@noop [0]{\@secondoftwo}%
\providecommand \href [0]{\begingroup \@sanitize@url \@href}%
\providecommand \@href[1]{\@@startlink{#1}\@@href}%
\providecommand \@@href[1]{\endgroup#1\@@endlink}%
\providecommand \@sanitize@url [0]{\catcode `\\12\catcode `\$12\catcode `\&12\catcode `\#12\catcode `\^12\catcode `\_12\catcode `\%12\relax}%
\providecommand \@@startlink[1]{}%
\providecommand \@@endlink[0]{}%
\providecommand \url  [0]{\begingroup\@sanitize@url \@url }%
\providecommand \@url [1]{\endgroup\@href {#1}{\urlprefix }}%
\providecommand \urlprefix  [0]{URL }%
\providecommand \Eprint [0]{\href }%
\providecommand \doibase [0]{https://doi.org/}%
\providecommand \selectlanguage [0]{\@gobble}%
\providecommand \bibinfo  [0]{\@secondoftwo}%
\providecommand \bibfield  [0]{\@secondoftwo}%
\providecommand \translation [1]{[#1]}%
\providecommand \BibitemOpen [0]{}%
\providecommand \bibitemStop [0]{}%
\providecommand \bibitemNoStop [0]{.\EOS\space}%
\providecommand \EOS [0]{\spacefactor3000\relax}%
\providecommand \BibitemShut  [1]{\csname bibitem#1\endcsname}%
\let\auto@bib@innerbib\@empty
\bibitem [{\citenamefont {Cheng}\ \emph {et~al.}(2015)\citenamefont {Cheng}, \citenamefont {Sa-Ngasoongsong}, \citenamefont {Beyca}, \citenamefont {Le}, \citenamefont {Yang}, \citenamefont {Kong},\ and\ \citenamefont {Bukkapatnam}}]{cheng2015time}%
  \BibitemOpen
  \bibfield  {author} {\bibinfo {author} {\bibfnamefont {C.}~\bibnamefont {Cheng}}, \bibinfo {author} {\bibfnamefont {A.}~\bibnamefont {Sa-Ngasoongsong}}, \bibinfo {author} {\bibfnamefont {O.}~\bibnamefont {Beyca}}, \bibinfo {author} {\bibfnamefont {T.}~\bibnamefont {Le}}, \bibinfo {author} {\bibfnamefont {H.}~\bibnamefont {Yang}}, \bibinfo {author} {\bibfnamefont {Z.}~\bibnamefont {Kong}},\ and\ \bibinfo {author} {\bibfnamefont {S.~T.}\ \bibnamefont {Bukkapatnam}},\ }\bibfield  {title} {\bibinfo {title} {Time series forecasting for nonlinear and non-stationary processes: a review and comparative study},\ }\href@noop {} {\bibfield  {journal} {\bibinfo  {journal} {Iie Transactions}\ }\textbf {\bibinfo {volume} {47}},\ \bibinfo {pages} {1053} (\bibinfo {year} {2015})}\BibitemShut {NoStop}%
\bibitem [{\citenamefont {Ghadami}\ and\ \citenamefont {Epureanu}(2022)}]{ghadami2022data}%
  \BibitemOpen
  \bibfield  {author} {\bibinfo {author} {\bibfnamefont {A.}~\bibnamefont {Ghadami}}\ and\ \bibinfo {author} {\bibfnamefont {B.~I.}\ \bibnamefont {Epureanu}},\ }\bibfield  {title} {\bibinfo {title} {Data-driven prediction in dynamical systems: recent developments},\ }\href@noop {} {\bibfield  {journal} {\bibinfo  {journal} {Philosophical Transactions of the Royal Society A}\ }\textbf {\bibinfo {volume} {380}},\ \bibinfo {pages} {20210213} (\bibinfo {year} {2022})}\BibitemShut {NoStop}%
\bibitem [{\citenamefont {Dadhich}\ \emph {et~al.}(2021)\citenamefont {Dadhich}, \citenamefont {Pathak}, \citenamefont {Mittal},\ and\ \citenamefont {Doshi}}]{holmstrom2016machine}%
  \BibitemOpen
  \bibfield  {author} {\bibinfo {author} {\bibfnamefont {S.}~\bibnamefont {Dadhich}}, \bibinfo {author} {\bibfnamefont {V.}~\bibnamefont {Pathak}}, \bibinfo {author} {\bibfnamefont {R.}~\bibnamefont {Mittal}},\ and\ \bibinfo {author} {\bibfnamefont {R.}~\bibnamefont {Doshi}},\ }\bibinfo {title} {Chapter 10 machine learning for weather forecasting},\ in\ \href {https://doi.org/doi:10.1515/9783110702514-010} {\emph {\bibinfo {booktitle} {Machine Learning for Sustainable Development}}},\ \bibinfo {editor} {edited by\ \bibinfo {editor} {\bibfnamefont {K.~K.}\ \bibnamefont {Hiran}}, \bibinfo {editor} {\bibfnamefont {D.}~\bibnamefont {Khazanchi}}, \bibinfo {editor} {\bibfnamefont {A.~K.}\ \bibnamefont {Vyas}},\ and\ \bibinfo {editor} {\bibfnamefont {S.}~\bibnamefont {Padmanaban}}}\ (\bibinfo  {publisher} {De Gruyter},\ \bibinfo {address} {Berlin, Boston},\ \bibinfo {year} {2021})\ pp.\ \bibinfo {pages} {161--174}\BibitemShut {NoStop}%
\bibitem [{\citenamefont {Rolnick}\ \emph {et~al.}(2022)\citenamefont {Rolnick}, \citenamefont {Donti}, \citenamefont {Kaack}, \citenamefont {Kochanski}, \citenamefont {Lacoste}, \citenamefont {Sankaran}, \citenamefont {Ross}, \citenamefont {Milojevic-Dupont}, \citenamefont {Jaques}, \citenamefont {Waldman-Brown} \emph {et~al.}}]{rolnick2022tackling}%
  \BibitemOpen
  \bibfield  {author} {\bibinfo {author} {\bibfnamefont {D.}~\bibnamefont {Rolnick}}, \bibinfo {author} {\bibfnamefont {P.~L.}\ \bibnamefont {Donti}}, \bibinfo {author} {\bibfnamefont {L.~H.}\ \bibnamefont {Kaack}}, \bibinfo {author} {\bibfnamefont {K.}~\bibnamefont {Kochanski}}, \bibinfo {author} {\bibfnamefont {A.}~\bibnamefont {Lacoste}}, \bibinfo {author} {\bibfnamefont {K.}~\bibnamefont {Sankaran}}, \bibinfo {author} {\bibfnamefont {A.~S.}\ \bibnamefont {Ross}}, \bibinfo {author} {\bibfnamefont {N.}~\bibnamefont {Milojevic-Dupont}}, \bibinfo {author} {\bibfnamefont {N.}~\bibnamefont {Jaques}}, \bibinfo {author} {\bibfnamefont {A.}~\bibnamefont {Waldman-Brown}}, \emph {et~al.},\ }\bibfield  {title} {\bibinfo {title} {Tackling climate change with machine learning},\ }\href@noop {} {\bibfield  {journal} {\bibinfo  {journal} {ACM Computing Surveys (CSUR)}\ }\textbf {\bibinfo {volume} {55}},\ \bibinfo {pages} {1} (\bibinfo {year} {2022})}\BibitemShut {NoStop}%
\bibitem [{\citenamefont {Sezer}\ \emph {et~al.}(2020)\citenamefont {Sezer}, \citenamefont {Gudelek},\ and\ \citenamefont {Ozbayoglu}}]{SEZER2020106181}%
  \BibitemOpen
  \bibfield  {author} {\bibinfo {author} {\bibfnamefont {O.~B.}\ \bibnamefont {Sezer}}, \bibinfo {author} {\bibfnamefont {M.~U.}\ \bibnamefont {Gudelek}},\ and\ \bibinfo {author} {\bibfnamefont {A.~M.}\ \bibnamefont {Ozbayoglu}},\ }\bibfield  {title} {\bibinfo {title} {Financial time series forecasting with deep learning : A systematic literature review: 2005–2019},\ }\href {https://doi.org/https://doi.org/10.1016/j.asoc.2020.106181} {\bibfield  {journal} {\bibinfo  {journal} {Applied Soft Computing}\ }\textbf {\bibinfo {volume} {90}},\ \bibinfo {pages} {106181} (\bibinfo {year} {2020})}\BibitemShut {NoStop}%
\bibitem [{\citenamefont {Huhn}\ and\ \citenamefont {Magri}(2020{\natexlab{a}})}]{huhn_stability_2020}%
  \BibitemOpen
  \bibfield  {author} {\bibinfo {author} {\bibfnamefont {F.}~\bibnamefont {Huhn}}\ and\ \bibinfo {author} {\bibfnamefont {L.}~\bibnamefont {Magri}},\ }\bibfield  {title} {\bibinfo {title} {Stability, sensitivity and optimisation of chaotic acoustic oscillations},\ }\href {https://doi.org/10.1017/jfm.2019.828} {\bibfield  {journal} {\bibinfo  {journal} {Journal of Fluid Mechanics}\ }\textbf {\bibinfo {volume} {882}},\ \bibinfo {pages} {A24} (\bibinfo {year} {2020}{\natexlab{a}})}\BibitemShut {NoStop}%
\bibitem [{\citenamefont {Srinivasan}\ \emph {et~al.}(2019)\citenamefont {Srinivasan}, \citenamefont {Guastoni}, \citenamefont {Azizpour}, \citenamefont {Schlatter},\ and\ \citenamefont {Vinuesa}}]{srinivasan_predictions_2019}%
  \BibitemOpen
  \bibfield  {author} {\bibinfo {author} {\bibfnamefont {P.~A.}\ \bibnamefont {Srinivasan}}, \bibinfo {author} {\bibfnamefont {L.}~\bibnamefont {Guastoni}}, \bibinfo {author} {\bibfnamefont {H.}~\bibnamefont {Azizpour}}, \bibinfo {author} {\bibfnamefont {P.}~\bibnamefont {Schlatter}},\ and\ \bibinfo {author} {\bibfnamefont {R.}~\bibnamefont {Vinuesa}},\ }\bibfield  {title} {\bibinfo {title} {Predictions of turbulent shear flows using deep neural networks},\ }\href {https://doi.org/10.1103/PhysRevFluids.4.054603} {\bibfield  {journal} {\bibinfo  {journal} {Physical Review Fluids}\ }\textbf {\bibinfo {volume} {4}},\ \bibinfo {pages} {054603} (\bibinfo {year} {2019})}\BibitemShut {NoStop}%
\bibitem [{\citenamefont {Doan}\ \emph {et~al.}(2019{\natexlab{a}})\citenamefont {Doan}, \citenamefont {Polifke},\ and\ \citenamefont {Magri}}]{doan2019physicsaware}%
  \BibitemOpen
  \bibfield  {author} {\bibinfo {author} {\bibfnamefont {N.~A.~K.}\ \bibnamefont {Doan}}, \bibinfo {author} {\bibfnamefont {W.}~\bibnamefont {Polifke}},\ and\ \bibinfo {author} {\bibfnamefont {L.}~\bibnamefont {Magri}},\ }\href@noop {} {\bibinfo {title} {A physics-aware machine to predict extreme events in turbulence}} (\bibinfo {year} {2019}{\natexlab{a}}),\ \Eprint {https://arxiv.org/abs/1912.10994} {arXiv:1912.10994 [physics.flu-dyn]} \BibitemShut {NoStop}%
\bibitem [{\citenamefont {Vlachas}\ \emph {et~al.}(2020)\citenamefont {Vlachas}, \citenamefont {Pathak}, \citenamefont {Hunt}, \citenamefont {Sapsis}, \citenamefont {Girvan}, \citenamefont {Ott},\ and\ \citenamefont {Koumoutsakos}}]{vlachas_backpropagation_2020}%
  \BibitemOpen
  \bibfield  {author} {\bibinfo {author} {\bibfnamefont {P.~R.}\ \bibnamefont {Vlachas}}, \bibinfo {author} {\bibfnamefont {J.}~\bibnamefont {Pathak}}, \bibinfo {author} {\bibfnamefont {B.~R.}\ \bibnamefont {Hunt}}, \bibinfo {author} {\bibfnamefont {T.~P.}\ \bibnamefont {Sapsis}}, \bibinfo {author} {\bibfnamefont {M.}~\bibnamefont {Girvan}}, \bibinfo {author} {\bibfnamefont {E.}~\bibnamefont {Ott}},\ and\ \bibinfo {author} {\bibfnamefont {P.}~\bibnamefont {Koumoutsakos}},\ }\bibfield  {title} {\bibinfo {title} {Backpropagation algorithms and {Reservoir} {Computing} in {Recurrent} {Neural} {Networks} for the forecasting of complex spatiotemporal dynamics},\ }\href {https://doi.org/10.1016/j.neunet.2020.02.016} {\bibfield  {journal} {\bibinfo  {journal} {Neural Networks}\ }\textbf {\bibinfo {volume} {126}},\ \bibinfo {pages} {191} (\bibinfo {year} {2020})}\BibitemShut {NoStop}%
\bibitem [{\citenamefont {Racca}\ and\ \citenamefont {Magri}(2022{\natexlab{a}})}]{racca_statistical_2022}%
  \BibitemOpen
  \bibfield  {author} {\bibinfo {author} {\bibfnamefont {A.}~\bibnamefont {Racca}}\ and\ \bibinfo {author} {\bibfnamefont {L.}~\bibnamefont {Magri}},\ }\bibfield  {title} {\bibinfo {title} {Statistical prediction of extreme events from small datasets},\ }in\ \href@noop {} {\emph {\bibinfo {booktitle} {Computational Science -- ICCS 2022}}},\ \bibinfo {editor} {edited by\ \bibinfo {editor} {\bibfnamefont {D.}~\bibnamefont {Groen}}, \bibinfo {editor} {\bibfnamefont {C.}~\bibnamefont {de~Mulatier}}, \bibinfo {editor} {\bibfnamefont {M.}~\bibnamefont {Paszynski}}, \bibinfo {editor} {\bibfnamefont {V.~V.}\ \bibnamefont {Krzhizhanovskaya}}, \bibinfo {editor} {\bibfnamefont {J.~J.}\ \bibnamefont {Dongarra}},\ and\ \bibinfo {editor} {\bibfnamefont {P.~M.~A.}\ \bibnamefont {Sloot}}}\ (\bibinfo  {publisher} {Springer International Publishing},\ \bibinfo {address} {Cham},\ \bibinfo {year} {2022})\ pp.\ \bibinfo {pages} {707--713}\BibitemShut {NoStop}%
\bibitem [{\citenamefont {Doan}\ \emph {et~al.}(2019{\natexlab{b}})\citenamefont {Doan}, \citenamefont {Polifke},\ and\ \citenamefont {Magri}}]{10.1007/978-3-030-22747-0_15}%
  \BibitemOpen
  \bibfield  {author} {\bibinfo {author} {\bibfnamefont {N.~A.~K.}\ \bibnamefont {Doan}}, \bibinfo {author} {\bibfnamefont {W.}~\bibnamefont {Polifke}},\ and\ \bibinfo {author} {\bibfnamefont {L.}~\bibnamefont {Magri}},\ }\bibfield  {title} {\bibinfo {title} {Physics-informed echo state networks for chaotic systems forecasting},\ }in\ \href@noop {} {\emph {\bibinfo {booktitle} {Computational Science -- ICCS 2019}}},\ \bibinfo {editor} {edited by\ \bibinfo {editor} {\bibfnamefont {J.~M.~F.}\ \bibnamefont {Rodrigues}}, \bibinfo {editor} {\bibfnamefont {P.~J.~S.}\ \bibnamefont {Cardoso}}, \bibinfo {editor} {\bibfnamefont {J.}~\bibnamefont {Monteiro}}, \bibinfo {editor} {\bibfnamefont {R.}~\bibnamefont {Lam}}, \bibinfo {editor} {\bibfnamefont {V.~V.}\ \bibnamefont {Krzhizhanovskaya}}, \bibinfo {editor} {\bibfnamefont {M.~H.}\ \bibnamefont {Lees}}, \bibinfo {editor} {\bibfnamefont {J.~J.}\ \bibnamefont {Dongarra}},\ and\ \bibinfo {editor} {\bibfnamefont {P.~M.}\ \bibnamefont {Sloot}}}\ (\bibinfo  {publisher}
  {Springer International Publishing},\ \bibinfo {address} {Cham},\ \bibinfo {year} {2019})\ pp.\ \bibinfo {pages} {192--198}\BibitemShut {NoStop}%
\bibitem [{\citenamefont {Margazoglou}\ and\ \citenamefont {Magri}(2023)}]{margazoglou_stability_2023}%
  \BibitemOpen
  \bibfield  {author} {\bibinfo {author} {\bibfnamefont {G.}~\bibnamefont {Margazoglou}}\ and\ \bibinfo {author} {\bibfnamefont {L.}~\bibnamefont {Magri}},\ }\bibfield  {title} {\bibinfo {title} {Stability analysis of chaotic systems from data},\ }\href {https://doi.org/10.1007/s11071-023-08285-1} {\bibfield  {journal} {\bibinfo  {journal} {Nonlinear Dynamics}\ }\textbf {\bibinfo {volume} {111}},\ \bibinfo {pages} {8799} (\bibinfo {year} {2023})}\BibitemShut {NoStop}%
\bibitem [{\citenamefont {Racca}\ and\ \citenamefont {Magri}(2021)}]{RACCA2021252}%
  \BibitemOpen
  \bibfield  {author} {\bibinfo {author} {\bibfnamefont {A.}~\bibnamefont {Racca}}\ and\ \bibinfo {author} {\bibfnamefont {L.}~\bibnamefont {Magri}},\ }\bibfield  {title} {\bibinfo {title} {Robust optimization and validation of echo state networks for learning chaotic dynamics},\ }\href {https://doi.org/https://doi.org/10.1016/j.neunet.2021.05.004} {\bibfield  {journal} {\bibinfo  {journal} {Neural Networks}\ }\textbf {\bibinfo {volume} {142}},\ \bibinfo {pages} {252} (\bibinfo {year} {2021})}\BibitemShut {NoStop}%
\bibitem [{\citenamefont {Werbos}(1990)}]{58337}%
  \BibitemOpen
  \bibfield  {author} {\bibinfo {author} {\bibfnamefont {P.}~\bibnamefont {Werbos}},\ }\bibfield  {title} {\bibinfo {title} {Backpropagation through time: what it does and how to do it},\ }\href {https://doi.org/10.1109/5.58337} {\bibfield  {journal} {\bibinfo  {journal} {Proceedings of the IEEE}\ }\textbf {\bibinfo {volume} {78}},\ \bibinfo {pages} {1550} (\bibinfo {year} {1990})}\BibitemShut {NoStop}%
\bibitem [{\citenamefont {Jaeger}(2001{\natexlab{a}})}]{jaeger__2001}%
  \BibitemOpen
  \bibfield  {author} {\bibinfo {author} {\bibfnamefont {H.}~\bibnamefont {Jaeger}},\ }\bibfield  {title} {\bibinfo {title} {The “echo state” approach to analysing and training recurrent neural networks-with an erratum note},\ }\href@noop {} {\bibfield  {journal} {\bibinfo  {journal} {Bonn, Germany: German National Research Center for Information Technology GMD Technical Report}\ }\textbf {\bibinfo {volume} {148}},\ \bibinfo {pages} {13} (\bibinfo {year} {2001}{\natexlab{a}})}\BibitemShut {NoStop}%
\bibitem [{\citenamefont {Maass}\ \emph {et~al.}(2002)\citenamefont {Maass}, \citenamefont {Natschläger},\ and\ \citenamefont {Markram}}]{10.1162/089976602760407955}%
  \BibitemOpen
  \bibfield  {author} {\bibinfo {author} {\bibfnamefont {W.}~\bibnamefont {Maass}}, \bibinfo {author} {\bibfnamefont {T.}~\bibnamefont {Natschläger}},\ and\ \bibinfo {author} {\bibfnamefont {H.}~\bibnamefont {Markram}},\ }\bibfield  {title} {\bibinfo {title} {{Real-Time Computing Without Stable States: A New Framework for Neural Computation Based on Perturbations}},\ }\href {https://doi.org/10.1162/089976602760407955} {\bibfield  {journal} {\bibinfo  {journal} {Neural Computation}\ }\textbf {\bibinfo {volume} {14}},\ \bibinfo {pages} {2531} (\bibinfo {year} {2002})},\ \Eprint {https://arxiv.org/abs/https://direct.mit.edu/neco/article-pdf/14/11/2531/815288/089976602760407955.pdf} {https://direct.mit.edu/neco/article-pdf/14/11/2531/815288/089976602760407955.pdf} \BibitemShut {NoStop}%
\bibitem [{\citenamefont {Boedecker}\ \emph {et~al.}(2012)\citenamefont {Boedecker}, \citenamefont {Obst}, \citenamefont {Lizier}, \citenamefont {Mayer},\ and\ \citenamefont {Asada}}]{boedecker_information_2012}%
  \BibitemOpen
  \bibfield  {author} {\bibinfo {author} {\bibfnamefont {J.}~\bibnamefont {Boedecker}}, \bibinfo {author} {\bibfnamefont {O.}~\bibnamefont {Obst}}, \bibinfo {author} {\bibfnamefont {J.~T.}\ \bibnamefont {Lizier}}, \bibinfo {author} {\bibfnamefont {N.~M.}\ \bibnamefont {Mayer}},\ and\ \bibinfo {author} {\bibfnamefont {M.}~\bibnamefont {Asada}},\ }\bibfield  {title} {\bibinfo {title} {Information processing in echo state networks at the edge of chaos},\ }\href {https://doi.org/10.1007/s12064-011-0146-8} {\bibfield  {journal} {\bibinfo  {journal} {Theory in Biosciences}\ }\textbf {\bibinfo {volume} {131}},\ \bibinfo {pages} {205} (\bibinfo {year} {2012})}\BibitemShut {NoStop}%
\bibitem [{\citenamefont {Racca}\ and\ \citenamefont {Magri}(2022{\natexlab{b}})}]{racca_data-driven_2022}%
  \BibitemOpen
  \bibfield  {author} {\bibinfo {author} {\bibfnamefont {A.}~\bibnamefont {Racca}}\ and\ \bibinfo {author} {\bibfnamefont {L.}~\bibnamefont {Magri}},\ }\bibfield  {title} {\bibinfo {title} {Data-driven prediction and control of extreme events in a chaotic flow},\ }\href {https://doi.org/10.1103/PhysRevFluids.7.104402} {\bibfield  {journal} {\bibinfo  {journal} {Physical Review Fluids}\ }\textbf {\bibinfo {volume} {7}},\ \bibinfo {pages} {104402} (\bibinfo {year} {2022}{\natexlab{b}})}\BibitemShut {NoStop}%
\bibitem [{\citenamefont {Doan}\ \emph {et~al.}(2021)\citenamefont {Doan}, \citenamefont {Polifke},\ and\ \citenamefont {Magri}}]{doan2021short}%
  \BibitemOpen
  \bibfield  {author} {\bibinfo {author} {\bibfnamefont {N.~A.~K.}\ \bibnamefont {Doan}}, \bibinfo {author} {\bibfnamefont {W.}~\bibnamefont {Polifke}},\ and\ \bibinfo {author} {\bibfnamefont {L.}~\bibnamefont {Magri}},\ }\bibfield  {title} {\bibinfo {title} {Short-and long-term predictions of chaotic flows and extreme events: a physics-constrained reservoir computing approach},\ }\href@noop {} {\bibfield  {journal} {\bibinfo  {journal} {Proceedings of the Royal Society A}\ }\textbf {\bibinfo {volume} {477}},\ \bibinfo {pages} {20210135} (\bibinfo {year} {2021})}\BibitemShut {NoStop}%
\bibitem [{\citenamefont {Feynman}(2018)}]{feynman2018simulating}%
  \BibitemOpen
  \bibfield  {author} {\bibinfo {author} {\bibfnamefont {R.~P.}\ \bibnamefont {Feynman}},\ }\bibfield  {title} {\bibinfo {title} {Simulating physics with computers},\ }in\ \href@noop {} {\emph {\bibinfo {booktitle} {Feynman and computation}}}\ (\bibinfo  {publisher} {CRC Press},\ \bibinfo {year} {2018})\ pp.\ \bibinfo {pages} {133--153}\BibitemShut {NoStop}%
\bibitem [{\citenamefont {Shor}(1994)}]{365700}%
  \BibitemOpen
  \bibfield  {author} {\bibinfo {author} {\bibfnamefont {P.}~\bibnamefont {Shor}},\ }\bibfield  {title} {\bibinfo {title} {Algorithms for quantum computation: discrete logarithms and factoring},\ }in\ \href {https://doi.org/10.1109/SFCS.1994.365700} {\emph {\bibinfo {booktitle} {Proceedings 35th Annual Symposium on Foundations of Computer Science}}}\ (\bibinfo {year} {1994})\ pp.\ \bibinfo {pages} {124--134}\BibitemShut {NoStop}%
\bibitem [{\citenamefont {Harrow}\ \emph {et~al.}(2009)\citenamefont {Harrow}, \citenamefont {Hassidim},\ and\ \citenamefont {Lloyd}}]{Harrow_2009}%
  \BibitemOpen
  \bibfield  {author} {\bibinfo {author} {\bibfnamefont {A.~W.}\ \bibnamefont {Harrow}}, \bibinfo {author} {\bibfnamefont {A.}~\bibnamefont {Hassidim}},\ and\ \bibinfo {author} {\bibfnamefont {S.}~\bibnamefont {Lloyd}},\ }\bibfield  {title} {\bibinfo {title} {Quantum algorithm for linear systems of equations},\ }\bibfield  {journal} {\bibinfo  {journal} {Physical Review Letters}\ }\textbf {\bibinfo {volume} {103}},\ \href {https://doi.org/10.1103/physrevlett.103.150502} {10.1103/physrevlett.103.150502} (\bibinfo {year} {2009})\BibitemShut {NoStop}%
\bibitem [{\citenamefont {Bravo-Prieto}\ \emph {et~al.}(2023)\citenamefont {Bravo-Prieto}, \citenamefont {LaRose}, \citenamefont {Cerezo}, \citenamefont {Subasi}, \citenamefont {Cincio},\ and\ \citenamefont {Coles}}]{bravoprieto2020variational}%
  \BibitemOpen
  \bibfield  {author} {\bibinfo {author} {\bibfnamefont {C.}~\bibnamefont {Bravo-Prieto}}, \bibinfo {author} {\bibfnamefont {R.}~\bibnamefont {LaRose}}, \bibinfo {author} {\bibfnamefont {M.}~\bibnamefont {Cerezo}}, \bibinfo {author} {\bibfnamefont {Y.}~\bibnamefont {Subasi}}, \bibinfo {author} {\bibfnamefont {L.}~\bibnamefont {Cincio}},\ and\ \bibinfo {author} {\bibfnamefont {P.~J.}\ \bibnamefont {Coles}},\ }\bibfield  {title} {\bibinfo {title} {Variational quantum linear solver},\ }\href@noop {} {\bibfield  {journal} {\bibinfo  {journal} {Quantum}\ }\textbf {\bibinfo {volume} {7}},\ \bibinfo {pages} {1188} (\bibinfo {year} {2023})}\BibitemShut {NoStop}%
\bibitem [{\citenamefont {Bharti}\ \emph {et~al.}(2022)\citenamefont {Bharti}, \citenamefont {Cervera-Lierta}, \citenamefont {Kyaw}, \citenamefont {Haug}, \citenamefont {Alperin-Lea}, \citenamefont {Anand}, \citenamefont {Degroote}, \citenamefont {Heimonen}, \citenamefont {Kottmann}, \citenamefont {Menke} \emph {et~al.}}]{bharti2022noisy}%
  \BibitemOpen
  \bibfield  {author} {\bibinfo {author} {\bibfnamefont {K.}~\bibnamefont {Bharti}}, \bibinfo {author} {\bibfnamefont {A.}~\bibnamefont {Cervera-Lierta}}, \bibinfo {author} {\bibfnamefont {T.~H.}\ \bibnamefont {Kyaw}}, \bibinfo {author} {\bibfnamefont {T.}~\bibnamefont {Haug}}, \bibinfo {author} {\bibfnamefont {S.}~\bibnamefont {Alperin-Lea}}, \bibinfo {author} {\bibfnamefont {A.}~\bibnamefont {Anand}}, \bibinfo {author} {\bibfnamefont {M.}~\bibnamefont {Degroote}}, \bibinfo {author} {\bibfnamefont {H.}~\bibnamefont {Heimonen}}, \bibinfo {author} {\bibfnamefont {J.~S.}\ \bibnamefont {Kottmann}}, \bibinfo {author} {\bibfnamefont {T.}~\bibnamefont {Menke}}, \emph {et~al.},\ }\bibfield  {title} {\bibinfo {title} {Noisy intermediate-scale quantum algorithms},\ }\href@noop {} {\bibfield  {journal} {\bibinfo  {journal} {Reviews of Modern Physics}\ }\textbf {\bibinfo {volume} {94}},\ \bibinfo {pages} {015004} (\bibinfo {year} {2022})}\BibitemShut {NoStop}%
\bibitem [{\citenamefont {Cerezo}\ \emph {et~al.}(2021)\citenamefont {Cerezo}, \citenamefont {Arrasmith}, \citenamefont {Babbush}, \citenamefont {Benjamin}, \citenamefont {Endo}, \citenamefont {Fujii}, \citenamefont {McClean}, \citenamefont {Mitarai}, \citenamefont {Yuan}, \citenamefont {Cincio},\ and\ \citenamefont {Coles}}]{cerezo_variational_2021}%
  \BibitemOpen
  \bibfield  {author} {\bibinfo {author} {\bibfnamefont {M.}~\bibnamefont {Cerezo}}, \bibinfo {author} {\bibfnamefont {A.}~\bibnamefont {Arrasmith}}, \bibinfo {author} {\bibfnamefont {R.}~\bibnamefont {Babbush}}, \bibinfo {author} {\bibfnamefont {S.~C.}\ \bibnamefont {Benjamin}}, \bibinfo {author} {\bibfnamefont {S.}~\bibnamefont {Endo}}, \bibinfo {author} {\bibfnamefont {K.}~\bibnamefont {Fujii}}, \bibinfo {author} {\bibfnamefont {J.~R.}\ \bibnamefont {McClean}}, \bibinfo {author} {\bibfnamefont {K.}~\bibnamefont {Mitarai}}, \bibinfo {author} {\bibfnamefont {X.}~\bibnamefont {Yuan}}, \bibinfo {author} {\bibfnamefont {L.}~\bibnamefont {Cincio}},\ and\ \bibinfo {author} {\bibfnamefont {P.~J.}\ \bibnamefont {Coles}},\ }\bibfield  {title} {\bibinfo {title} {Variational quantum algorithms},\ }\href {https://doi.org/10.1038/s42254-021-00348-9} {\bibfield  {journal} {\bibinfo  {journal} {Nature Reviews Physics}\ }\textbf {\bibinfo {volume} {3}},\ \bibinfo {pages} {625} (\bibinfo {year} {2021})},\ \bibinfo {note}
  {number: 9 Publisher: Nature Publishing Group}\BibitemShut {NoStop}%
\bibitem [{\citenamefont {Mitarai}\ \emph {et~al.}(2018)\citenamefont {Mitarai}, \citenamefont {Negoro}, \citenamefont {Kitagawa},\ and\ \citenamefont {Fujii}}]{Mitarai_2018}%
  \BibitemOpen
  \bibfield  {author} {\bibinfo {author} {\bibfnamefont {K.}~\bibnamefont {Mitarai}}, \bibinfo {author} {\bibfnamefont {M.}~\bibnamefont {Negoro}}, \bibinfo {author} {\bibfnamefont {M.}~\bibnamefont {Kitagawa}},\ and\ \bibinfo {author} {\bibfnamefont {K.}~\bibnamefont {Fujii}},\ }\bibfield  {title} {\bibinfo {title} {Quantum circuit learning},\ }\bibfield  {journal} {\bibinfo  {journal} {Physical Review A}\ }\textbf {\bibinfo {volume} {98}},\ \href {https://doi.org/10.1103/physreva.98.032309} {10.1103/physreva.98.032309} (\bibinfo {year} {2018})\BibitemShut {NoStop}%
\bibitem [{\citenamefont {Fujii}\ and\ \citenamefont {Nakajima}(2017)}]{fujii_harnessing_2017}%
  \BibitemOpen
  \bibfield  {author} {\bibinfo {author} {\bibfnamefont {K.}~\bibnamefont {Fujii}}\ and\ \bibinfo {author} {\bibfnamefont {K.}~\bibnamefont {Nakajima}},\ }\bibfield  {title} {\bibinfo {title} {Harnessing {Disordered}-{Ensemble} {Quantum} {Dynamics} for {Machine} {Learning}},\ }\href {https://doi.org/10.1103/PhysRevApplied.8.024030} {\bibfield  {journal} {\bibinfo  {journal} {Physical Review Applied}\ }\textbf {\bibinfo {volume} {8}},\ \bibinfo {pages} {024030} (\bibinfo {year} {2017})},\ \bibinfo {note} {publisher: American Physical Society}\BibitemShut {NoStop}%
\bibitem [{\citenamefont {Fujii}\ and\ \citenamefont {Nakajima}(2021)}]{Fujii2021}%
  \BibitemOpen
  \bibfield  {author} {\bibinfo {author} {\bibfnamefont {K.}~\bibnamefont {Fujii}}\ and\ \bibinfo {author} {\bibfnamefont {K.}~\bibnamefont {Nakajima}},\ }\bibinfo {title} {Quantum reservoir computing: A reservoir approach toward quantum machine learning on near-term quantum devices},\ in\ \href {https://doi.org/10.1007/978-981-13-1687-6_18} {\emph {\bibinfo {booktitle} {Reservoir Computing: Theory, Physical Implementations, and Applications}}},\ \bibinfo {editor} {edited by\ \bibinfo {editor} {\bibfnamefont {K.}~\bibnamefont {Nakajima}}\ and\ \bibinfo {editor} {\bibfnamefont {I.}~\bibnamefont {Fischer}}}\ (\bibinfo  {publisher} {Springer Singapore},\ \bibinfo {address} {Singapore},\ \bibinfo {year} {2021})\ pp.\ \bibinfo {pages} {423--450}\BibitemShut {NoStop}%
\bibitem [{\citenamefont {Chen}\ \emph {et~al.}(2022)\citenamefont {Chen}, \citenamefont {Fry}, \citenamefont {Deshmukh}, \citenamefont {Rastunkov},\ and\ \citenamefont {Stefanski}}]{chen2022reservoir}%
  \BibitemOpen
  \bibfield  {author} {\bibinfo {author} {\bibfnamefont {S.~Y.-C.}\ \bibnamefont {Chen}}, \bibinfo {author} {\bibfnamefont {D.}~\bibnamefont {Fry}}, \bibinfo {author} {\bibfnamefont {A.}~\bibnamefont {Deshmukh}}, \bibinfo {author} {\bibfnamefont {V.}~\bibnamefont {Rastunkov}},\ and\ \bibinfo {author} {\bibfnamefont {C.}~\bibnamefont {Stefanski}},\ }\href@noop {} {\bibinfo {title} {Reservoir computing via quantum recurrent neural networks}} (\bibinfo {year} {2022}),\ \Eprint {https://arxiv.org/abs/2211.02612} {arXiv:2211.02612 [cs.NE]} \BibitemShut {NoStop}%
\bibitem [{\citenamefont {Pfeffer}\ \emph {et~al.}(2022)\citenamefont {Pfeffer}, \citenamefont {Heyder},\ and\ \citenamefont {Schumacher}}]{pfeffer_hybrid_2022}%
  \BibitemOpen
  \bibfield  {author} {\bibinfo {author} {\bibfnamefont {P.}~\bibnamefont {Pfeffer}}, \bibinfo {author} {\bibfnamefont {F.}~\bibnamefont {Heyder}},\ and\ \bibinfo {author} {\bibfnamefont {J.}~\bibnamefont {Schumacher}},\ }\bibfield  {title} {\bibinfo {title} {Hybrid quantum-classical reservoir computing of thermal convection flow},\ }\href {https://doi.org/10.1103/PhysRevResearch.4.033176} {\bibfield  {journal} {\bibinfo  {journal} {Physical Review Research}\ }\textbf {\bibinfo {volume} {4}},\ \bibinfo {pages} {033176} (\bibinfo {year} {2022})}\BibitemShut {NoStop}%
\bibitem [{\citenamefont {Suzuki}\ \emph {et~al.}(2022)\citenamefont {Suzuki}, \citenamefont {Gao}, \citenamefont {Pradel}, \citenamefont {Yasuoka},\ and\ \citenamefont {Yamamoto}}]{suzuki_natural_2022}%
  \BibitemOpen
  \bibfield  {author} {\bibinfo {author} {\bibfnamefont {Y.}~\bibnamefont {Suzuki}}, \bibinfo {author} {\bibfnamefont {Q.}~\bibnamefont {Gao}}, \bibinfo {author} {\bibfnamefont {K.~C.}\ \bibnamefont {Pradel}}, \bibinfo {author} {\bibfnamefont {K.}~\bibnamefont {Yasuoka}},\ and\ \bibinfo {author} {\bibfnamefont {N.}~\bibnamefont {Yamamoto}},\ }\bibfield  {title} {\bibinfo {title} {Natural quantum reservoir computing for temporal information processing},\ }\href {https://doi.org/10.1038/s41598-022-05061-w} {\bibfield  {journal} {\bibinfo  {journal} {Scientific Reports}\ }\textbf {\bibinfo {volume} {12}},\ \bibinfo {pages} {1353} (\bibinfo {year} {2022})}\BibitemShut {NoStop}%
\bibitem [{\citenamefont {Fry}\ \emph {et~al.}(2023)\citenamefont {Fry}, \citenamefont {Deshmukh}, \citenamefont {Chen}, \citenamefont {Rastunkov},\ and\ \citenamefont {Markov}}]{fry2023optimizing}%
  \BibitemOpen
  \bibfield  {author} {\bibinfo {author} {\bibfnamefont {D.}~\bibnamefont {Fry}}, \bibinfo {author} {\bibfnamefont {A.}~\bibnamefont {Deshmukh}}, \bibinfo {author} {\bibfnamefont {S.~Y.-C.}\ \bibnamefont {Chen}}, \bibinfo {author} {\bibfnamefont {V.}~\bibnamefont {Rastunkov}},\ and\ \bibinfo {author} {\bibfnamefont {V.}~\bibnamefont {Markov}},\ }\href@noop {} {\bibinfo {title} {Optimizing quantum noise-induced reservoir computing for nonlinear and chaotic time series prediction}} (\bibinfo {year} {2023}),\ \Eprint {https://arxiv.org/abs/2303.05488} {arXiv:2303.05488 [quant-ph]} \BibitemShut {NoStop}%
\bibitem [{\citenamefont {Wudarski}\ \emph {et~al.}(2023)\citenamefont {Wudarski}, \citenamefont {O`Connor}, \citenamefont {Geaney}, \citenamefont {Asanjan}, \citenamefont {Wilson}, \citenamefont {Strbac}, \citenamefont {Lott},\ and\ \citenamefont {Venturelli}}]{wudarski2023hybrid}%
  \BibitemOpen
  \bibfield  {author} {\bibinfo {author} {\bibfnamefont {F.}~\bibnamefont {Wudarski}}, \bibinfo {author} {\bibfnamefont {D.}~\bibnamefont {O`Connor}}, \bibinfo {author} {\bibfnamefont {S.}~\bibnamefont {Geaney}}, \bibinfo {author} {\bibfnamefont {A.~A.}\ \bibnamefont {Asanjan}}, \bibinfo {author} {\bibfnamefont {M.}~\bibnamefont {Wilson}}, \bibinfo {author} {\bibfnamefont {E.}~\bibnamefont {Strbac}}, \bibinfo {author} {\bibfnamefont {P.~A.}\ \bibnamefont {Lott}},\ and\ \bibinfo {author} {\bibfnamefont {D.}~\bibnamefont {Venturelli}},\ }\href@noop {} {\bibinfo {title} {Hybrid quantum-classical reservoir computing for simulating chaotic systems}} (\bibinfo {year} {2023}),\ \Eprint {https://arxiv.org/abs/2311.14105} {arXiv:2311.14105 [quant-ph]} \BibitemShut {NoStop}%
\bibitem [{\citenamefont {Dudas}\ \emph {et~al.}(2023)\citenamefont {Dudas}, \citenamefont {Carles}, \citenamefont {Plouet}, \citenamefont {Mizrahi}, \citenamefont {Grollier},\ and\ \citenamefont {Marković}}]{dudas_quantum_2023}%
  \BibitemOpen
  \bibfield  {author} {\bibinfo {author} {\bibfnamefont {J.}~\bibnamefont {Dudas}}, \bibinfo {author} {\bibfnamefont {B.}~\bibnamefont {Carles}}, \bibinfo {author} {\bibfnamefont {E.}~\bibnamefont {Plouet}}, \bibinfo {author} {\bibfnamefont {F.~A.}\ \bibnamefont {Mizrahi}}, \bibinfo {author} {\bibfnamefont {J.}~\bibnamefont {Grollier}},\ and\ \bibinfo {author} {\bibfnamefont {D.}~\bibnamefont {Marković}},\ }\bibfield  {title} {\bibinfo {title} {Quantum reservoir computing implementation on coherently coupled quantum oscillators},\ }\href {https://doi.org/10.1038/s41534-023-00734-4} {\bibfield  {journal} {\bibinfo  {journal} {npj Quantum Information}\ }\textbf {\bibinfo {volume} {9}},\ \bibinfo {pages} {1} (\bibinfo {year} {2023})},\ \bibinfo {note} {number: 1 Publisher: Nature Publishing Group}\BibitemShut {NoStop}%
\bibitem [{\citenamefont {Tran}\ and\ \citenamefont {Nakajima}(2020)}]{tran_higher-order_2020}%
  \BibitemOpen
  \bibfield  {author} {\bibinfo {author} {\bibfnamefont {Q.~H.}\ \bibnamefont {Tran}}\ and\ \bibinfo {author} {\bibfnamefont {K.}~\bibnamefont {Nakajima}},\ }\href {http://arxiv.org/abs/2006.08999} {\bibinfo {title} {Higher-{Order} {Quantum} {Reservoir} {Computing}}} (\bibinfo {year} {2020}),\ \bibinfo {note} {arXiv:2006.08999 [nlin, physics:quant-ph]}\BibitemShut {NoStop}%
\bibitem [{\citenamefont {Lorenz}(1995)}]{75462}%
  \BibitemOpen
  \bibfield  {author} {\bibinfo {author} {\bibfnamefont {E.}~\bibnamefont {Lorenz}},\ }\emph {\bibinfo {title} {Predictability: a problem partly solved}},\ \href@noop {} {Ph.D. thesis},\ \bibinfo {address} {Shinfield Park, Reading} (\bibinfo {year} {1995})\BibitemShut {NoStop}%
\bibitem [{\citenamefont {Moehlis}\ \emph {et~al.}(2004)\citenamefont {Moehlis}, \citenamefont {Faisst},\ and\ \citenamefont {Eckhardt}}]{moehlis_low-dimensional_2004}%
  \BibitemOpen
  \bibfield  {author} {\bibinfo {author} {\bibfnamefont {J.}~\bibnamefont {Moehlis}}, \bibinfo {author} {\bibfnamefont {H.}~\bibnamefont {Faisst}},\ and\ \bibinfo {author} {\bibfnamefont {B.}~\bibnamefont {Eckhardt}},\ }\bibfield  {title} {\bibinfo {title} {A low-dimensional model for turbulent shear flows},\ }\href {https://doi.org/10.1088/1367-2630/6/1/056} {\bibfield  {journal} {\bibinfo  {journal} {New Journal of Physics}\ }\textbf {\bibinfo {volume} {6}},\ \bibinfo {pages} {56} (\bibinfo {year} {2004})}\BibitemShut {NoStop}%
\bibitem [{\citenamefont {Boffetta}\ \emph {et~al.}(2002)\citenamefont {Boffetta}, \citenamefont {Cencini}, \citenamefont {Falcioni},\ and\ \citenamefont {Vulpiani}}]{BOFFETTA2002367}%
  \BibitemOpen
  \bibfield  {author} {\bibinfo {author} {\bibfnamefont {G.}~\bibnamefont {Boffetta}}, \bibinfo {author} {\bibfnamefont {M.}~\bibnamefont {Cencini}}, \bibinfo {author} {\bibfnamefont {M.}~\bibnamefont {Falcioni}},\ and\ \bibinfo {author} {\bibfnamefont {A.}~\bibnamefont {Vulpiani}},\ }\bibfield  {title} {\bibinfo {title} {Predictability: a way to characterize complexity},\ }\href {https://doi.org/https://doi.org/10.1016/S0370-1573(01)00025-4} {\bibfield  {journal} {\bibinfo  {journal} {Physics Reports}\ }\textbf {\bibinfo {volume} {356}},\ \bibinfo {pages} {367} (\bibinfo {year} {2002})}\BibitemShut {NoStop}%
\bibitem [{\citenamefont {lukosevicius}(2012)}]{lukosevicius_practical_2012}%
  \BibitemOpen
  \bibfield  {author} {\bibinfo {author} {\bibfnamefont {M.}~\bibnamefont {lukosevicius}},\ }\bibfield  {title} {\bibinfo {title} {A {Practical} {Guide} to {Applying} {Echo} {State} {Networks}},\ }in\ \href {https://doi.org/10.1007/978-3-642-35289-8_36} {\emph {\bibinfo {booktitle} {Neural Networks: Tricks of the trade, Second Edition}}},\ \bibinfo {series and number} {Lecture {Notes} in {Computer} {Science}},\ \bibinfo {editor} {edited by\ \bibinfo {editor} {\bibfnamefont {G.}~\bibnamefont {Montavon}}, \bibinfo {editor} {\bibfnamefont {G.~B.}\ \bibnamefont {Orr}},\ and\ \bibinfo {editor} {\bibfnamefont {K.-R.}\ \bibnamefont {Muller}}}\ (\bibinfo  {publisher} {Springer},\ \bibinfo {address} {Berlin, Heidelberg},\ \bibinfo {year} {2012})\ pp.\ \bibinfo {pages} {659--686}\BibitemShut {NoStop}%
\bibitem [{\citenamefont {Huhn}\ and\ \citenamefont {Magri}(2020{\natexlab{b}})}]{10.1007/978-3-030-50433-5_10}%
  \BibitemOpen
  \bibfield  {author} {\bibinfo {author} {\bibfnamefont {F.}~\bibnamefont {Huhn}}\ and\ \bibinfo {author} {\bibfnamefont {L.}~\bibnamefont {Magri}},\ }\bibfield  {title} {\bibinfo {title} {Learning ergodic averages in chaotic systems},\ }in\ \href@noop {} {\emph {\bibinfo {booktitle} {Computational Science -- ICCS 2020}}},\ \bibinfo {editor} {edited by\ \bibinfo {editor} {\bibfnamefont {V.~V.}\ \bibnamefont {Krzhizhanovskaya}}, \bibinfo {editor} {\bibfnamefont {G.}~\bibnamefont {Z{\'a}vodszky}}, \bibinfo {editor} {\bibfnamefont {M.~H.}\ \bibnamefont {Lees}}, \bibinfo {editor} {\bibfnamefont {J.~J.}\ \bibnamefont {Dongarra}}, \bibinfo {editor} {\bibfnamefont {P.~M.~A.}\ \bibnamefont {Sloot}}, \bibinfo {editor} {\bibfnamefont {S.}~\bibnamefont {Brissos}},\ and\ \bibinfo {editor} {\bibfnamefont {J.}~\bibnamefont {Teixeira}}}\ (\bibinfo  {publisher} {Springer International Publishing},\ \bibinfo {address} {Cham},\ \bibinfo {year} {2020})\ pp.\ \bibinfo {pages} {124--132}\BibitemShut {NoStop}%
\bibitem [{\citenamefont {Herteux}\ and\ \citenamefont {R{\"a}th}(2020)}]{herteux2020breaking}%
  \BibitemOpen
  \bibfield  {author} {\bibinfo {author} {\bibfnamefont {J.}~\bibnamefont {Herteux}}\ and\ \bibinfo {author} {\bibfnamefont {C.}~\bibnamefont {R{\"a}th}},\ }\bibfield  {title} {\bibinfo {title} {Breaking symmetries of the reservoir equations in echo state networks},\ }\href@noop {} {\bibfield  {journal} {\bibinfo  {journal} {Chaos: An Interdisciplinary Journal of Nonlinear Science}\ }\textbf {\bibinfo {volume} {30}} (\bibinfo {year} {2020})}\BibitemShut {NoStop}%
\bibitem [{\citenamefont {Snoek}\ \emph {et~al.}(2012)\citenamefont {Snoek}, \citenamefont {Larochelle},\ and\ \citenamefont {Adams}}]{snoek2012practical}%
  \BibitemOpen
  \bibfield  {author} {\bibinfo {author} {\bibfnamefont {J.}~\bibnamefont {Snoek}}, \bibinfo {author} {\bibfnamefont {H.}~\bibnamefont {Larochelle}},\ and\ \bibinfo {author} {\bibfnamefont {R.~P.}\ \bibnamefont {Adams}},\ }\bibfield  {title} {\bibinfo {title} {Practical bayesian optimization of machine learning algorithms},\ }in\ \href {https://proceedings.neurips.cc/paper_files/paper/2012/file/05311655a15b75fab86956663e1819cd-Paper.pdf} {\emph {\bibinfo {booktitle} {Advances in Neural Information Processing Systems}}},\ Vol.~\bibinfo {volume} {25},\ \bibinfo {editor} {edited by\ \bibinfo {editor} {\bibfnamefont {F.}~\bibnamefont {Pereira}}, \bibinfo {editor} {\bibfnamefont {C.}~\bibnamefont {Burges}}, \bibinfo {editor} {\bibfnamefont {L.}~\bibnamefont {Bottou}},\ and\ \bibinfo {editor} {\bibfnamefont {K.}~\bibnamefont {Weinberger}}}\ (\bibinfo  {publisher} {Curran Associates, Inc.},\ \bibinfo {year} {2012})\BibitemShut {NoStop}%
\bibitem [{\citenamefont {Pedregosa}\ \emph {et~al.}(2011)\citenamefont {Pedregosa}, \citenamefont {Varoquaux}, \citenamefont {Gramfort}, \citenamefont {Michel}, \citenamefont {Thirion}, \citenamefont {Grisel}, \citenamefont {Blondel}, \citenamefont {Prettenhofer}, \citenamefont {Weiss}, \citenamefont {Dubourg}, \citenamefont {Vanderplas}, \citenamefont {Passos}, \citenamefont {Cournapeau}, \citenamefont {Brucher}, \citenamefont {Perrot},\ and\ \citenamefont {Duchesnay}}]{scikit-learn}%
  \BibitemOpen
  \bibfield  {author} {\bibinfo {author} {\bibfnamefont {F.}~\bibnamefont {Pedregosa}}, \bibinfo {author} {\bibfnamefont {G.}~\bibnamefont {Varoquaux}}, \bibinfo {author} {\bibfnamefont {A.}~\bibnamefont {Gramfort}}, \bibinfo {author} {\bibfnamefont {V.}~\bibnamefont {Michel}}, \bibinfo {author} {\bibfnamefont {B.}~\bibnamefont {Thirion}}, \bibinfo {author} {\bibfnamefont {O.}~\bibnamefont {Grisel}}, \bibinfo {author} {\bibfnamefont {M.}~\bibnamefont {Blondel}}, \bibinfo {author} {\bibfnamefont {P.}~\bibnamefont {Prettenhofer}}, \bibinfo {author} {\bibfnamefont {R.}~\bibnamefont {Weiss}}, \bibinfo {author} {\bibfnamefont {V.}~\bibnamefont {Dubourg}}, \bibinfo {author} {\bibfnamefont {J.}~\bibnamefont {Vanderplas}}, \bibinfo {author} {\bibfnamefont {A.}~\bibnamefont {Passos}}, \bibinfo {author} {\bibfnamefont {D.}~\bibnamefont {Cournapeau}}, \bibinfo {author} {\bibfnamefont {M.}~\bibnamefont {Brucher}}, \bibinfo {author} {\bibfnamefont {M.}~\bibnamefont {Perrot}},\ and\ \bibinfo {author} {\bibfnamefont
  {E.}~\bibnamefont {Duchesnay}},\ }\bibfield  {title} {\bibinfo {title} {Scikit-learn: Machine learning in python},\ }\href@noop {} {\bibfield  {journal} {\bibinfo  {journal} {J. Mach. Learn. Res.}\ }\textbf {\bibinfo {volume} {12}},\ \bibinfo {pages} {2825–2830} (\bibinfo {year} {2011})}\BibitemShut {NoStop}%
\bibitem [{\citenamefont {Nielsen}\ and\ \citenamefont {Chuang}(2011)}]{9781107002173}%
  \BibitemOpen
  \bibfield  {author} {\bibinfo {author} {\bibfnamefont {M.~A.}\ \bibnamefont {Nielsen}}\ and\ \bibinfo {author} {\bibfnamefont {I.~L.}\ \bibnamefont {Chuang}},\ }\href@noop {} {\emph {\bibinfo {title} {Quantum Computation and Quantum Information: 10th Anniversary Edition}}}\ (\bibinfo  {publisher} {Cambridge University Press},\ \bibinfo {year} {2011})\BibitemShut {NoStop}%
\bibitem [{\citenamefont {G\"otting}\ \emph {et~al.}(2023)\citenamefont {G\"otting}, \citenamefont {Lohof},\ and\ \citenamefont {Gies}}]{PhysRevA.108.052427}%
  \BibitemOpen
  \bibfield  {author} {\bibinfo {author} {\bibfnamefont {N.}~\bibnamefont {G\"otting}}, \bibinfo {author} {\bibfnamefont {F.}~\bibnamefont {Lohof}},\ and\ \bibinfo {author} {\bibfnamefont {C.}~\bibnamefont {Gies}},\ }\bibfield  {title} {\bibinfo {title} {Exploring quantumness in quantum reservoir computing},\ }\href {https://doi.org/10.1103/PhysRevA.108.052427} {\bibfield  {journal} {\bibinfo  {journal} {Phys. Rev. A}\ }\textbf {\bibinfo {volume} {108}},\ \bibinfo {pages} {052427} (\bibinfo {year} {2023})}\BibitemShut {NoStop}%
\bibitem [{\citenamefont {Pfeffer}\ \emph {et~al.}(2023)\citenamefont {Pfeffer}, \citenamefont {Heyder},\ and\ \citenamefont {Schumacher}}]{pfeffer_reduced-order_2023}%
  \BibitemOpen
  \bibfield  {author} {\bibinfo {author} {\bibfnamefont {P.}~\bibnamefont {Pfeffer}}, \bibinfo {author} {\bibfnamefont {F.}~\bibnamefont {Heyder}},\ and\ \bibinfo {author} {\bibfnamefont {J.}~\bibnamefont {Schumacher}},\ }\bibfield  {title} {\bibinfo {title} {Reduced-order modeling of two-dimensional turbulent rayleigh-b{\'e}nard flow by hybrid quantum-classical reservoir computing},\ }\href@noop {} {\bibfield  {journal} {\bibinfo  {journal} {Physical Review Research}\ }\textbf {\bibinfo {volume} {5}},\ \bibinfo {pages} {043242} (\bibinfo {year} {2023})}\BibitemShut {NoStop}%
\bibitem [{\citenamefont {Havlíček}\ \emph {et~al.}(2019)\citenamefont {Havlíček}, \citenamefont {Córcoles}, \citenamefont {Temme}, \citenamefont {Harrow}, \citenamefont {Kandala}, \citenamefont {Chow},\ and\ \citenamefont {Gambetta}}]{havlicek_supervised_2019}%
  \BibitemOpen
  \bibfield  {author} {\bibinfo {author} {\bibfnamefont {V.}~\bibnamefont {Havlíček}}, \bibinfo {author} {\bibfnamefont {A.~D.}\ \bibnamefont {Córcoles}}, \bibinfo {author} {\bibfnamefont {K.}~\bibnamefont {Temme}}, \bibinfo {author} {\bibfnamefont {A.~W.}\ \bibnamefont {Harrow}}, \bibinfo {author} {\bibfnamefont {A.}~\bibnamefont {Kandala}}, \bibinfo {author} {\bibfnamefont {J.~M.}\ \bibnamefont {Chow}},\ and\ \bibinfo {author} {\bibfnamefont {J.~M.}\ \bibnamefont {Gambetta}},\ }\bibfield  {title} {\bibinfo {title} {Supervised learning with quantum-enhanced feature spaces},\ }\href {https://doi.org/10.1038/s41586-019-0980-2} {\bibfield  {journal} {\bibinfo  {journal} {Nature}\ }\textbf {\bibinfo {volume} {567}},\ \bibinfo {pages} {209} (\bibinfo {year} {2019})},\ \bibinfo {note} {number: 7747 Publisher: Nature Publishing Group}\BibitemShut {NoStop}%
\bibitem [{\citenamefont {Abbas}\ \emph {et~al.}(2021)\citenamefont {Abbas}, \citenamefont {Sutter}, \citenamefont {Zoufal}, \citenamefont {Lucchi}, \citenamefont {Figalli},\ and\ \citenamefont {Woerner}}]{abbas_power_2021}%
  \BibitemOpen
  \bibfield  {author} {\bibinfo {author} {\bibfnamefont {A.}~\bibnamefont {Abbas}}, \bibinfo {author} {\bibfnamefont {D.}~\bibnamefont {Sutter}}, \bibinfo {author} {\bibfnamefont {C.}~\bibnamefont {Zoufal}}, \bibinfo {author} {\bibfnamefont {A.}~\bibnamefont {Lucchi}}, \bibinfo {author} {\bibfnamefont {A.}~\bibnamefont {Figalli}},\ and\ \bibinfo {author} {\bibfnamefont {S.}~\bibnamefont {Woerner}},\ }\bibfield  {title} {\bibinfo {title} {The power of quantum neural networks},\ }\href {https://doi.org/10.1038/s43588-021-00084-1} {\bibfield  {journal} {\bibinfo  {journal} {Nature Computational Science}\ }\textbf {\bibinfo {volume} {1}},\ \bibinfo {pages} {403} (\bibinfo {year} {2021})},\ \bibinfo {note} {number: 6 Publisher: Nature Publishing Group}\BibitemShut {NoStop}%
\bibitem [{\citenamefont {Mujal}\ \emph {et~al.}(2021)\citenamefont {Mujal}, \citenamefont {Mart{\'\i}nez-Pe{\~n}a}, \citenamefont {Nokkala}, \citenamefont {Garc{\'\i}a-Beni}, \citenamefont {Giorgi}, \citenamefont {Soriano},\ and\ \citenamefont {Zambrini}}]{mujal_opportunities_2021}%
  \BibitemOpen
  \bibfield  {author} {\bibinfo {author} {\bibfnamefont {P.}~\bibnamefont {Mujal}}, \bibinfo {author} {\bibfnamefont {R.}~\bibnamefont {Mart{\'\i}nez-Pe{\~n}a}}, \bibinfo {author} {\bibfnamefont {J.}~\bibnamefont {Nokkala}}, \bibinfo {author} {\bibfnamefont {J.}~\bibnamefont {Garc{\'\i}a-Beni}}, \bibinfo {author} {\bibfnamefont {G.~L.}\ \bibnamefont {Giorgi}}, \bibinfo {author} {\bibfnamefont {M.~C.}\ \bibnamefont {Soriano}},\ and\ \bibinfo {author} {\bibfnamefont {R.}~\bibnamefont {Zambrini}},\ }\bibfield  {title} {\bibinfo {title} {Opportunities in quantum reservoir computing and extreme learning machines},\ }\href@noop {} {\bibfield  {journal} {\bibinfo  {journal} {Advanced Quantum Technologies}\ }\textbf {\bibinfo {volume} {4}},\ \bibinfo {pages} {2100027} (\bibinfo {year} {2021})}\BibitemShut {NoStop}%
\bibitem [{\citenamefont {Jaeger}\ \emph {et~al.}(2007)\citenamefont {Jaeger}, \citenamefont {Mantas}, \citenamefont {Popovici},\ and\ \citenamefont {Siewert}}]{JAEGER2007335}%
  \BibitemOpen
  \bibfield  {author} {\bibinfo {author} {\bibfnamefont {H.}~\bibnamefont {Jaeger}}, \bibinfo {author} {\bibnamefont {Mantas}}, \bibinfo {author} {\bibfnamefont {D.}~\bibnamefont {Popovici}},\ and\ \bibinfo {author} {\bibfnamefont {U.}~\bibnamefont {Siewert}},\ }\bibfield  {title} {\bibinfo {title} {Optimization and applications of echo state networks with leaky- integrator neurons},\ }\href {https://doi.org/https://doi.org/10.1016/j.neunet.2007.04.016} {\bibfield  {journal} {\bibinfo  {journal} {Neural Networks}\ }\textbf {\bibinfo {volume} {20}},\ \bibinfo {pages} {335} (\bibinfo {year} {2007})},\ \bibinfo {note} {echo State Networks and Liquid State Machines}\BibitemShut {NoStop}%
\bibitem [{\citenamefont {{Qiskit contributors}}(2023)}]{Qiskit}%
  \BibitemOpen
  \bibfield  {author} {\bibinfo {author} {\bibnamefont {{Qiskit contributors}}},\ }\href {https://doi.org/10.5281/zenodo.2573505} {\bibinfo {title} {Qiskit: An open-source framework for quantum computing}} (\bibinfo {year} {2023})\BibitemShut {NoStop}%
\bibitem [{\citenamefont {Mujal}\ \emph {et~al.}(2023)\citenamefont {Mujal}, \citenamefont {Martínez-Peña}, \citenamefont {Giorgi}, \citenamefont {Soriano},\ and\ \citenamefont {Zambrini}}]{mujal_time-series_2023}%
  \BibitemOpen
  \bibfield  {author} {\bibinfo {author} {\bibfnamefont {P.}~\bibnamefont {Mujal}}, \bibinfo {author} {\bibfnamefont {R.}~\bibnamefont {Martínez-Peña}}, \bibinfo {author} {\bibfnamefont {G.~L.}\ \bibnamefont {Giorgi}}, \bibinfo {author} {\bibfnamefont {M.~C.}\ \bibnamefont {Soriano}},\ and\ \bibinfo {author} {\bibfnamefont {R.}~\bibnamefont {Zambrini}},\ }\bibfield  {title} {\bibinfo {title} {Time-series quantum reservoir computing with weak and projective measurements},\ }\href {https://doi.org/10.1038/s41534-023-00682-z} {\bibfield  {journal} {\bibinfo  {journal} {npj Quantum Information}\ }\textbf {\bibinfo {volume} {9}},\ \bibinfo {pages} {16} (\bibinfo {year} {2023})}\BibitemShut {NoStop}%
\bibitem [{\citenamefont {Huang}\ \emph {et~al.}(2020)\citenamefont {Huang}, \citenamefont {Kueng},\ and\ \citenamefont {Preskill}}]{huang2020predicting}%
  \BibitemOpen
  \bibfield  {author} {\bibinfo {author} {\bibfnamefont {H.-Y.}\ \bibnamefont {Huang}}, \bibinfo {author} {\bibfnamefont {R.}~\bibnamefont {Kueng}},\ and\ \bibinfo {author} {\bibfnamefont {J.}~\bibnamefont {Preskill}},\ }\bibfield  {title} {\bibinfo {title} {Predicting many properties of a quantum system from very few measurements},\ }\href@noop {} {\bibfield  {journal} {\bibinfo  {journal} {Nature Physics}\ }\textbf {\bibinfo {volume} {16}},\ \bibinfo {pages} {1050} (\bibinfo {year} {2020})}\BibitemShut {NoStop}%
\bibitem [{\citenamefont {Hu}\ \emph {et~al.}(2023)\citenamefont {Hu}, \citenamefont {Angelatos}, \citenamefont {Khan}, \citenamefont {Vives}, \citenamefont {T{\"u}reci}, \citenamefont {Bello}, \citenamefont {Rowlands}, \citenamefont {Ribeill},\ and\ \citenamefont {T{\"u}reci}}]{hu2023tackling}%
  \BibitemOpen
  \bibfield  {author} {\bibinfo {author} {\bibfnamefont {F.}~\bibnamefont {Hu}}, \bibinfo {author} {\bibfnamefont {G.}~\bibnamefont {Angelatos}}, \bibinfo {author} {\bibfnamefont {S.~A.}\ \bibnamefont {Khan}}, \bibinfo {author} {\bibfnamefont {M.}~\bibnamefont {Vives}}, \bibinfo {author} {\bibfnamefont {E.}~\bibnamefont {T{\"u}reci}}, \bibinfo {author} {\bibfnamefont {L.}~\bibnamefont {Bello}}, \bibinfo {author} {\bibfnamefont {G.~E.}\ \bibnamefont {Rowlands}}, \bibinfo {author} {\bibfnamefont {G.~J.}\ \bibnamefont {Ribeill}},\ and\ \bibinfo {author} {\bibfnamefont {H.~E.}\ \bibnamefont {T{\"u}reci}},\ }\bibfield  {title} {\bibinfo {title} {Tackling sampling noise in physical systems for machine learning applications: Fundamental limits and eigentasks},\ }\href@noop {} {\bibfield  {journal} {\bibinfo  {journal} {Physical Review X}\ }\textbf {\bibinfo {volume} {13}},\ \bibinfo {pages} {041020} (\bibinfo {year} {2023})}\BibitemShut {NoStop}%
\bibitem [{\citenamefont {{\v{C}}indrak}\ \emph {et~al.}(2024)\citenamefont {{\v{C}}indrak}, \citenamefont {Donvil}, \citenamefont {L{\"u}dge},\ and\ \citenamefont {Jaurigue}}]{vcindrak2024enhancing}%
  \BibitemOpen
  \bibfield  {author} {\bibinfo {author} {\bibfnamefont {S.}~\bibnamefont {{\v{C}}indrak}}, \bibinfo {author} {\bibfnamefont {B.}~\bibnamefont {Donvil}}, \bibinfo {author} {\bibfnamefont {K.}~\bibnamefont {L{\"u}dge}},\ and\ \bibinfo {author} {\bibfnamefont {L.}~\bibnamefont {Jaurigue}},\ }\bibfield  {title} {\bibinfo {title} {Enhancing the performance of quantum reservoir computing and solving the time-complexity problem by artificial memory restriction},\ }\href@noop {} {\bibfield  {journal} {\bibinfo  {journal} {Physical Review Research}\ }\textbf {\bibinfo {volume} {6}},\ \bibinfo {pages} {013051} (\bibinfo {year} {2024})}\BibitemShut {NoStop}%
\bibitem [{\citenamefont {Pathak}\ \emph {et~al.}(2018)\citenamefont {Pathak}, \citenamefont {Wikner}, \citenamefont {Fussell}, \citenamefont {Chandra}, \citenamefont {Hunt}, \citenamefont {Girvan},\ and\ \citenamefont {Ott}}]{10.1063/1.5028373}%
  \BibitemOpen
  \bibfield  {author} {\bibinfo {author} {\bibfnamefont {J.}~\bibnamefont {Pathak}}, \bibinfo {author} {\bibfnamefont {A.}~\bibnamefont {Wikner}}, \bibinfo {author} {\bibfnamefont {R.}~\bibnamefont {Fussell}}, \bibinfo {author} {\bibfnamefont {S.}~\bibnamefont {Chandra}}, \bibinfo {author} {\bibfnamefont {B.~R.}\ \bibnamefont {Hunt}}, \bibinfo {author} {\bibfnamefont {M.}~\bibnamefont {Girvan}},\ and\ \bibinfo {author} {\bibfnamefont {E.}~\bibnamefont {Ott}},\ }\bibfield  {title} {\bibinfo {title} {{Hybrid forecasting of chaotic processes: Using machine learning in conjunction with a knowledge-based model}},\ }\href {https://doi.org/10.1063/1.5028373} {\bibfield  {journal} {\bibinfo  {journal} {Chaos: An Interdisciplinary Journal of Nonlinear Science}\ }\textbf {\bibinfo {volume} {28}},\ \bibinfo {pages} {041101} (\bibinfo {year} {2018})}\BibitemShut {NoStop}%
\bibitem [{\citenamefont {Goutte}\ and\ \citenamefont {Gaussier}(2005)}]{10.1007/978-3-540-31865-1_25}%
  \BibitemOpen
  \bibfield  {author} {\bibinfo {author} {\bibfnamefont {C.}~\bibnamefont {Goutte}}\ and\ \bibinfo {author} {\bibfnamefont {E.}~\bibnamefont {Gaussier}},\ }\bibfield  {title} {\bibinfo {title} {A probabilistic interpretation of precision, recall and f-score, with implication for evaluation},\ }in\ \href@noop {} {\emph {\bibinfo {booktitle} {Advances in Information Retrieval}}},\ \bibinfo {editor} {edited by\ \bibinfo {editor} {\bibfnamefont {D.~E.}\ \bibnamefont {Losada}}\ and\ \bibinfo {editor} {\bibfnamefont {J.~M.}\ \bibnamefont {Fernandez-Luna}}}\ (\bibinfo  {publisher} {Springer Berlin Heidelberg},\ \bibinfo {address} {Berlin, Heidelberg},\ \bibinfo {year} {2005})\ pp.\ \bibinfo {pages} {345--359}\BibitemShut {NoStop}%
\bibitem [{\citenamefont {Boyd}\ and\ \citenamefont {Chua}(1985)}]{boyd_fading_1985}%
  \BibitemOpen
  \bibfield  {author} {\bibinfo {author} {\bibfnamefont {S.}~\bibnamefont {Boyd}}\ and\ \bibinfo {author} {\bibfnamefont {L.}~\bibnamefont {Chua}},\ }\bibfield  {title} {\bibinfo {title} {Fading memory and the problem of approximating nonlinear operators with {Volterra} series},\ }\href {https://doi.org/10.1109/TCS.1985.1085649} {\bibfield  {journal} {\bibinfo  {journal} {IEEE Transactions on Circuits and Systems}\ }\textbf {\bibinfo {volume} {32}},\ \bibinfo {pages} {1150} (\bibinfo {year} {1985})},\ \bibinfo {note} {conference Name: IEEE Transactions on Circuits and Systems}\BibitemShut {NoStop}%
\bibitem [{\citenamefont {Carroll}(2022)}]{carroll_optimizing_2022}%
  \BibitemOpen
  \bibfield  {author} {\bibinfo {author} {\bibfnamefont {T.~L.}\ \bibnamefont {Carroll}},\ }\bibfield  {title} {\bibinfo {title} {Optimizing {Memory} in {Reservoir} {Computers}},\ }\href {https://doi.org/10.1063/5.0078151} {\bibfield  {journal} {\bibinfo  {journal} {Chaos: An Interdisciplinary Journal of Nonlinear Science}\ }\textbf {\bibinfo {volume} {32}},\ \bibinfo {pages} {023123} (\bibinfo {year} {2022})},\ \bibinfo {note} {arXiv:2201.01605 [cs]}\BibitemShut {NoStop}%
\bibitem [{\citenamefont {Farkaš}\ \emph {et~al.}(2016)\citenamefont {Farkaš}, \citenamefont {Bosák},\ and\ \citenamefont {Gergeľ}}]{farkas_computational_2016}%
  \BibitemOpen
  \bibfield  {author} {\bibinfo {author} {\bibfnamefont {I.}~\bibnamefont {Farkaš}}, \bibinfo {author} {\bibfnamefont {R.}~\bibnamefont {Bosák}},\ and\ \bibinfo {author} {\bibfnamefont {P.}~\bibnamefont {Gergeľ}},\ }\bibfield  {title} {\bibinfo {title} {Computational analysis of memory capacity in echo state networks},\ }\href {https://doi.org/10.1016/j.neunet.2016.07.012} {\bibfield  {journal} {\bibinfo  {journal} {Neural Networks}\ }\textbf {\bibinfo {volume} {83}},\ \bibinfo {pages} {109} (\bibinfo {year} {2016})}\BibitemShut {NoStop}%
\bibitem [{\citenamefont {Han}\ \emph {et~al.}(2021)\citenamefont {Han}, \citenamefont {Zhao},\ and\ \citenamefont {Small}}]{han2021revisiting}%
  \BibitemOpen
  \bibfield  {author} {\bibinfo {author} {\bibfnamefont {X.}~\bibnamefont {Han}}, \bibinfo {author} {\bibfnamefont {Y.}~\bibnamefont {Zhao}},\ and\ \bibinfo {author} {\bibfnamefont {M.}~\bibnamefont {Small}},\ }\bibfield  {title} {\bibinfo {title} {Revisiting the memory capacity in reservoir computing of directed acyclic network},\ }\href@noop {} {\bibfield  {journal} {\bibinfo  {journal} {Chaos: An Interdisciplinary Journal of Nonlinear Science}\ }\textbf {\bibinfo {volume} {31}} (\bibinfo {year} {2021})}\BibitemShut {NoStop}%
\bibitem [{\citenamefont {Jaeger}(2001{\natexlab{b}})}]{Jaeger_2001}%
  \BibitemOpen
  \bibfield  {author} {\bibinfo {author} {\bibfnamefont {H.}~\bibnamefont {Jaeger}},\ }\href {https://doi.org/10.24406/publica-fhg-291107} {\emph {\bibinfo {title} {Short term memory in echo state networks}}}\ (\bibinfo  {publisher} {GMD Forschungszentrum Informationstechnik},\ \bibinfo {year} {2001})\BibitemShut {NoStop}%
\bibitem [{\citenamefont {Storm}\ \emph {et~al.}(2022)\citenamefont {Storm}, \citenamefont {Gustavsson},\ and\ \citenamefont {Mehlig}}]{storm_constraints_2022}%
  \BibitemOpen
  \bibfield  {author} {\bibinfo {author} {\bibfnamefont {L.}~\bibnamefont {Storm}}, \bibinfo {author} {\bibfnamefont {K.}~\bibnamefont {Gustavsson}},\ and\ \bibinfo {author} {\bibfnamefont {B.}~\bibnamefont {Mehlig}},\ }\bibfield  {title} {\bibinfo {title} {Constraints on parameter choices for successful time-series prediction with echo-state networks},\ }\href {https://doi.org/10.1088/2632-2153/aca1f6} {\bibfield  {journal} {\bibinfo  {journal} {Machine Learning: Science and Technology}\ }\textbf {\bibinfo {volume} {3}},\ \bibinfo {pages} {045021} (\bibinfo {year} {2022})},\ \bibinfo {note} {publisher: IOP Publishing}\BibitemShut {NoStop}%
\end{thebibliography}%

\end{document}